\documentclass{aa}
\usepackage{graphicx}
\usepackage{txfonts}
\usepackage{color}
\begin{document}

   \title{Unsupervised feature-learning for galaxy SEDs with denoising autoencoders}

   \author{J. Frontera-Pons\inst{1}, F. Sureau\inst{1}, J. Bobin\inst{1}, E. Le Floc'h\inst{1}}

   \institute{IRFU - Service d'Astrophysique/SEDI, CEA Saclay, 91191 Gif-sur-Yvette, France.\\
   \email{joana.frontera-pons@cea.fr}}

   %\date{Received September 15, 1996; accepted March 16, 1997}

\abstract{With the increasing number of deep multi-wavelength galaxy surveys, the spectral energy distribution (SED) of galaxies has become an invaluable tool for studying the formation of their structures and their evolution. In this context, standard analysis relies on simple spectro-photometric selection criteria based on a few SED colors. If this fully supervised classification already yielded clear achievements, it is not optimal to extract relevant information from the data. In this article, we propose to employ very recent advances in machine learning, and more precisely in feature learning, to derive a data-driven diagram. We show that the proposed approach based on denoising autoencoders recovers the bi-modality in the galaxy population in an unsupervised manner, without using any prior knowledge on galaxy SED classification. This technique has been compared to principal component analysis (PCA) and to standard color/color representations. In addition, preliminary results illustrate that this enables the capturing of extra physically meaningful information, such as redshift dependence, galaxy mass evolution and variation over the specific star formation rate. PCA also results in an unsupervised representation with physical properties, such as mass and sSFR, although this representation separates out. less other characteristics (bimodality, redshift evolution) than denoising autoencoders.}
  
\keywords{methods : unsupervised feature learning /  methods: denoising autoencoders / galaxies : classification}

\maketitle

%-------------------------------------------------------------------

\section{Introduction}

Over the last decade, the increasing number of large and deep multi-wavelength galaxy surveys has led to a dramatic improvement of our understanding of structure formation  and galaxy evolution. In particular, deep observations at optical and near-IR wavelengths have enabled the probe of galaxy stellar emission with unprecedented details across a wide range of stellar mass, galaxy environment, and cosmic epochs, bringing more and more stringent constraints on the build-up of baryonic matter throughout cosmic history \citep{Ilbert13,Madau14,Grazian15}.
However, the vast majority of galaxy studies carried out so far have mostly relied on supervised techniques, where the galaxy emission throughout the electro-magnetic spectrum is analyzed and interpreted based on an {a priori} knowledge of the range of properties that extragalactic sources can exhibit. Furthermore,  galaxies  in the distant Universe are very often identified and  classified using simplistic spectro-photometric criteria, like standard color selections that consider the variations of the galaxy spectral energy distribution (SED) between two (or three at most) different wavelengths \citep[{e.g.}][]{Franx03,Daddi04,Adelberger04}. Although the largest possible range of frequencies is used in the physical interpretation of the SEDs,
% when constraining their spectral energy distributions and their physical properties, 
most of the available information is therefore ignored in the very first step, i.e., in  the definition of  the different galaxy populations itself. 

 As an illustration, a major result of modern extragalactic astrophysics has been the discovery of a very clear bimodality in the distribution of galaxy properties and galaxy colors, which emerged soon after the first billion years after Big Bang \citep{Baldry04,Brammer09,Whitaker11}: on the one hand, star-forming galaxies with a dominance of blue and young massive stars mostly show  blue optical colors; on the other hand, passive galaxies where star formation has been quenched  display much redder colors owing to the dominant contribution of more evolved stellar populations. This finding has set the basis of numerous investigations of galaxy evolution throughout cosmic time, with the goal of constraining the properties of these two populations as a function of redshift, and determining the  different processes triggering and quenching star formation in galaxies \citep{Faber07,Ilbert13}.  However, this distinction of galaxy classes, which is mostly based on  optical colors,  only offers a very reductive description of all the information encoded in galaxy SEDs. %Besides, it can be severely impacted by degeneracies among key physical parameters, like the galaxy stellar age and the effect of dust extinction which both makes galaxies redder.

Going further therefore requires employing unsupervised learning methods that can efficiently extract relevant information from the full data. To the best of our knowledge, unsupervised learning for galaxy SEDs classification has mainly focused on employing the principal component analysis (PCA) method. In \cite{Wild14}, PCA is mainly been used as a dimensionality reduction technique to extract a low number of so-called super-colors. In both  \cite{Wild14} and \cite{Chen12}, PCA is performed on a large dataset of simulated galaxy SEDs, which enables a direct connection between the model parameters and the estimated principal components. In these articles, super-sampled SEDs are used to derive the principal components. The goal of the present paper is more challenging: i) unsupervised learning has to be performed on galaxy rest-frame SEDs, which are derived from observed data ii) unsupervised learning  also needs to be applied to photometric data, which are made of a low number of broad bands ({e.g.}, $11$ bands in the next). Moreover, the PCA performs a linear decomposition of the data, which is not well adapted to capture physical characteristics ({e.g.}, galaxy mass, age), which are more likely related to the observed SED in a more intricate way.

Very recent advances in machine learning, such as deep learning methods, e.g., \cite{hinton2006fast}, introduced highly sophisticated data analysis tools that are promising means of building unsupervised data-driven alternatives to SED color diagrams. These statistical methods (see for {e.g.}, \cite{larochelle2007empirical, bengio2009learning, szegedy2013intriguing} and references therein) have already proved their efficiency in solving supervised data classification tasks in applications as diverse as computer vision ({e.g.}, \cite{krizhevsky2012imagenet, ciresan2012multi}), speech recognition ({e.g. }\cite{hinton2012deep, dahl2012context}), natural language processing ({e.g.}, \cite{collobert2011natural}), to  name just a few.
Machine learning techniques have recently been advocated as a powerful tool to derive useful features straight from the data. This aspect of learning, called representation learning \cite{bengio2013representation}, is expected to provide an efficient re-parameterization of the data, composed of more salient features. The main interest of the methods detailed here is their ability to extract information in a unsupervised manner. In other words, the aim is to design features that enable us to efficiently unfold complex underlying structures in the data without including prior information or labelled examples.\\
In the sequel, we will make use of denoising autoencoders (DAE) introduced in \cite{vincent2008extracting}, which are built by connecting computational nodes similar to those used by neural networks. However, these two structures are essentially different. Classical neural networks are forward networks that are trained to obtain the outcome for training samples and make predictions for the test samples according to the learnt model. On the other hand, denoising autoencoders are forth and backward networks that attempt to reconstruct the original input vector from a corrupted version of it. Furthermore, the architecture may be constructed by stacking several layers of denoising autoencoders, which leads to more complex and ultimately more useful representations.\\

\paragraph*{{Contribution:}}
In this article, we investigate the use of one recently introduced machine learning method, which has been coined denoising autoencoder, \cite{vincent2008extracting}, for unsupervised feature learning from galaxy SEDs. In the spirit of SED color diagrams, the proposed approach enables  a new galaxy SEDs' representation to be derived. We evaluate how the resulting DAE diagram can recover the standard star-forming/quiescent galaxy bimodality. At the same time, we show that, according to the current understanding of autoencoders, DAE yields a diagram that extracts astrophysically relevant information from the data that standard SED color diagrams do not exhibit. We compare DAE techniques for feature extraction with PCA and standard color-color diagrams. This work therefore illustrates the interest of these methods for galaxy SEDs' representation and paves the way for the design of more sophisticated models, which is beyond the scope of this study.

%--------------------------------------------------------------------
\section{Unsupervised learning with denoising autoencoders}
The goal of this study is to investigate an alternative to SED color diagrams that is based on features learned straight from the data in an unsupervised manner, which enables us to exploit the information contained in all the bands. Amongst available machine learning methods, some are particularly well-suited for  unsupervised representation or feature learning; this is the case of the autoencoders in \cite{vincent2008extracting}.\\
Similar to standard neural networks (NN), autoencoders define a direct mapping function from the inputs to their representations or feature space, which is called encoding. In contrast to NN, it further implements a decoding function that maps back to the original input space. In the autoencoders framework, the parametric maps are specified as a neural network forth and back where the hidden units are treated as computational nodes (see \cite{hinton1989connectionist}). The goal is to create a good task-guided representation of the input by preserving a significant amount of information. This is carried out in the autoencoder learning process by minimizing some error between the inputs and their decoded reconstruction.

\subsection{Autoencoders}
In the following, we will define that a single SED is a vector $\mathbf x = [x_1, \dots, x_m]^T \in \mathbb R^m $, which is made of $m$ rest-frame magnitudes. The dataset $\{\mathbf x_1, \dots \mathbf x_N \}$ is composed of $N$ SED samples. In the framework of a single-layer autoencoder, one starts defining a parametric map that allows to compute the representation vector from the input. This function, usually called \emph{encoder}, $f_{\boldsymbol{\theta}}$, provides the representation as an output:
$\mathbf h_i = f_{\boldsymbol{\theta}} (\mathbf x_i) $,
where $\mathbf h_i \in \mathbb R^{nhid}$ is the feature vector or code, and $nhid$ is the number of hidden units or the dimension of the representation vector. Analogously, we also define another parameterized function, $g_{\boldsymbol{\theta}}$ called \emph{decoder}, that projects from the code space back into the input space, yielding a reconstruction of the input vector:
$\hat{\mathbf x}_i = g_{\boldsymbol{\theta}} (\mathbf h_i) $.\\
Therefore, the encoder and the decoder are completely characterized by the set of parameters ${\boldsymbol{\theta}}$, which are jointly learned to reconstruct as accurately as possible the original input vector. In practice, this is performed by minimizing the reconstruction error, $L(\mathbf x,\hat{\mathbf x})$, which measures the dissimilarity between $\mathbf x_i$ and its reconstruction $\hat{\mathbf x}_i$ over all the samples in the training set. Training a basic autoencoder corresponds to computing a value of the parameter vector ${\boldsymbol{\theta}}$ by minimizing the reconstruction error :
\begin{equation}
 J_{AE} (\boldsymbol{\theta}) = \displaystyle \sum_{i = 1}^N L ( \mathbf x_i, g_{\boldsymbol{\theta}}(f_{\boldsymbol{\theta}}(\mathbf x_i)))  
 .\end{equation}
This minimization is generally carried out by stochastic gradient descent.\\
Similar to standard neural networks, the encoder and decoder usually take the form of affine mappings, typically followed by a non-linearity,
\begin{equation}  f_{\boldsymbol{\theta}} (\mathbf x) = s_f(\mathbf b_f + \mathbf W_f \,\mathbf x) \end{equation} 
\begin{equation} g_{\boldsymbol{\theta}} (\mathbf h) = s_g (\mathbf b_g + \mathbf W_g \,\mathbf h),\end{equation} 
where $s_f$ and $s_g$ are the encoder and decoder activation functions. Common options for the activation functions are the element-wise sigmoid, hyperbolic tangent non-linearity, or identity function if staying linear, among others. Thus, the parameter vector of the model is $\boldsymbol{\theta} = \{ \mathbf W_f, \mathbf b_f, \mathbf W_g, \mathbf b_g\},$ where $\mathbf b_f$ and $\mathbf b_g$ are the bias vectors of the encoder and decoder, respectively, and $\mathbf W_f$ and $\mathbf W_g$ are the encoder and decoder weight matrices. It is customary to resort to weight-tying, in which one defines $\mathbf W_g = \mathbf W_f^T$. This is what will be implemented below. The choice of the decoder activation function $s_g$ and the reconstruction error measure $L(\cdot)$ depends on the input data domain range and nature. They are selected so that $L(\cdot)$ returns a negative log-likelihood for the observed value of $\mathbf x$, for instance through the squared reconstruction error, 
\begin{equation}L(\mathbf x, \hat{\mathbf x}) = ||\mathbf x - \hat{\mathbf x}||^2 \end{equation} 
or the binary cross-entropy, 
\begin{equation} L(\mathbf x,\hat{\mathbf x}) = -\sum _{j = 1} ^ m  x_j \log (\hat{x}_j) + (1 - x_j) \log (1 - \hat{x}_j). \end{equation} 

We note that, with this configuration, basic autoencoders may learn the identity function to minimize the reconstruction error. Therefore, some regularization should be included in the training criterion. Recent studies, like \cite{rifai2011contractive, alain2014regularized}, support the improvement brought by regularized autoencoders that impose some constraint on the code. The purpose of this constraint is to force the representation to be as unchanging as possible with respect to local variations in the input. The denoising autoencoders detailed in the next section make the whole transformation robust and insensitive to small random perturbations in the input, which is highly desireable to robustly capture features from the input data.  

\subsection{Denoising autoencoders}
Denoising autoencoders were proposed in \cite{vincent2008extracting}. The main idea consists on modifying the training objective of basic autoencoders to retrieve a clean input from an artificially corrupted version of it, denoted $\tilde{\mathbf x}$. In this case, the retrieved signal does not correspond to a perfect reconstruction of the input $\mathbf x$, but to the mean of the distribution that may generate $\mathbf x$. Specifically, the training objective to be minimized becomes
\begin{equation}
 J_{DAE} (\boldsymbol{\theta}) = \displaystyle \sum_{i = 1}^N  \mathbb E_{q(\tilde{\mathbf x}| {\mathbf x_i})} [ L ( \mathbf x_i, g_{\boldsymbol{\theta}}(f_{\boldsymbol{\theta}}(\tilde{\mathbf x}))) ],
\end{equation}
where $\mathbb E_{q(\tilde{\mathbf x}| {\mathbf x_i})}[\cdot]$ performs an expectation over all the corrupted samples in the data set pulled out the corruption process $q(\tilde{\mathbf x}| {\mathbf x_i})$. Common forms for the corruptors analyzed in \cite{vincent2008extracting, vincent2010stacked} include additive isotropic Gaussian noise, salt and pepper noise, and masking noise. These methods are general and can be used in most scenarios. However, one could incorporate prior knowledge about the kind of perturbation the data could encounter and make the model robust against it adapting the corruption process. The underlying structure of the data has to be retained by the scheme to undo the effect of the corruption process, {i.e}. to perform denoising. The cost funcion $J_{DAE}$ is again optimized by stochastic gradient descent. \\

\section{The data}
As a  first exploration of how the approach described above can supplement  the more traditional studies on galaxy evolution, we tested the DAE methodology with   a population of galaxies at redshift 0\,$<$\,z\,$<$\,1, characterized by well-constrained rest-frame SEDs at  optical and near-IR wavelengths and taken from 
%here was applied to the galaxy population identified in 
the COSMOS2015 photometric catalog that is publicly released by the COSMOS consortium \citep{Laigle16}.  
COSMOS is a galaxy survey combining deep multi-wavelength observations over a relatively large area of 2 square degrees \citep{Scoville07}, and it has become one of best observed and most commonly used cosmological fields considered for galaxy evolution studies these days. In particular, COSMOS has recently obtained deep  images in the $Y$, $J$, $H,$ and $K_s$ near-IR bands from the UltraVista-DR2 survey, new optical images using the Hyper-Suprime-Cam instrument on the Subaru Telescope, as well as deep observations at 3.6 and 4.5$\mu$m obtained with the {\it Spitzer Space Telescope}. The multi-band photometry inferred from this unique data set enabled the determination of spectral energy distributions (SED) for roughly half a million galaxies over a broad range of redshift (0<z<6),  which eventually led to accurate estimates of key physical parameters, such as their photometric redshifts, absolute magnitudes, stellar masses, and star formation rates. In the COSMOS2015 catalog, all these quantities were determined with standard SED fitting techniques, using a set of appropriate galaxy templates and considering, for each case, the median value of the associated probability distribution functions. 
 
 To minimize possible degeneracies of SED fits over the redshift range covered by the COSMOS galaxy population, photometric redshifts were determined using a limited set of 45 galaxy templates, which were fitted to the observed photometry leaving the redshift as a free parameter. These 45 templates include empirical models of Elliptical and Spiral galaxies, supplemented by  physical models of young blue star-forming sources.  The mean dust extinction value in each galaxy was left as an additional free parameter and three different extinction laws were considered. Comparisons with $>$10 000 spectroscopic redshifts, which were obtained as part of various campaigns of spectroscopic follow-up in the COSMOS field, show that the photometric redshift accuracy reaches $\sigma_{\Delta z / (1+z)} = 0.007$ for bright galaxies ($i^{+}<21$) at z$<$1.5, with a fraction of catastrophic failures of only $\eta=0.5$\%. For fainter sources at higher redshifts (3<z<6), the precision is still impressive with $\sigma_{\Delta z / (1+z)} = 0.021$ and $\eta = 13.2$\% \citep{Laigle16}.
Absolute magnitudes, stellar masses and star formation rates were then computed following the prescriptions of \citet{Ilbert13}, considering this time the best fit template from a much larger library of physical galaxy models and fixing the distance of the galaxy to the photometric redshift derived earlier. These galaxy synthetic spectra were generated using the stellar population synthesis model of \citet{Bruzual03}, assuming a large range of star formation histories ({e.g.}, exponentially-declining ($e^{-t/\tau}$) and delayed ($\tau^{-2} e^{-t/\tau}$) star formation) sampled at different galaxy ages, two different metallicities (solar and half-solar) and two different extinction laws. Because the COSMOS2015 catalog is photometrically selected from the  $Y$, $J$, $H,$ and $K_s$  UltraVista near-IR bands, the galaxy population identified at each redshift  closely follows a selection in stellar mass, with a completeness reaching almost 10$^{9}$\,M$_\odot$ at z\,$\sim$\,1 and 10$^{10}$\,M$_\odot$ at z$\sim$\,4 \citep{Laigle16}. We note that each galaxy in the catalog is also classified as star-forming or passive, depending on its $NUV-r^+$ and $r^+-J$ rest-frame colors \citep{Ilbert13}.
~

To minimize the uncertainty in the derived absolute magnitudes and to better interpret learned features, in this work we focus on a clean subset of this COSMOS dataset. This subset comprises all galaxies in the UltraVista area with observed magnitude $m_{i+}<24$ and with estimated redshift in the $[0,1]$ range. We removed galaxies with redshift uncertainty estimates ${\Delta \hat{z}_{68} / (1+\hat{z})}>0.1$, where $\Delta \hat{z}_{68}$ is the size of the estimated $68\%$ confidence interval around the estimated photometric redshift $\hat{z}$ since this uncertainty would greatly impact the dispersion in both color diagrams and learned representation. Finally, we also removed galaxies with no reliable estimate of the absolute magnitude in the $NUV$ band ($M_{NUV}<-70$) or in specific star formation rate ($\log \, (sSFR) <-30$), and we avoided areas masked in the optical because of bright saturated stars. The resulting catalog contains 63 646 galaxies, with absolute magnitudes provided in 11 bands ({i.e.}, NUV, $u^*$, $B$, $V$, $r^+$, $i^+$, $z$, $Y$, $J$, $H,$ and $K$ filters).

\section{DAE diagram}
The model was developed using Pylearn2  \cite{goodfellow2013pylearn2}. Pylearn2 is a deep learning research library, built on top of Theano \cite{bergstra2010theano}, which compiles Python to native code. The library contains a wealth of deep learning models, notably the denoising autoencoders analyzed here.\\

\begin{figure}[htb]
\begin{tabular}{cc}
\centering
\includegraphics[width=0.24\textwidth]{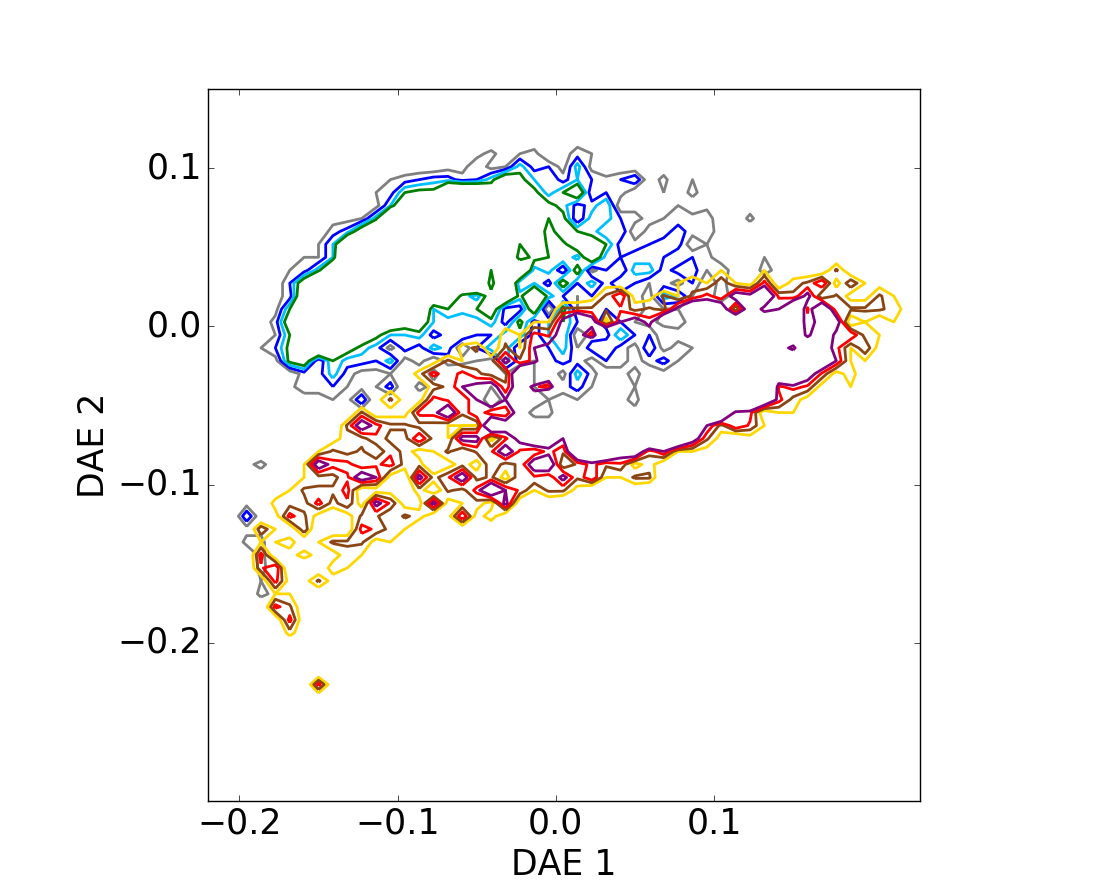}&
\includegraphics[width=0.24\textwidth]{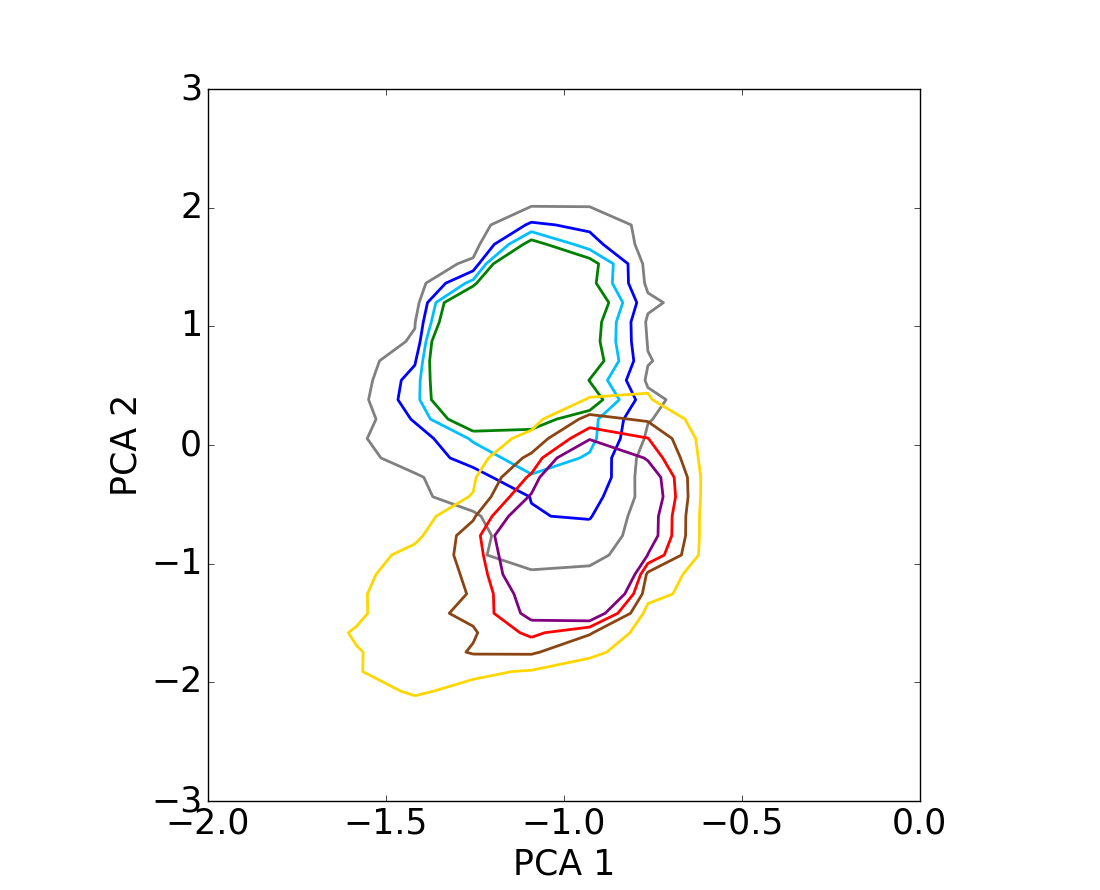}\\
\small{(a) DAE diagram} & \small{(b) PCA diagram}\\
\end{tabular}
\caption{Contour levels for the test set on the DAE diagram (a) and on the PCA diagram (b). The curves are depicted from $95\%$ of the population in grey to $70\%$ in green for star-forming galaxies, and from yellow to purple for quiescent sources. These figures illustrate that the separation between the two classes of populations is  clearer on the DAE diagram compared to the PCA diagram.}
\label{fig:contour_levels}
\end{figure}

The denoising autoencoder is trained with a training set that contains N = 10000 samples and then, inspected with a test set built with 10 000 samples. To consider the largest possible diversity of galaxy physical properties explored with the DAE, each of these sets is composed of 5 000 star forming galaxies and  5 000 quiescent galaxies that were randomly picked over the catalog. Each sample is an $m = 11$ band SED corresponding to a galaxy. \\
The number of visible units is fixed to $m = 11,$ in agreement with the size of the input data and the number of hidden units is set to $nhid = 11$. The weights are randomly initialized from a uniform distribution. Moreover, the activation functions for both the encoder and the decoder are chosen to be a sigmoid and the weights are tied across all the experiments. The input is artificially contaminated with a Gaussian corruption. High noise levels lead to greater diversity of the weights and, therefore, filters will be more sensitive to discriminative features. In other words, filters obtained with larger corruption levels will retain  more information yielding more invariant representations. On the other hand, higher noise levels also translate to higher reconstruction errors. In these  investigations, fixing the noise standard deviation to $\sigma_n = 0.05$ yielded a good trade-off between reconstruction error criteria and invariance of the learned representations. Furthermore, the optimization of the parameters is performed through stochastic gradient descent; and the reconstruction cost criteria to be minimized is the binary cross entropy averaged over all the training set. The learning rate was set to 0.1 and the batch size to 7; the training stops after 1 000 epochs to ensure the convergence of the algorithm. Fixing the parameters of a DAE is generally tricky. In this investigation the parameters of the algorithm have been selected empirically; a deeper study on the optimization step is required, but this is beyond  the scope of this  article.
Once the model is trained, the codes are obtained by simply applying the encoder to the new samples. Therefore, all the computational effort is in the training stage and the code calculation just requires a matrix multiplication followed by a non-linearity.\\
\begin{figure}[htb]
\begin{tabular}{cc}
\centering
\includegraphics[width=0.24\textwidth]{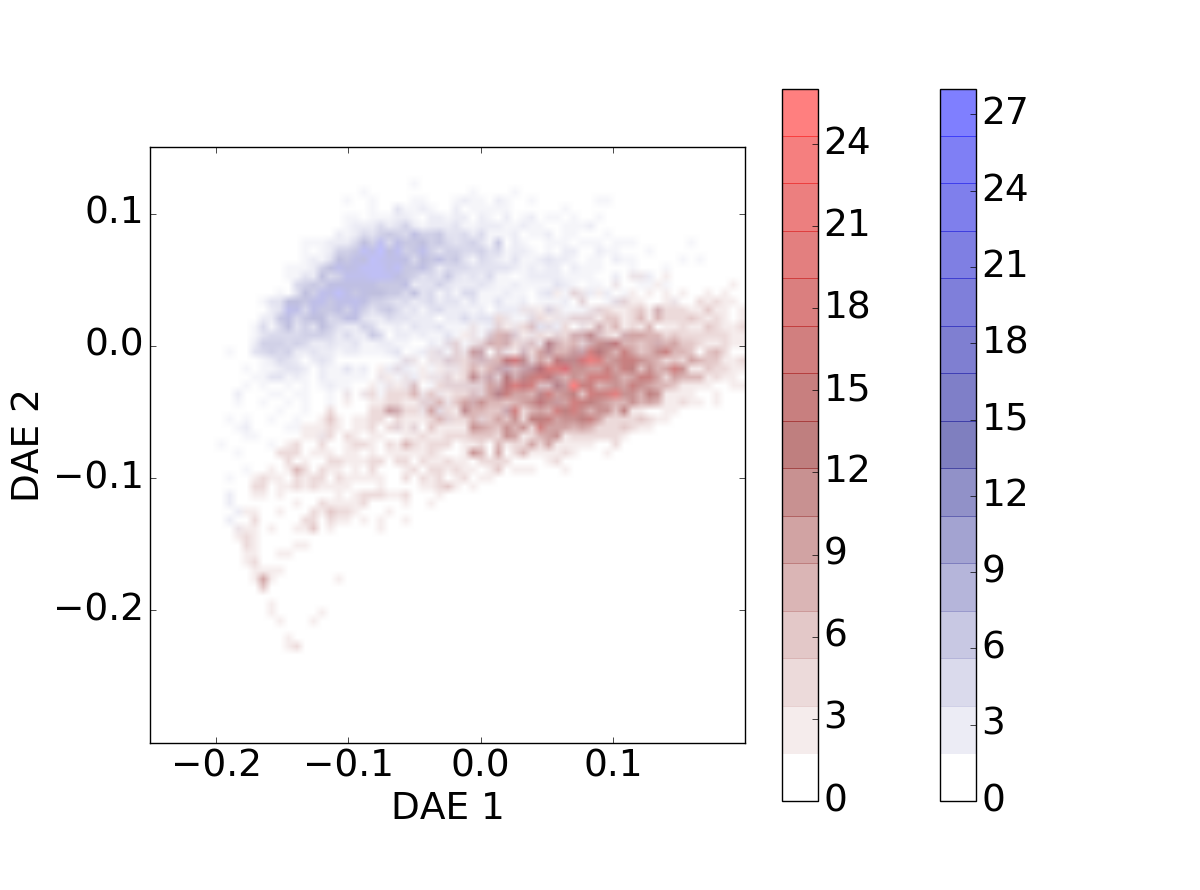}&
\includegraphics[width=0.24\textwidth]{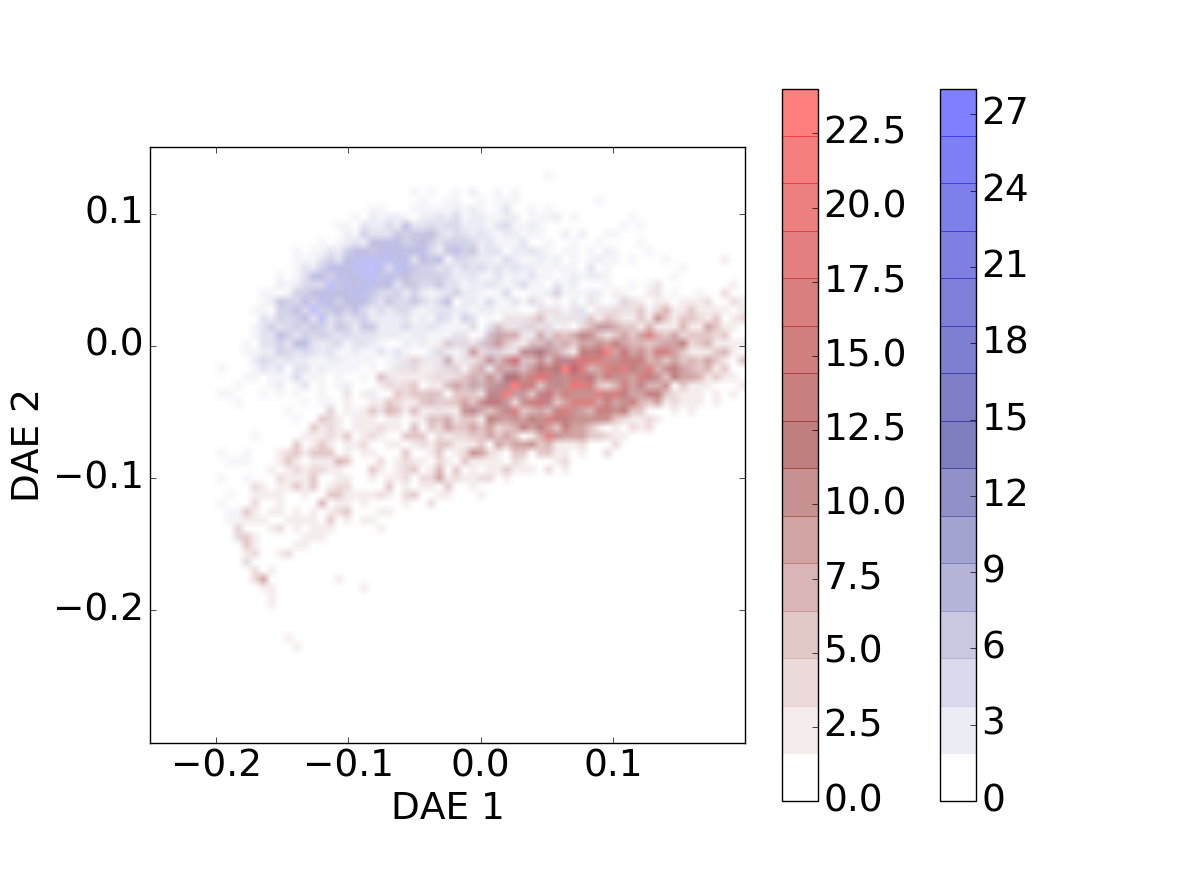}\\
\small{(a) Training set} & \small{(b) Test set}\\
\end{tabular}
\caption{Histogram of the training set and the test set on the DAE diagram. In the new plane, galaxies classified as star-forming galaxies in the COSMOS2015 catalog are depicted in blue, while the quiescent galaxies are shown in red.}
\label{fig:training}
\end{figure}
\begin{figure}[htb]
\begin{tabular}{cc}
\centering
\includegraphics[width=0.24\textwidth]{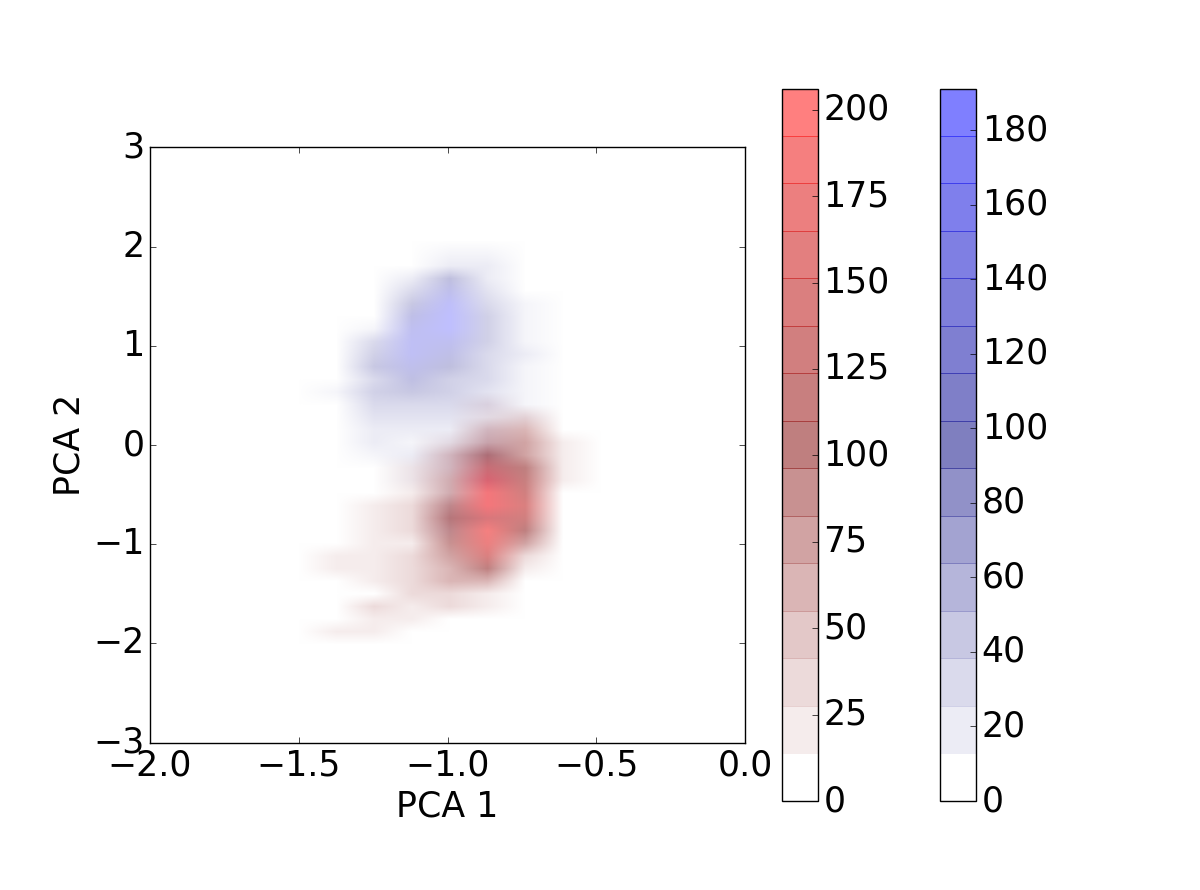}&
\includegraphics[width=0.24\textwidth]{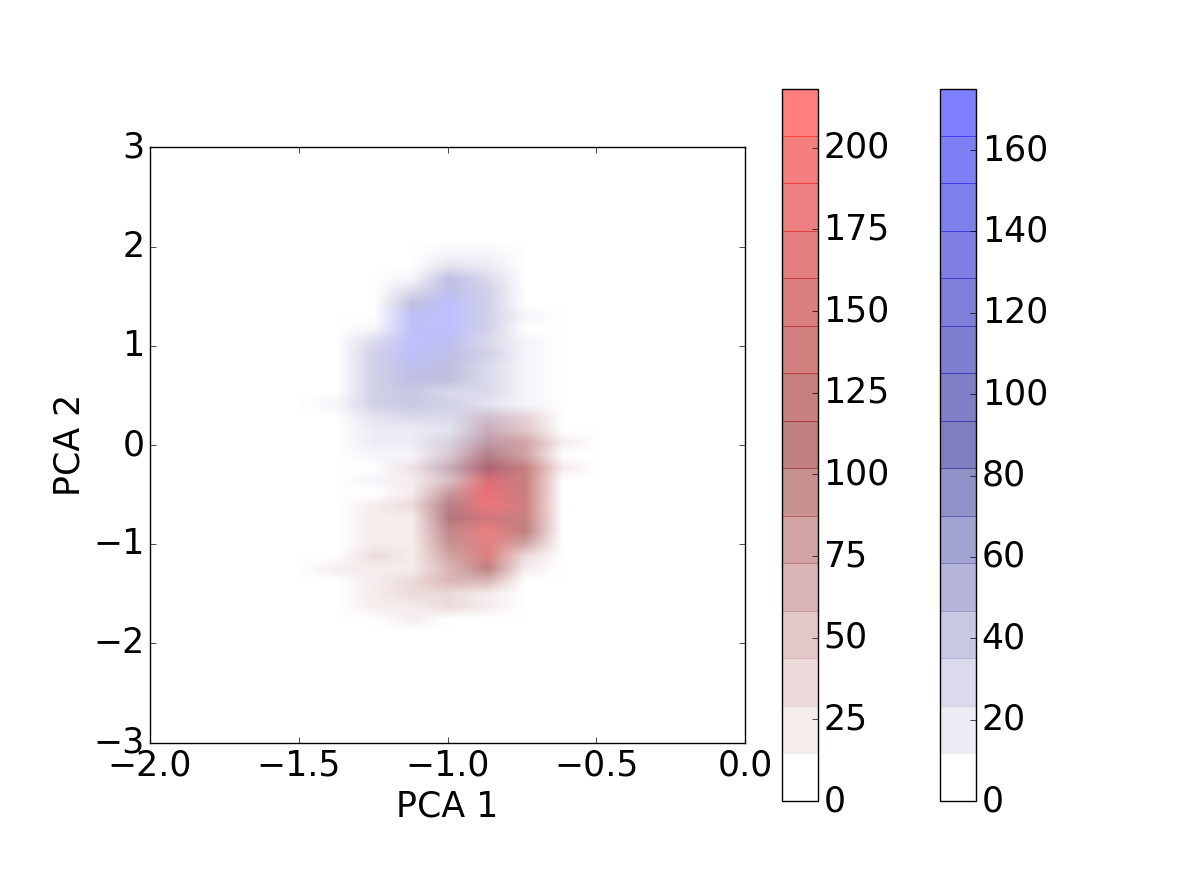}\\
\small{(a) Training set} & \small{(b) Test set}\\
\end{tabular}
\caption{Histogram of the training set and the test set projected on the two PCA components.}
\label{fig:PCAtraining}
\end{figure}
Figure \ref{fig:training} displays a histogram of the new DAE diagram for both the training set and the test set.  The model is optimized to minimize the reconstruction error over the training set. The $x-axis$ and the $y-axis$ correspond to the second and third component of the code's PCA respectively. In the proposed scheme, sources classified as star-forming galaxies in the COSMOS2015 catalog are represented in blue, while the quiescent sources are shown in red. Inspecting the histogram of the test set, it is clear that the DAE is remarkably efficient in separating the two galaxy populations, while no information on this  classification is provided as input to the model. We note that the proposed architecture does not claim optimality in any sense and the focus is on emphasizing its interest for unsupervised feature extraction. \\
For comparison, Fig. \ref{fig:PCAtraining} shows the histogram of the representation space built directly from the first two components of the PCA for the same training set and test set. We note that the PCA diagram also enables us to distinguish the two classes of populations, although the separation is less clear than on the DAE diagram. This fact is also illustrated in Fig. \ref{fig:contour_levels}. The curves depict the contour plots of the test's histograms displayed in Fig. \ref{fig:training} (b) and Fig. \ref{fig:PCAtraining} (b), respectively. The contour's colors grow from grey at $95 \%$ of the population to green at $70\%$ for star-forming galaxies, and from yellow to purple again from $95\%$ to $70\%$ for quiescent galaxies. From these figures, it is important to note that, the galaxies' distribution is more concentrated and the overlap between star-forming and passive sources is smaller for the DAE diagram than for the PCA diagram.\\
In the following sections, we  investigate in more detail how the different physical properties that characterize distant galaxies can impact their distribution in the different diagrams.
\subsection{Redshift bins}

\begin{figure}[htb]
\begin{tabular}{cc}
\centering
\includegraphics[width=0.24\textwidth]{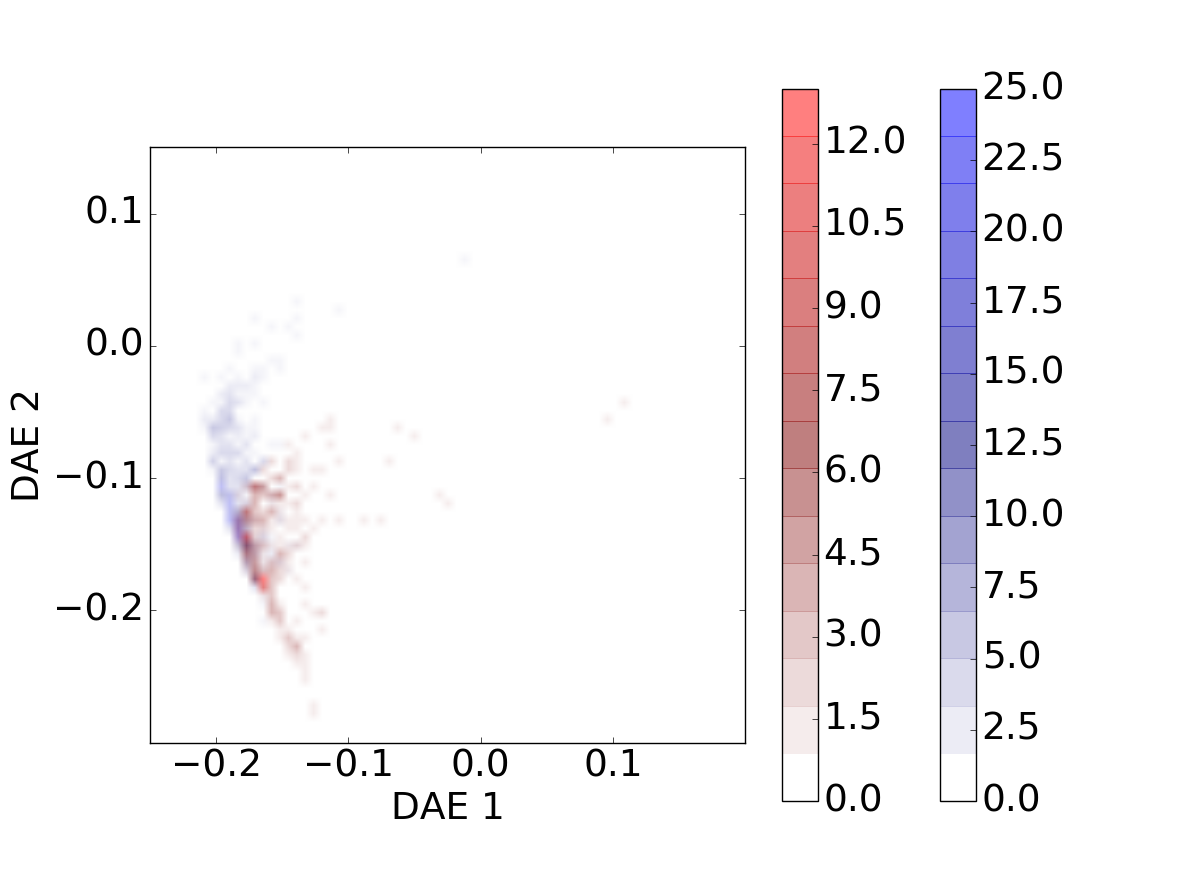}&
\includegraphics[width=0.24\textwidth]{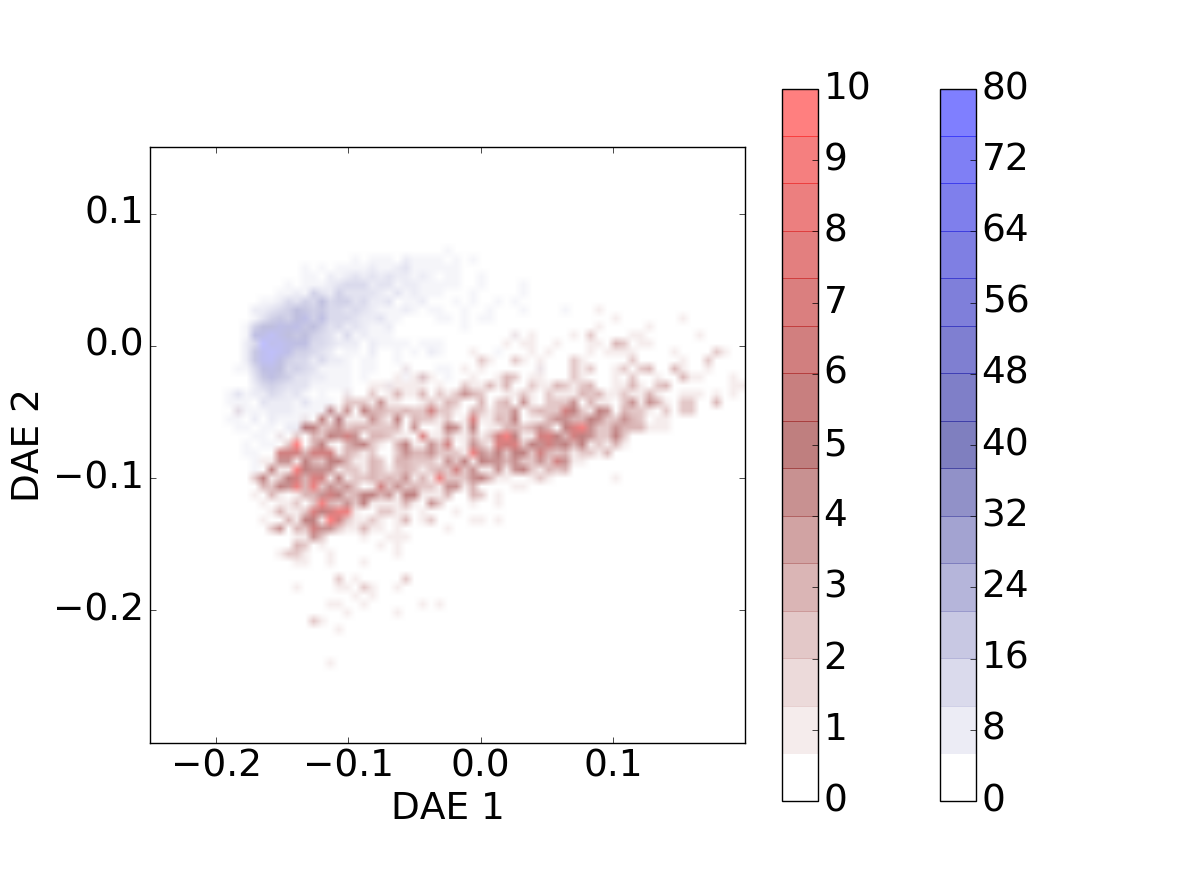}\\
\small{(a)  $z \in (0, 0.2]$} & \small{(b) $z \in (0.2, 0.4]$}\\
\includegraphics[width=0.24\textwidth]{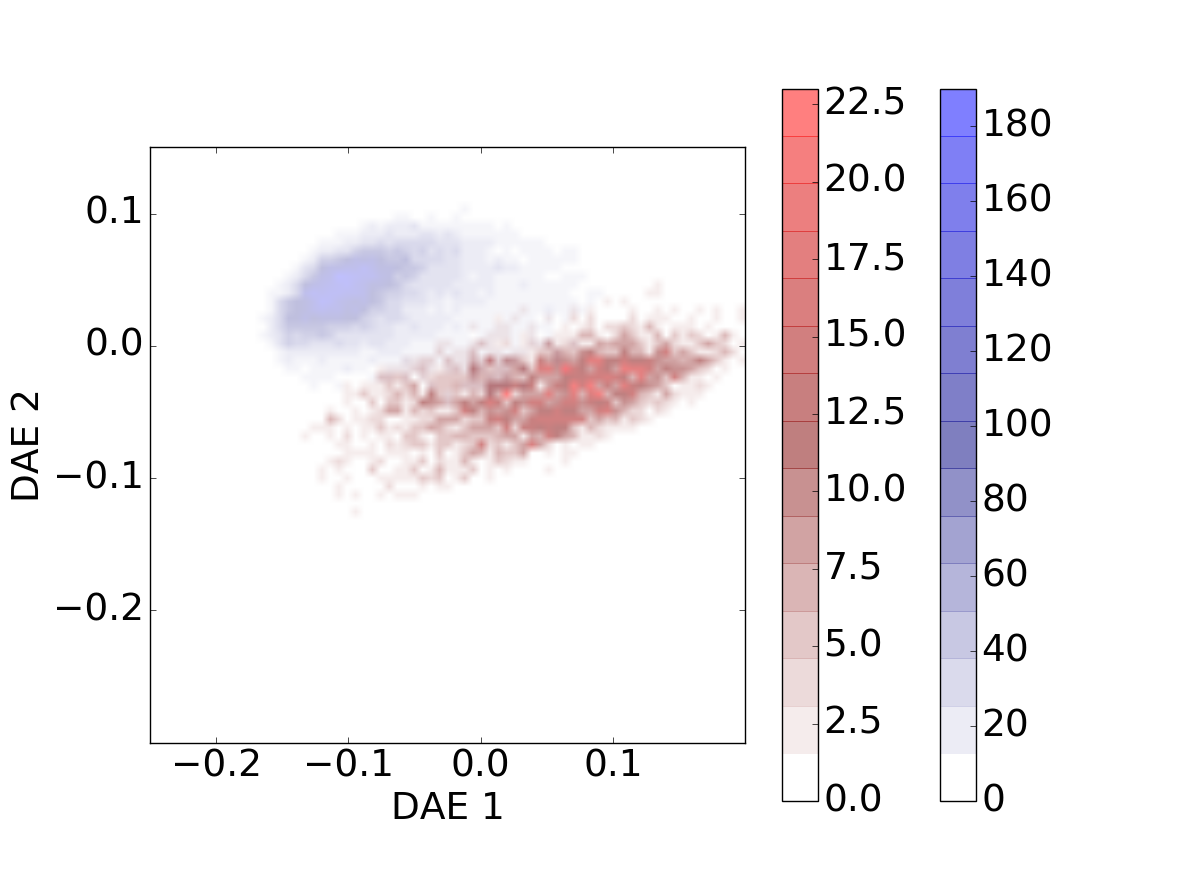}&
\includegraphics[width=0.24\textwidth]{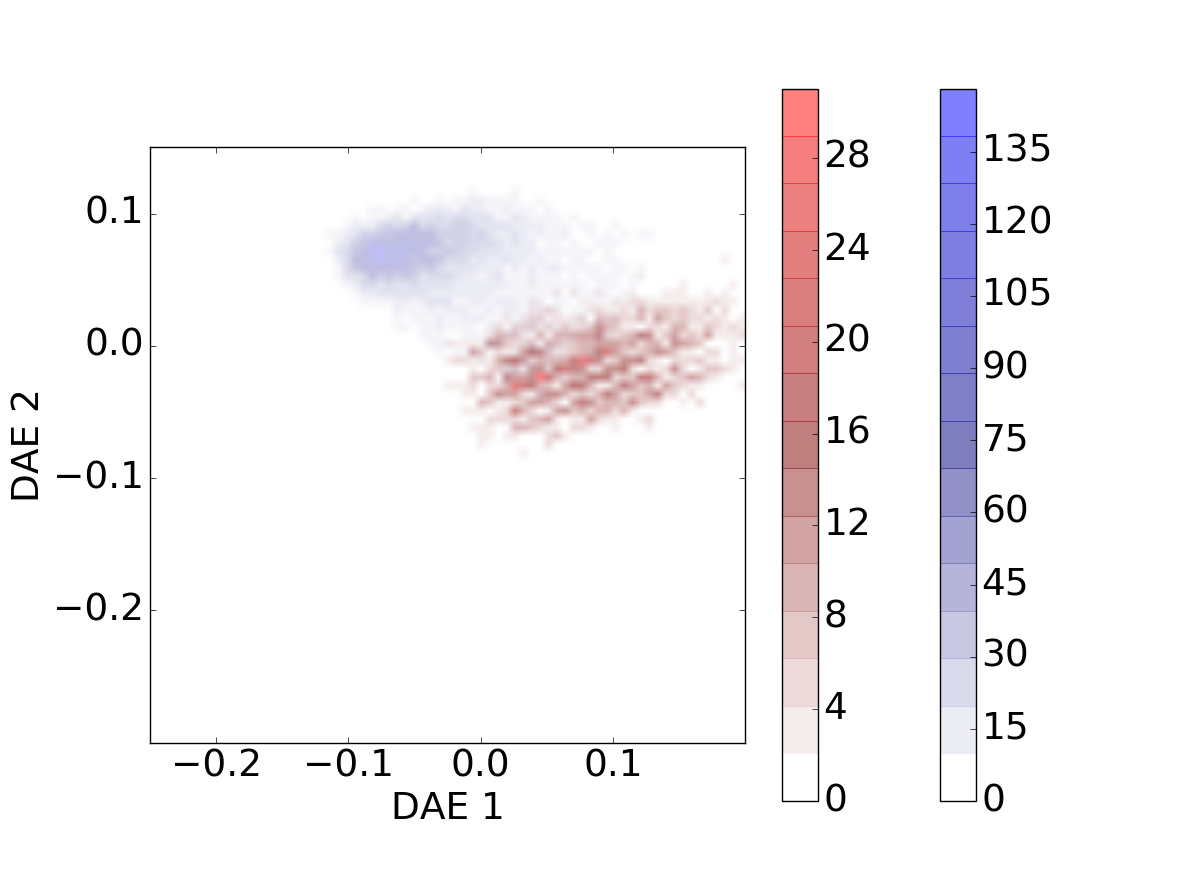}\\
 \small{(c) $z \in (0.4, 0.8]$} & \small{(d) $z \in (0.8, 1.0]$}\\
\end{tabular}
\caption{Evolution of the galaxies' distribution for different redshift bins on the DAE diagram. The galaxies' cloud progresses along the diagram as the redshift increases. The star-forming sources are represented in blue and quiescent sources in red.}
\label{fig:z_AE}
\end{figure}

\begin{figure}[htb]
\begin{tabular}{cc}
\centering
\includegraphics[width=0.24\textwidth]{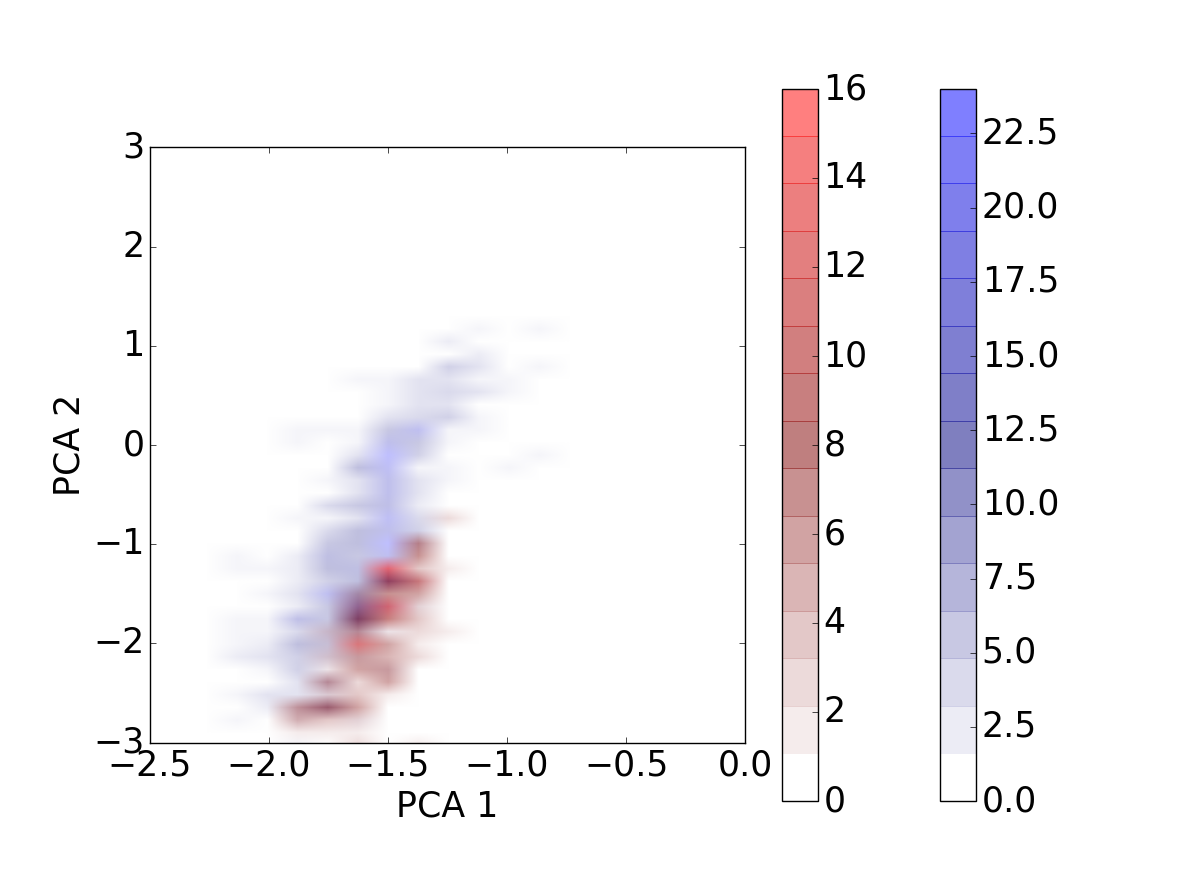}&
\includegraphics[width=0.24\textwidth]{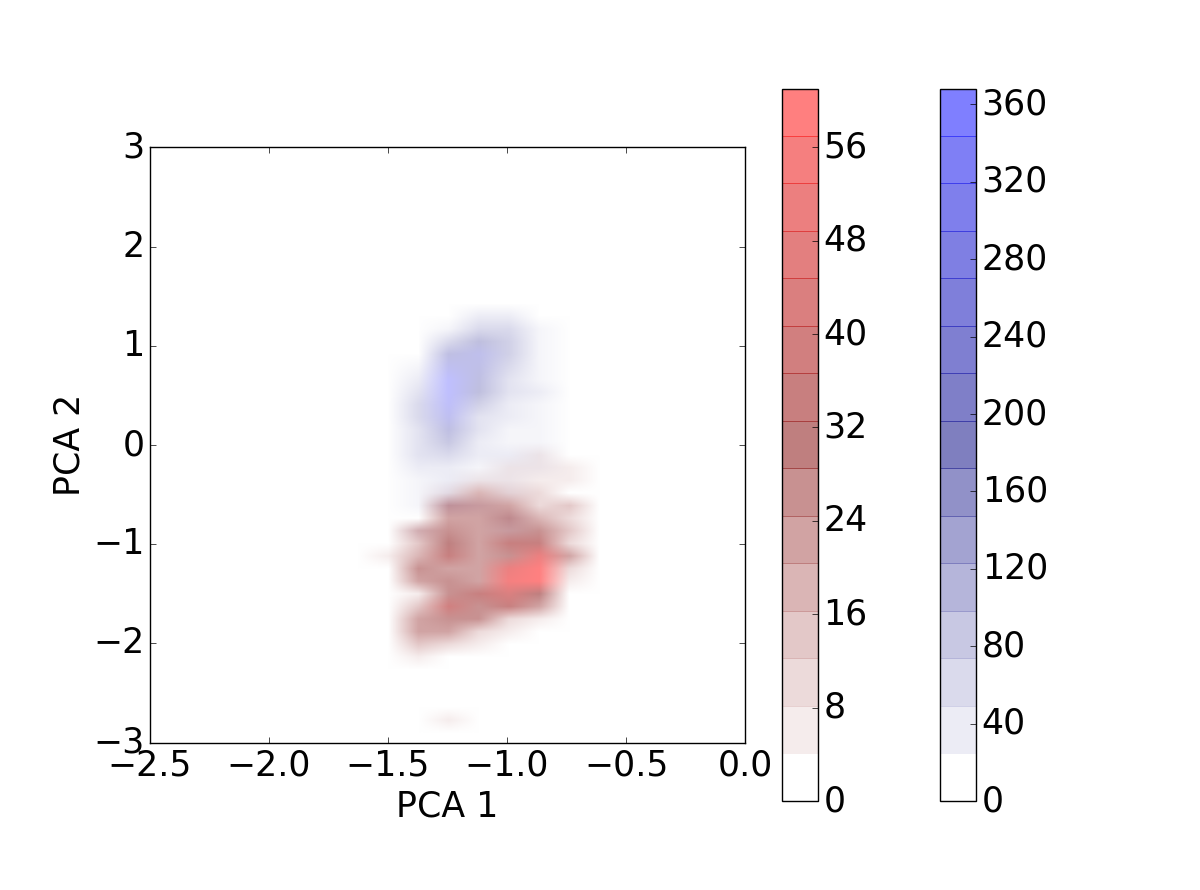}\\
\small{(a)  $z \in (0, 0.2]$} & \small{(b) $z \in (0.2, 0.4]$}\\
\includegraphics[width=0.24\textwidth]{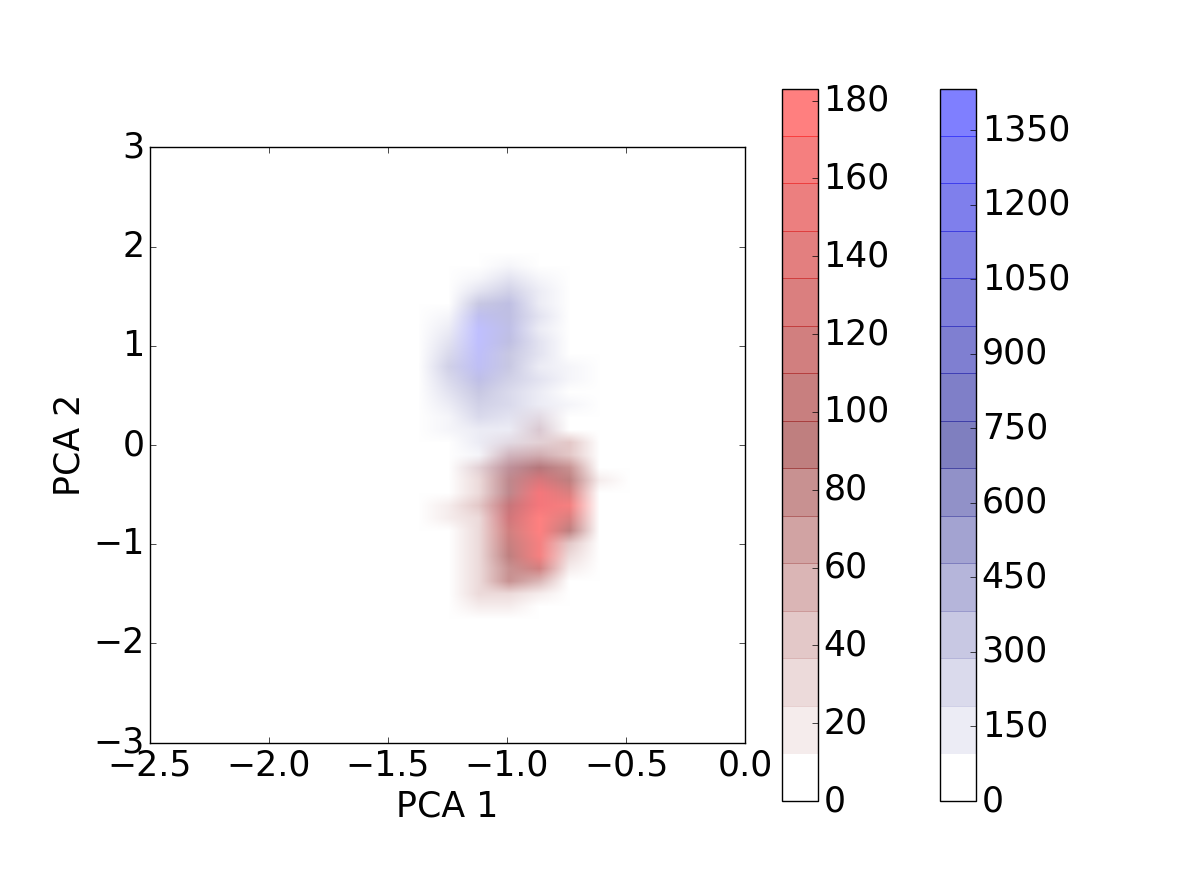}&
\includegraphics[width=0.24\textwidth]{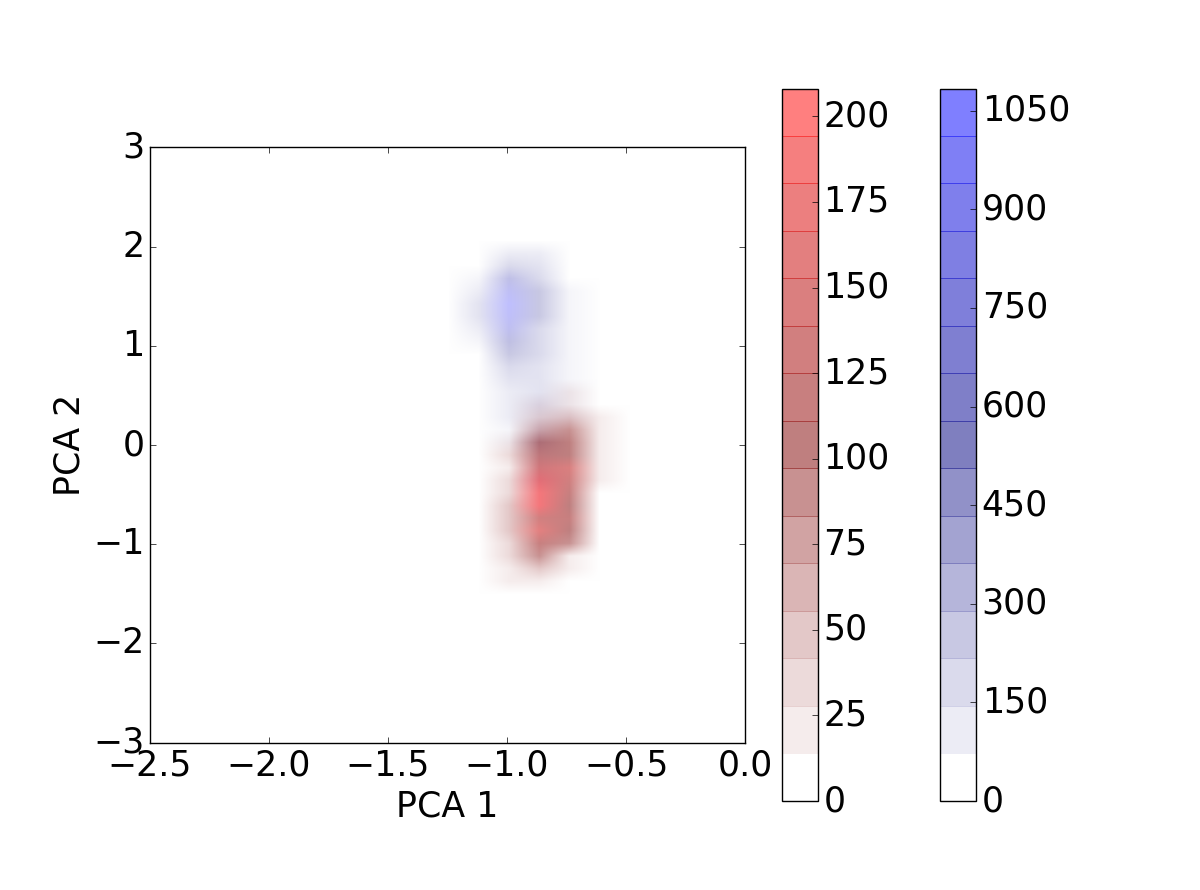}\\
 \small{(c) $z \in (0.4, 0.8]$} & \small{(d) $z \in (0.8, 1.0]$}\\
\end{tabular}
\caption{Evolution of the galaxies' distribution for different redshift bins on the PCA diagram. The position of the galaxies in the diagram evolves as the redshift value increases.}
\label{fig:z_PCA}
\end{figure}

\begin{figure}[h!]
\begin{tabular}{cc}
\centering
\includegraphics[width=0.24\textwidth]{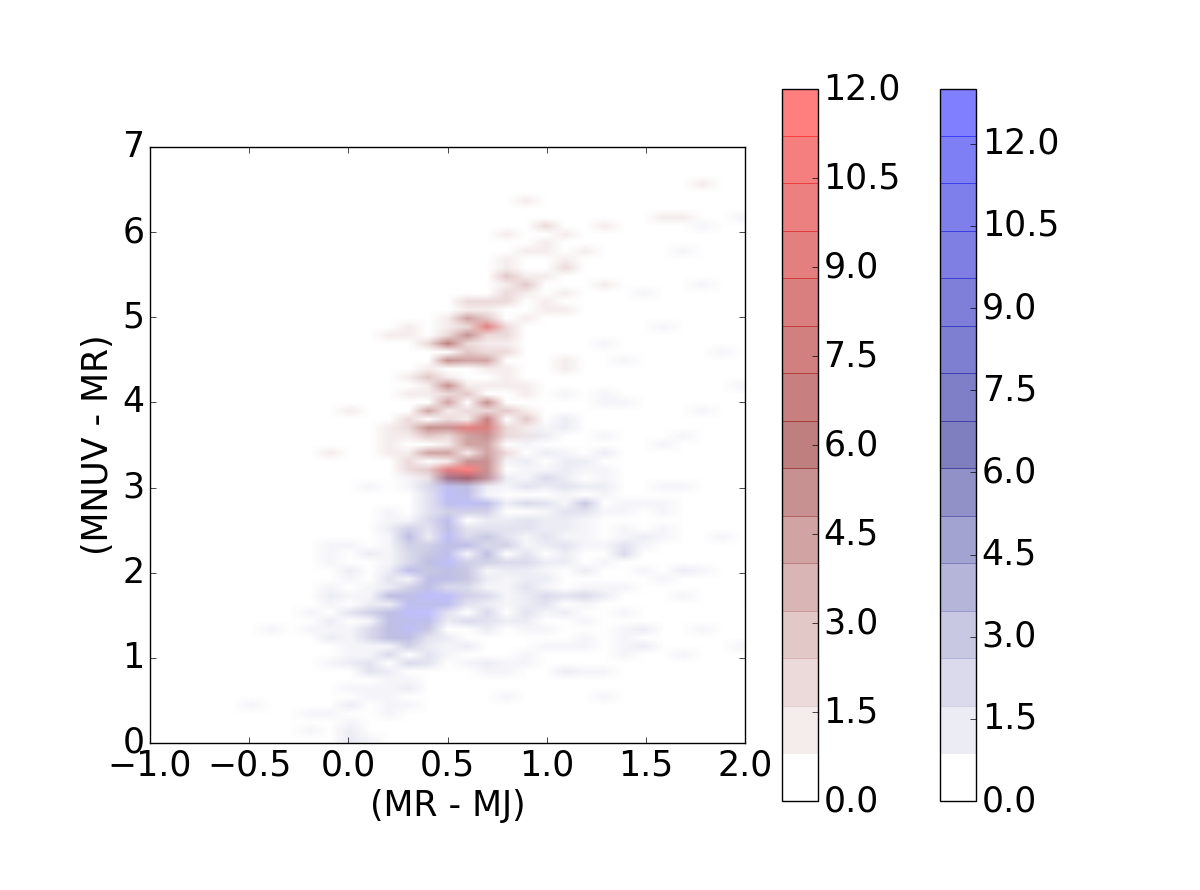}&
\includegraphics[width=0.24\textwidth]{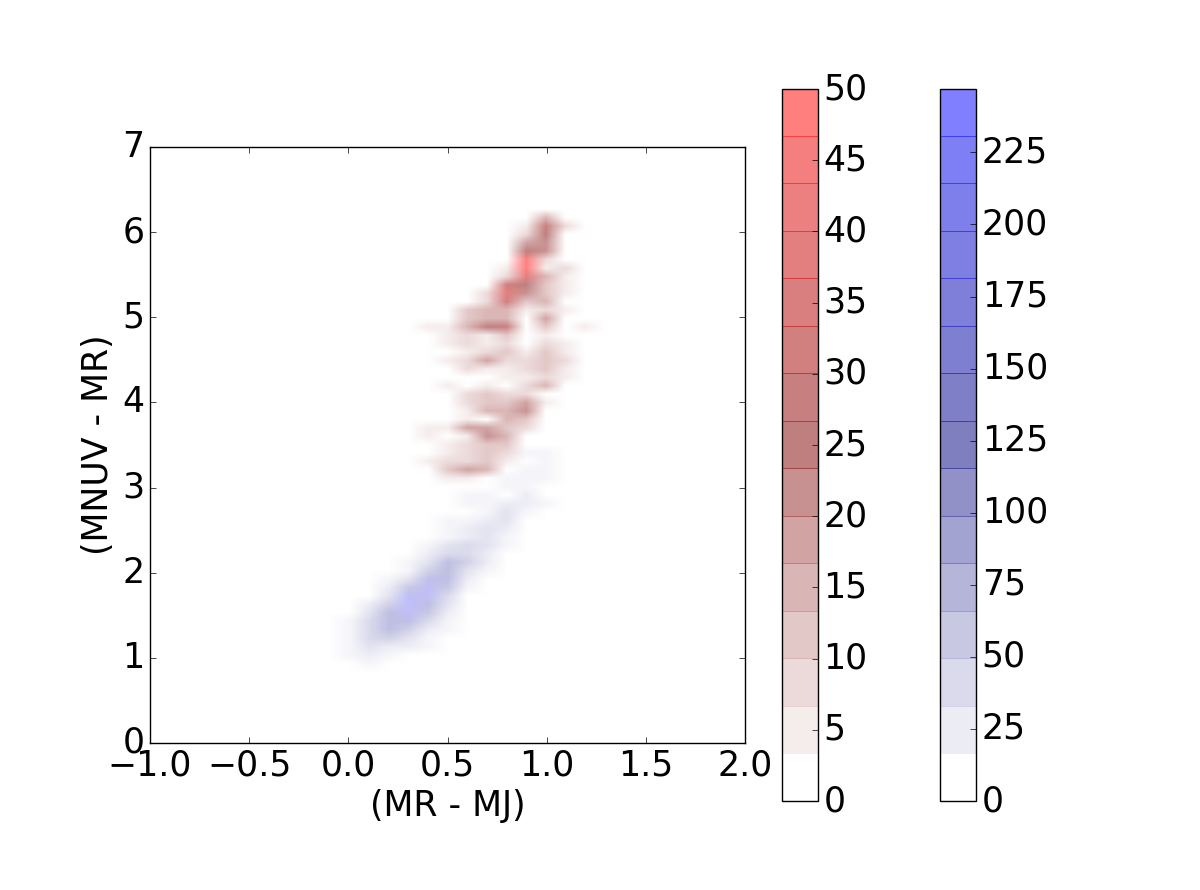}\\
\small{(a)  $z \in (0, 0.2]$} & \small{(b) $z \in (0.2, 0.4]$} \\
\includegraphics[width=0.24\textwidth]{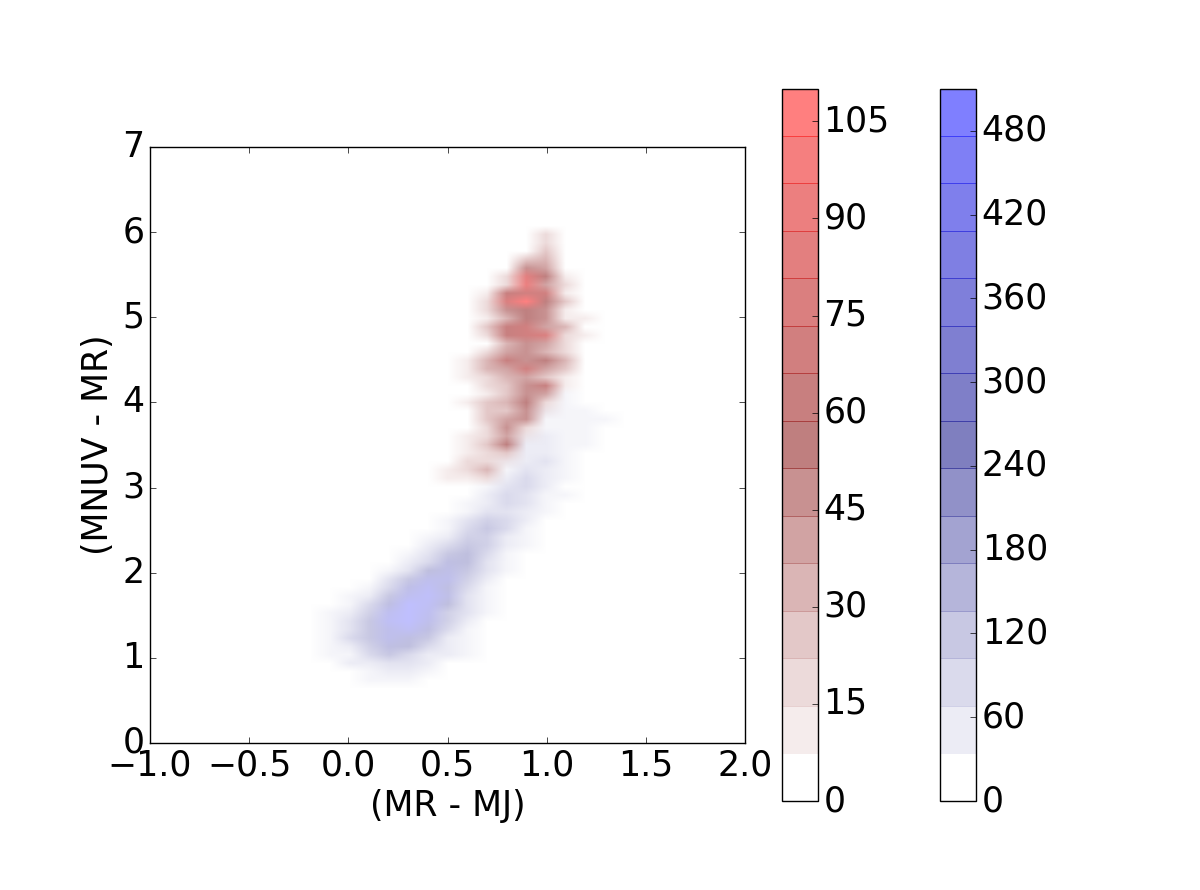}&
\includegraphics[width=0.24\textwidth]{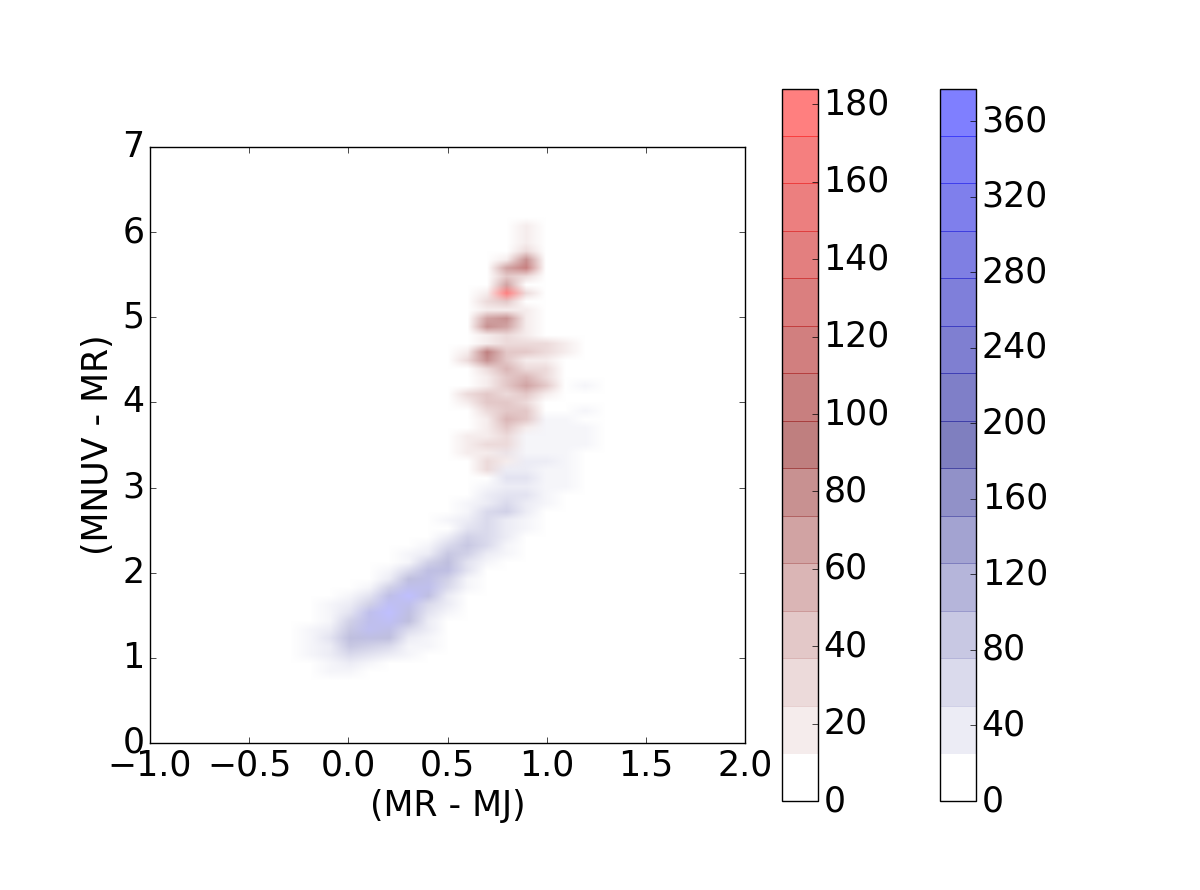}\\
 \small{(c) $z \in (0.4, 0.8]$} & \small{(d) $z \in (0.8, 1.0]$}\\
\end{tabular}
\caption{Evolution of the galaxies' distribution for different redshift bins on the  RJ - NUVR plane. The galaxies are disposed similarly in the diagram for different redshift bins. Therefore, no information concerning the redshift can be identified directly in the RJ - NUVR diagram.}
\label{fig:z_cc}
\end{figure}

In Fig.\,\ref{fig:z_AE}, we show the distribution of the galaxies on the DAE diagram, for different estimated redshift bins. And Fig. \ref{fig:z_PCA} shows the corresponding galaxies placed on the PCA diagram. Moreover, Fig.\,\ref{fig:z_cc} presents the same galaxies on the classical  
 $NUV$$-$$r^+$ versus $r^+-J$ color diagram (hereafter referred as the RJ-NUVR diagram), which corresponds to the diagnostic that was initially used to disentangle star-forming and quiescent passive sources in the COSMOS2015 catalog.
 %  has now become a standard approach for separating star-forming and passive galaxies in cosmological photometric surveys \citep{Williams09,Patel12}.} RJ - NUVR diagram. 
We note that the three diagrams enable us to distinguish between the two galaxy populations. However, the representations proposed in Figs.\,\ref{fig:z_AE} and\,\ref{fig:z_PCA} have been obtained in an unsupervised manner, while the separation in the RJ-NUVR diagrams has been hand-designed.\\

\begin{figure}[h]
\begin{tabular}{cc}
\centering
\includegraphics[width=0.24\textwidth]{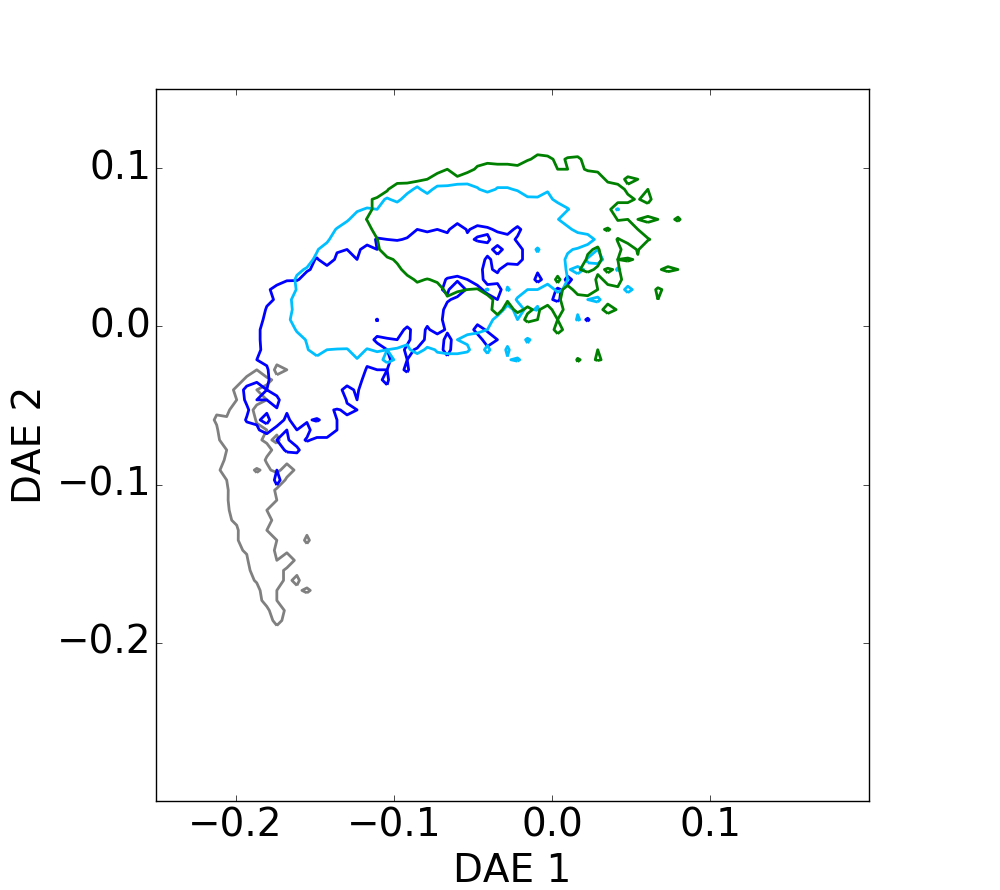}&
\includegraphics[width=0.24\textwidth]{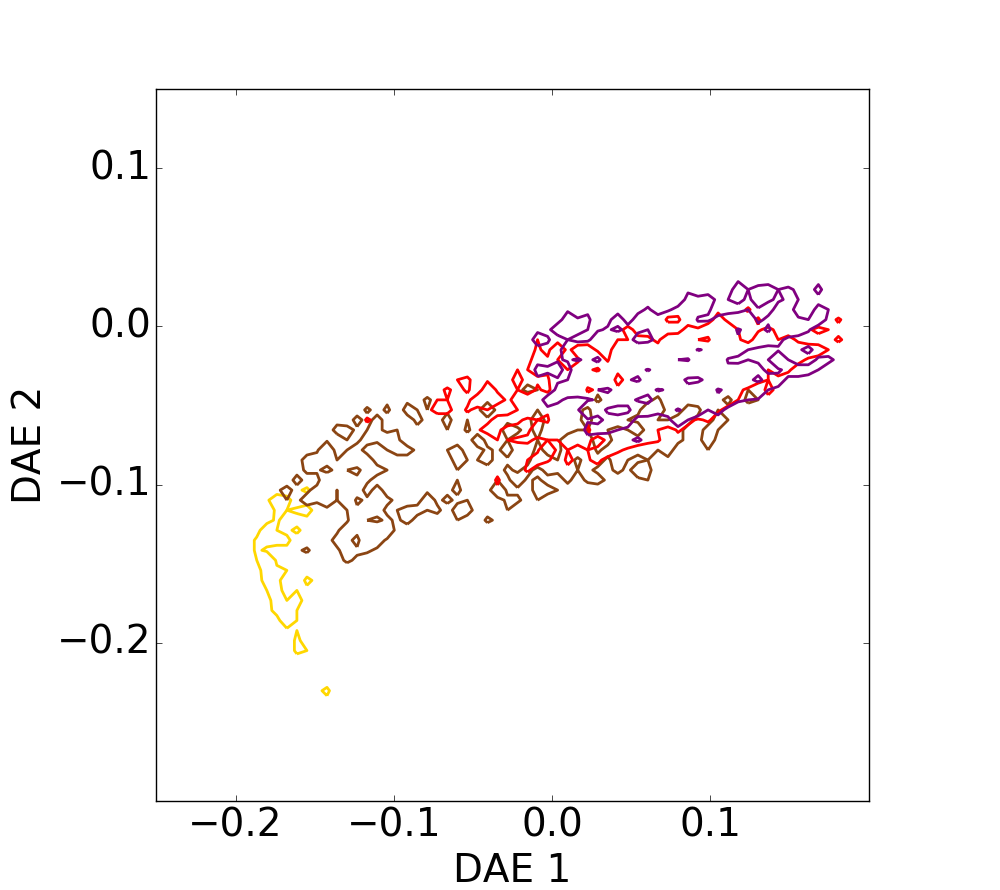}\\
\small{(a) Star forming - DAE}& \small{(b) Quiescent - DAE} \\
\includegraphics[width=0.24\textwidth]{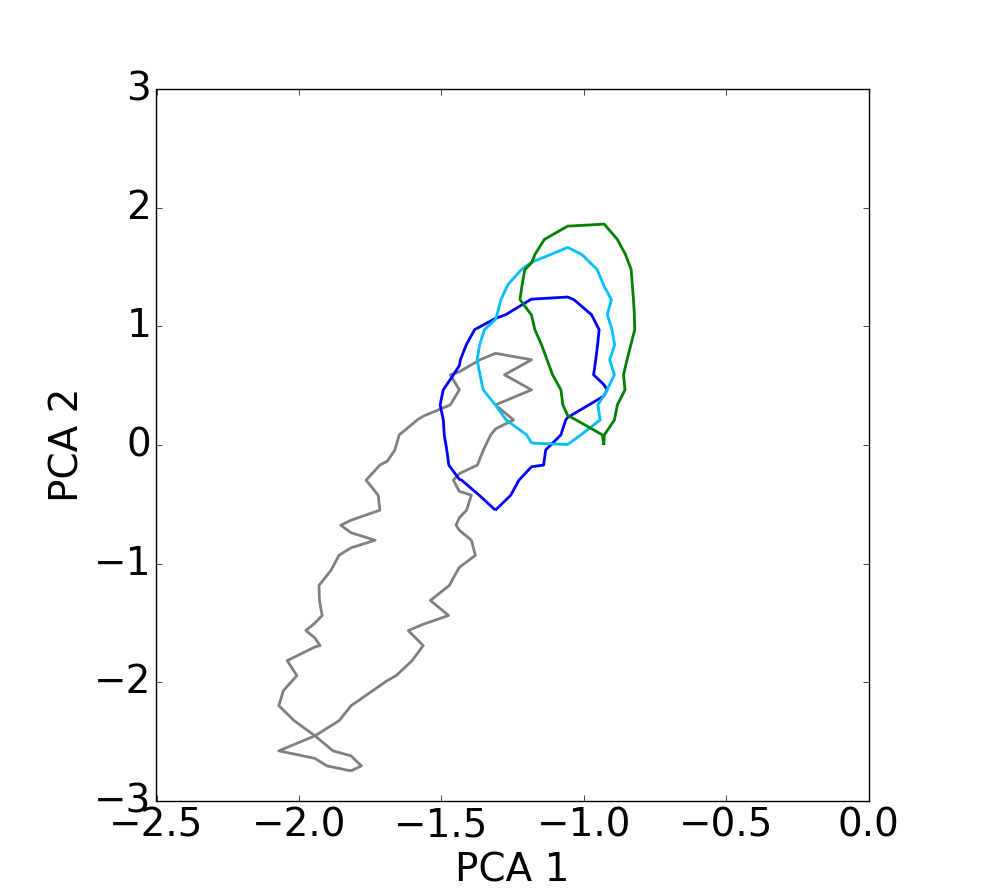}&
\includegraphics[width=0.24\textwidth]{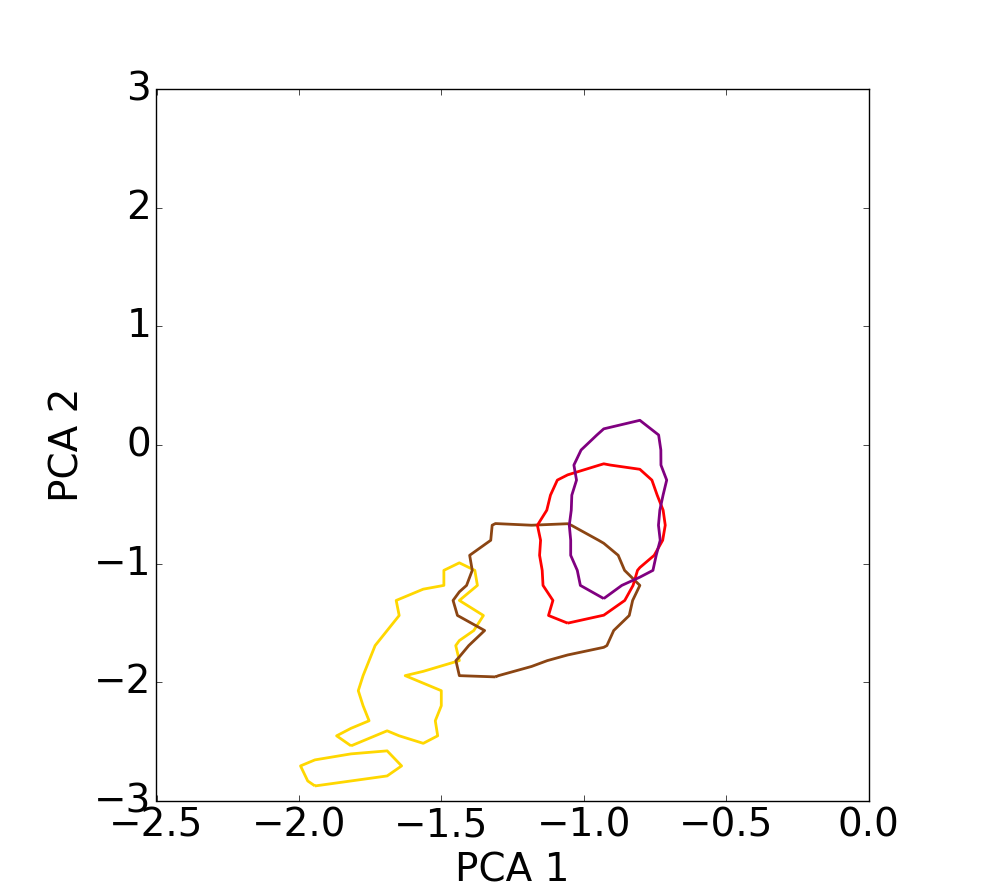}\\
 \small{(c) Star forming - PCA} & \small{(d) Quiescent - PCA}\\
\includegraphics[width=0.24\textwidth]{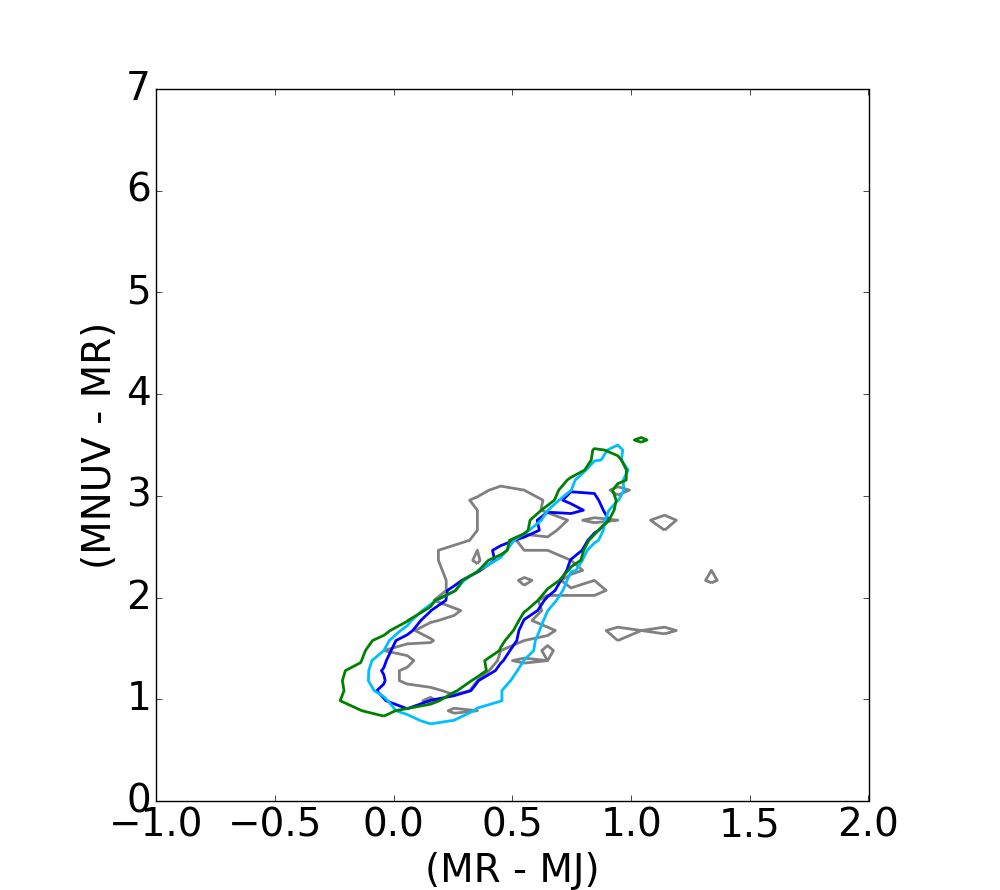}&
\includegraphics[width=0.24\textwidth]{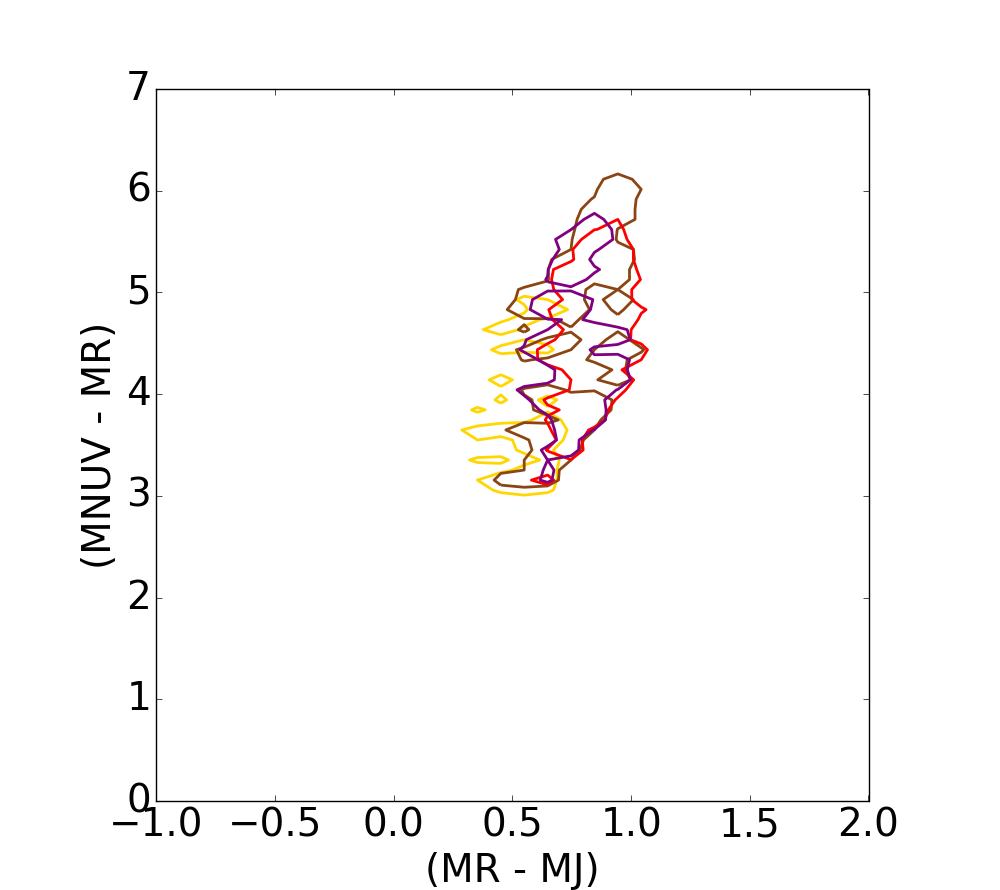}\\
 \small{(e) Star forming - RJNUV} & \small{(f) Quiescent - RJNUV}\\
\end{tabular}
\caption{Contour plots for the four redshift bins displayed in Figs. \ref{fig:z_AE},   \ref{fig:z_PCA}, and \ref{fig:z_cc} for the DAE diagram, the PCA  diagram and the the RJ - NUVR diagram, respectively. The contour plots are displayed at $75\%$ of the samples in the corresponding histogram for the different redshift bins. The contour's colors grow from grey to green for star-forming galaxies, and from yellow to purple for quiescent galaxies respectively. Accordingly, top figures illustrate the evolution of the galaxies with the redshift for the DAE diagram, and bottom figures evidence the overlap in the RJ - NUVR plane. Results obtained with the PCA, at the center, are slightly worse than those exhibited by the DAE diagram, in the sense that the overlap is smaller with the DAE diagram. And PCA performs slightly better than the RJ - NUVR diagram since no redshift variation is observed in the latter.}
\label{fig:z_contour}
\end{figure}

In addition, the DAE and the PCA diagrams provide visualization of the evolution of the galaxies distribution through the different redshift bins. In other words, the clouds, which indicate the concentration of the galaxies, move across the diagram as the redshift increases, which depicts a clear evolution of the galaxy SED properties with cosmic time. This fact is also highlighted in Fig.\,\ref{fig:z_contour}, where the $75\%$ contours computed from the histograms for the different redshift bins are depicted. The top figures show the DAE diagrams for the star-forming galaxies on the left and for the quiescent galaxies on the right; the center figures display the same data for the PCA diagrams; and the bottom figures correspond to the RJ-NUVR diagram. The contour's colors grow from grey to green for star-forming populations, and from yellow to purple for quiescent galaxies, respectively. From these results, it is clear that the overlap between the different clouds is lower in the DAE and in the PCA diagrams than in the classical RJ-NUVR diagram. %{Although, one can easily notice that intersection between star-forming and quiescent galaxies is significant in the PCA diagram}. 
The dynamics, describing the behaviour of the population through different variation factors, are illustrated by the DAE and PCA diagrams.
Therefore, data-based features reveal information and structures contained in the data that more standard color diagrams
% classical representations 
cannot  unfold easily. In fact, the traditional diagnostics solely based on galaxy colors use a very restricted piece of the available information. On the contrary, the unsupervised learning techniques output not only depends on the relative variations of each galaxy SED across the $11$ visible units considered in the model (see Sect.\,2), but it is also sensitive to the range of luminosities that galaxies encompass in a given filter. Since the typical luminosities that can be probed in a flux-limited galaxy survey vary with redshift, it  could thus explain the evolution of the properties  observed in Fig.\,\ref{fig:z_contour}. 

\begin{figure}[htb]
\begin{tabular}{cc}
\centering
\includegraphics[width=0.245\textwidth]{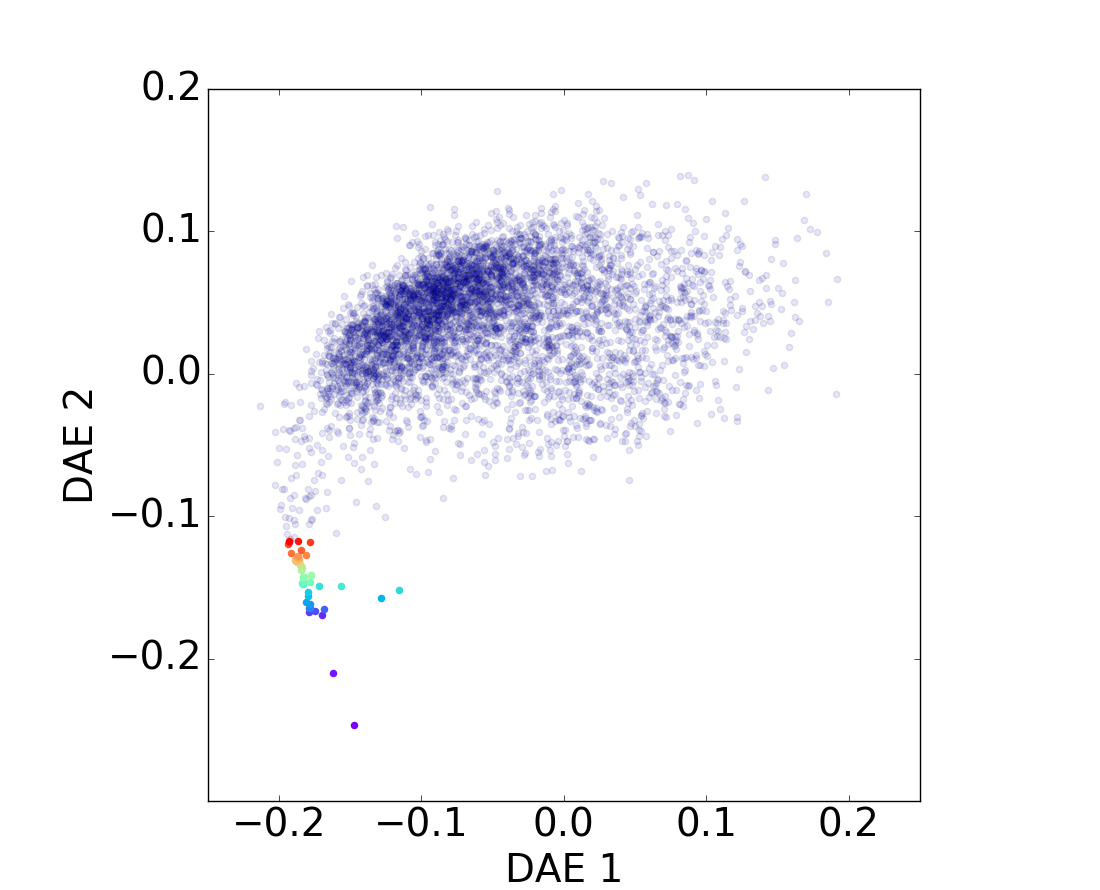}&
\includegraphics[width=0.245\textwidth]{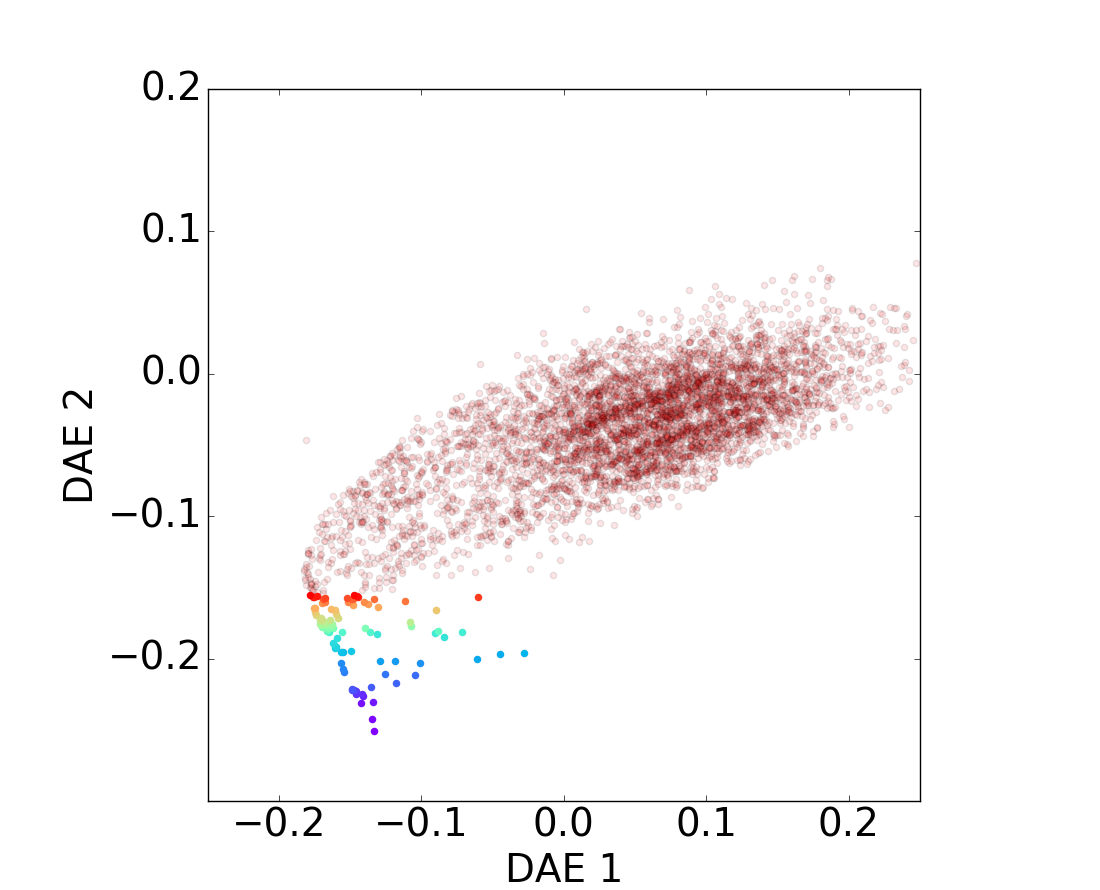}\\
\small{(a) Star forming - DAE}& \small{(b) Quiescent - DAE} \\
\includegraphics[width=0.245\textwidth]{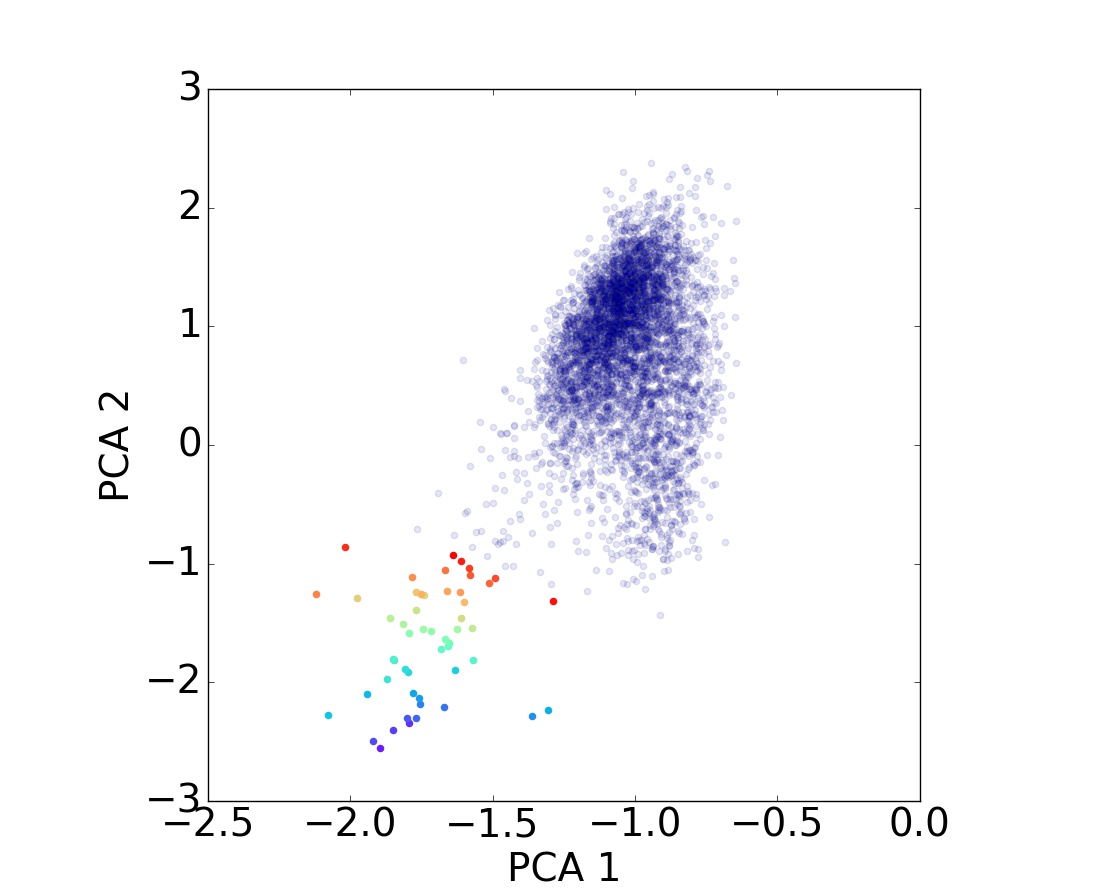}&
\includegraphics[width=0.245\textwidth]{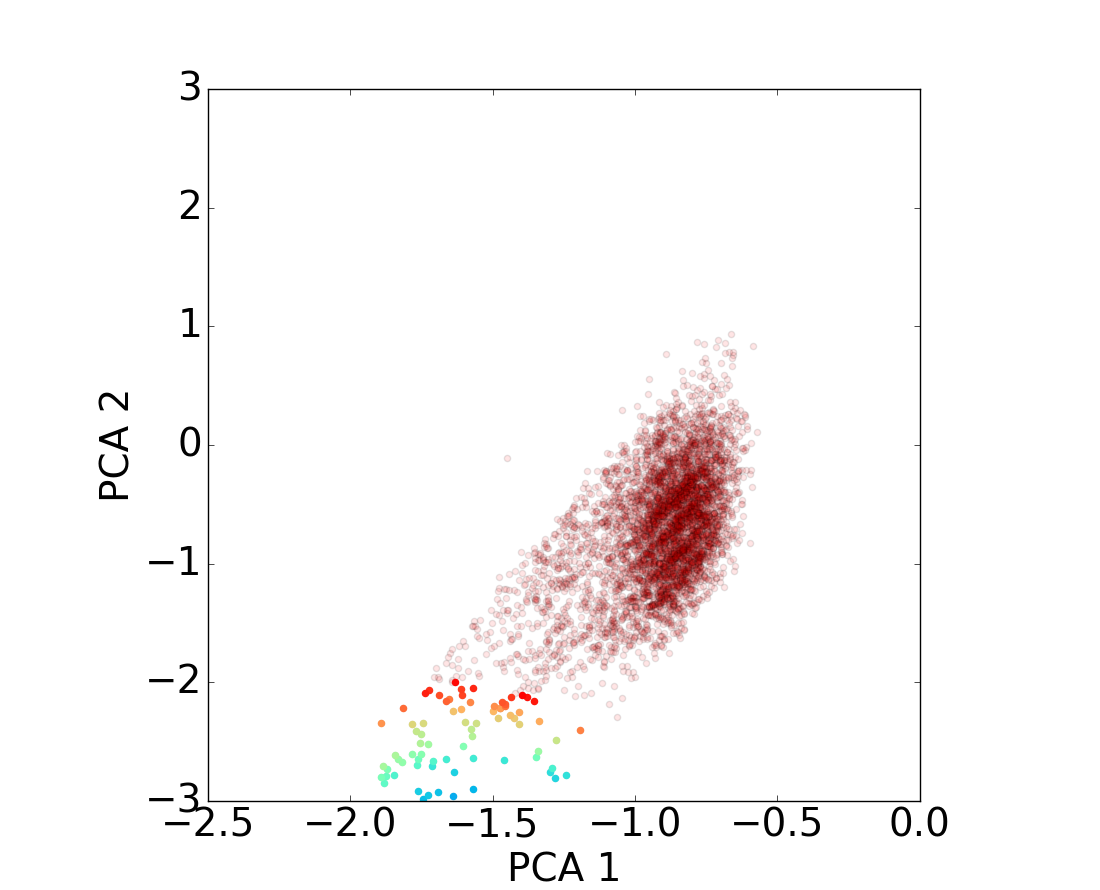}\\
 \small{(c) Star forming - PCA} & \small{(d) Quiescent - PCA}\\
\includegraphics[width=0.245\textwidth]{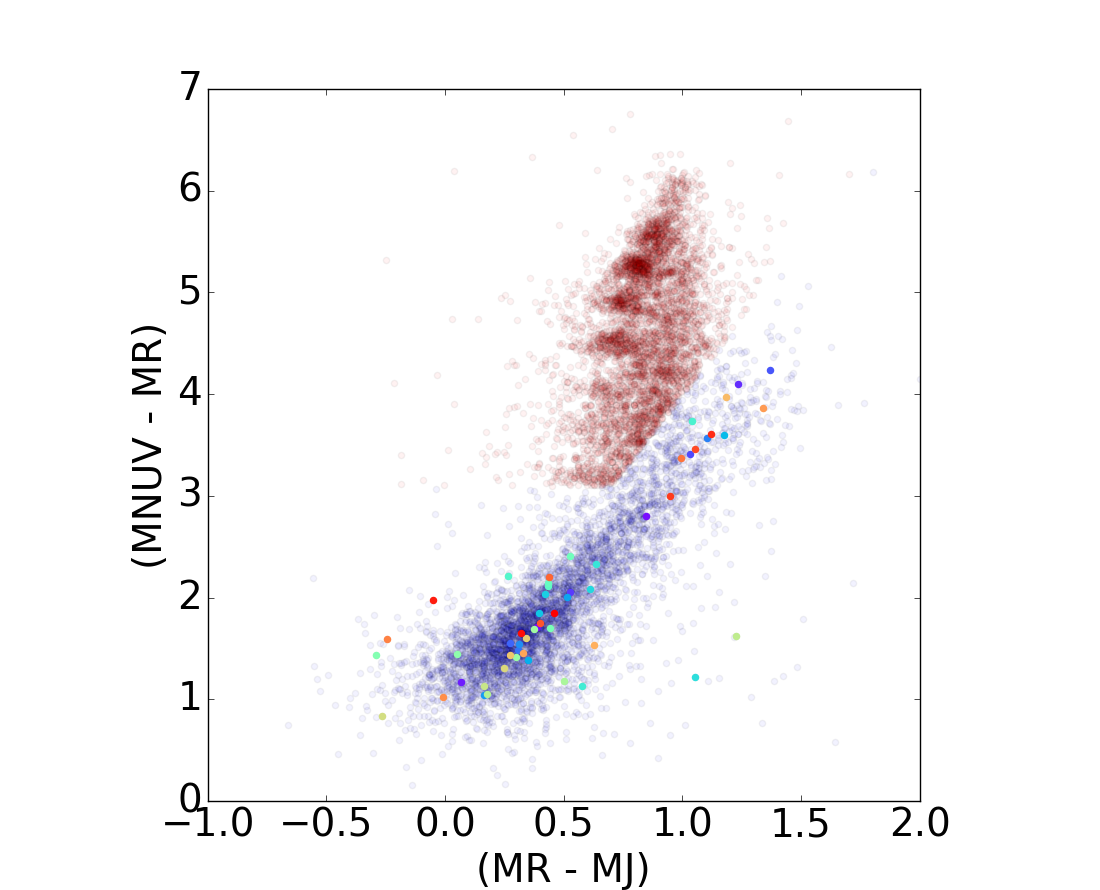}&
\includegraphics[width=0.245\textwidth]{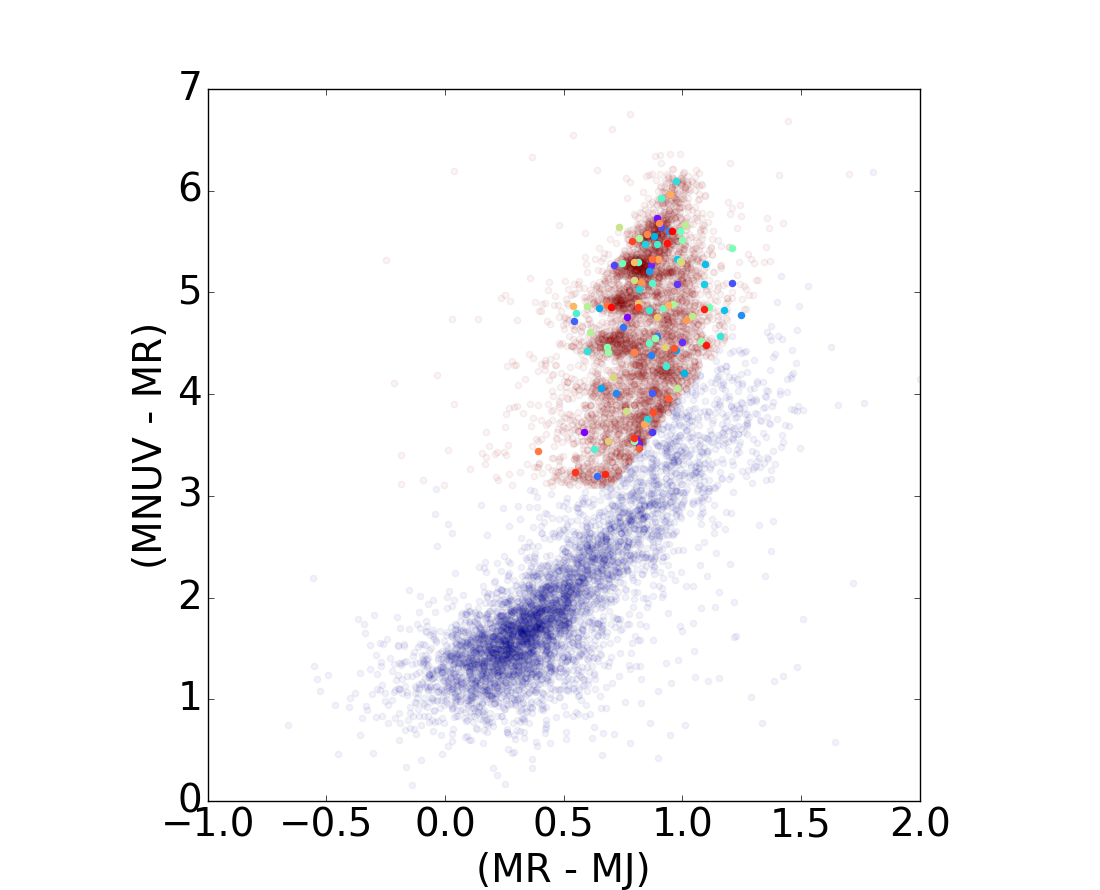}\\
 \small{(e) Star forming - RJNUV} & \small{(f) Quiescent - RJNUV}\\
\end{tabular}
\caption{Projection of the structured queues identified in the DAE diagram, (a) and (b), back to the PCA diagram, (c) and (d), and to the RJ - NUVR diagram, (e) and (f). Top figures display star-forming populations in blue (a) and quiescent sources in red (b) for the test set. Galaxies located in the queues are highlighted on the DAE diagram and projected back to the PCA diagram and to the RJ-NUVR plane following the same color code. It is clear that these structures do not correspond to a particular disposition on the RJ-NUVR diagram and that they present a similar pattern on the PCA and DAE diagrams.}
\label{fig:queue}
\end{figure}

Furthermore, the inspection of the DAE diagram reveals a specific elongated structure arising at the bottom left of the galaxy distribution, that strikingly deviates from the overall trend displayed by  the  main  galaxy sample. This structure is shown in Fig.\,\ref{fig:queue} (a) for the star-forming galaxies and in Fig.\,\ref{fig:queue} (b) for the quiescent galaxies contained in the test set.
% When inspecting the new DAE diagram, a specific elongated structure arises on the bottom left of the galaxies' distribution as shown in Fig. \ref{fig:queue} (a) for the star forming galaxies and in Fig. \ref{fig:queue} (b) for the quiescent galaxies contained in the test set. 
The galaxies falling in this particular structure are projected back to the PCA plane and to the classical RJ-NUVR diagram keeping the same color code, see Fig. \ref{fig:queue} (c)-(f). We note that the ordered pattern identified in the DAE diagram also appears in the PCA diagram, while it has no particular disposition in the RJ-NUVR diagram. This shows that unsupervised learning approaches can potentially unveil SED features that do not appear as obviously in the more standard color diagrams. Addressing the physical origin of this structure goes beyond the scope of this paper, but we already point out that it
%Moreover, these samples 
can be matched to low redshift galaxies, as suggested by Fig. \ref{fig:z_AE} (a). This feature of the DAE and PCA diagrams could thus be produced by a class of sources with very low luminosities, only detectable in the nearby Universe and characterized by
SEDs that are markedly different from those observed at higher redshift. On the other hand, it could also be related to mathematical artefacts owing to the initial SED fitting to obtain rest-frame magnitudes that arise, for instance, from the galaxy magnitudes at short wavelengths ({e.g. }, near-UV) where the SED of  faint sources at low redshift is generally less constrained than 
%that are generally less constrained at shot twaelengrth low-redshift galaxy SEDs at short wavelengths are generally not constrained as tightly as  
the SEDs of more distant sources\footnote{Large-field imaging carried out in space at near-UV wavelengths is typically shallower than the observations obtained from the ground. For the vast majority of low-redshift galaxies, absolute magnitudes in the near-UV thus rely on uncertain extrapolations of the best fit SED from the detection in the $U$-band. Galaxies at higher redshift are less affected by this bias, since their rest-frame near-UV emission is redshifted to the optical bands and thus better constrained.}.
% However, other physical magnitudes may also contribute to the formation of these queues distributions, or they might even be artifacts related to measurement instruments. 

\subsection{Mass distribution}
We  show in this section that the features extracted with the PCA and the denoising autoencoder are strongly linked with the  stellar mass of the galaxies.\\  

\begin{figure*}[htb]
\centering
\begin{tabular}{cc}
\includegraphics[width=0.03\textwidth]{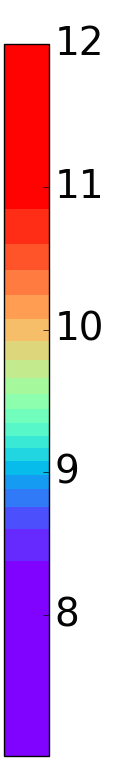}&
\begin{tabular}{cccc}
\includegraphics[width=0.22\textwidth]{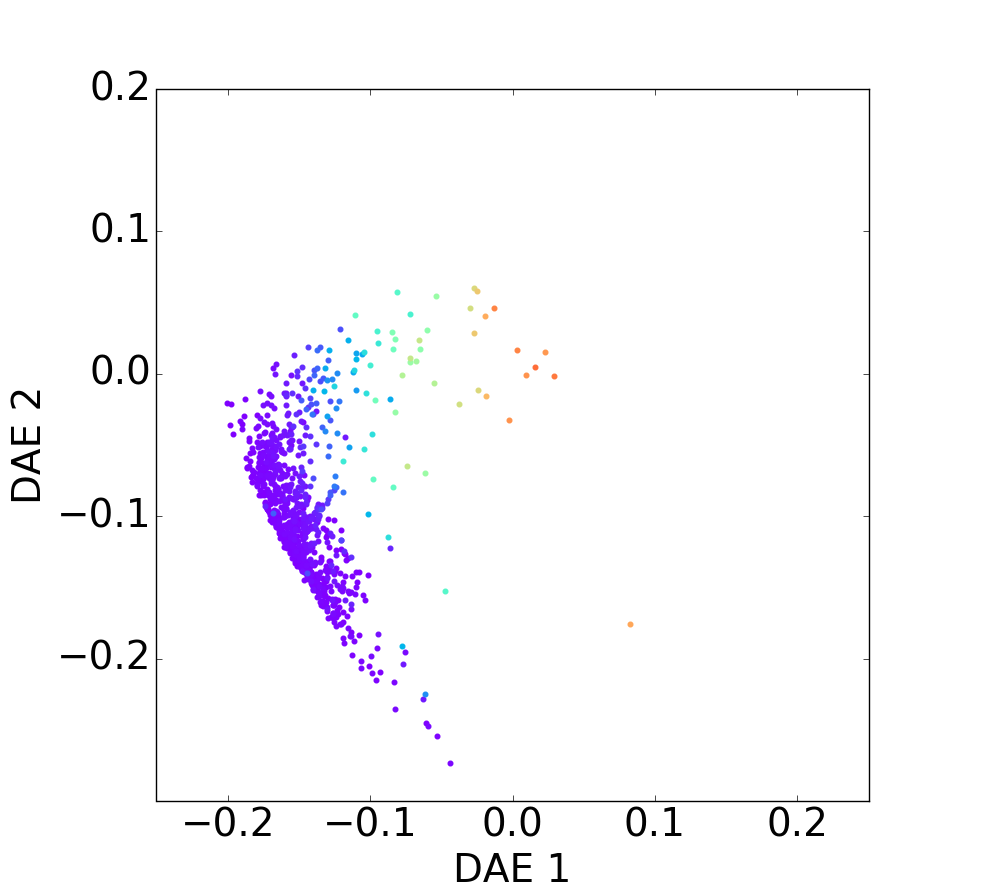}&
\includegraphics[width=0.22\textwidth]{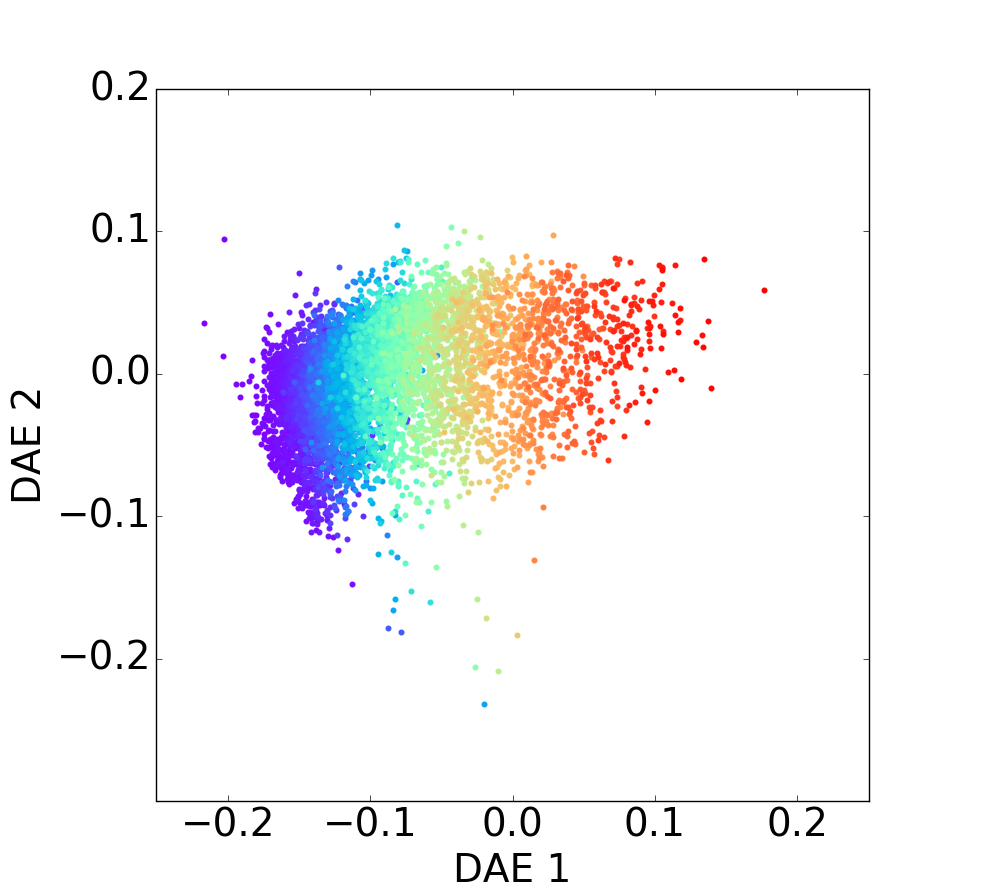}&
\includegraphics[width=0.22\textwidth]{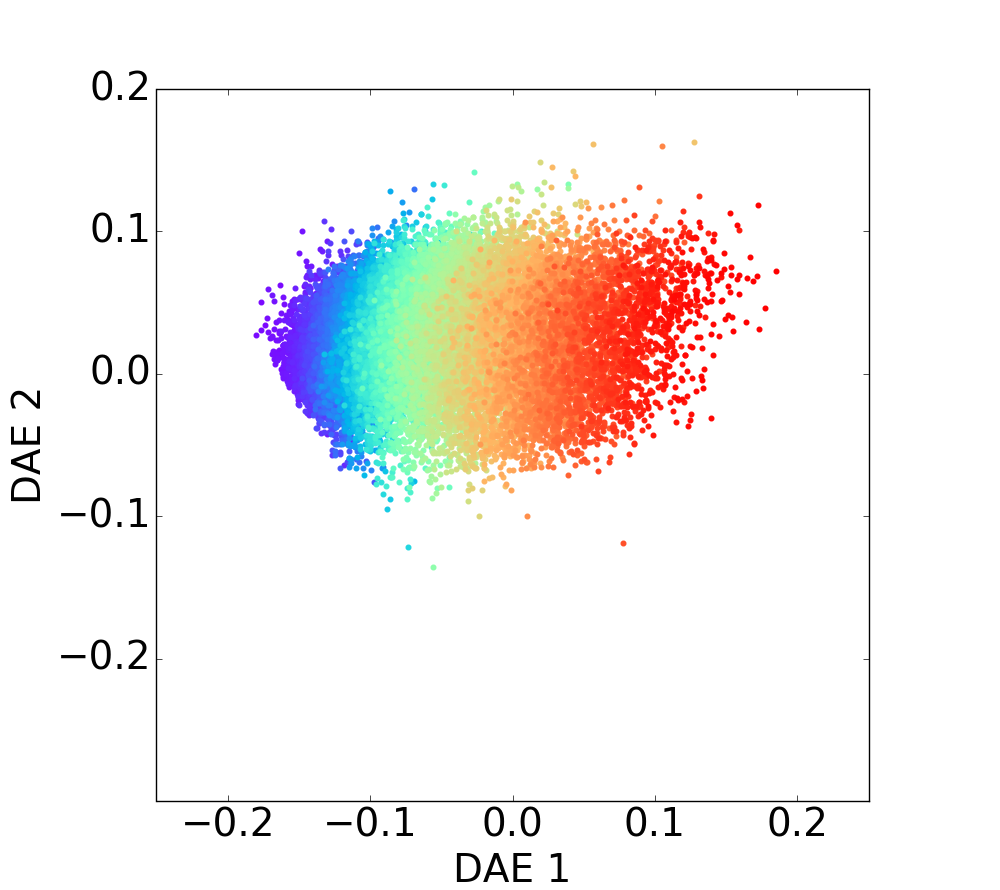}&
\includegraphics[width=0.22\textwidth]{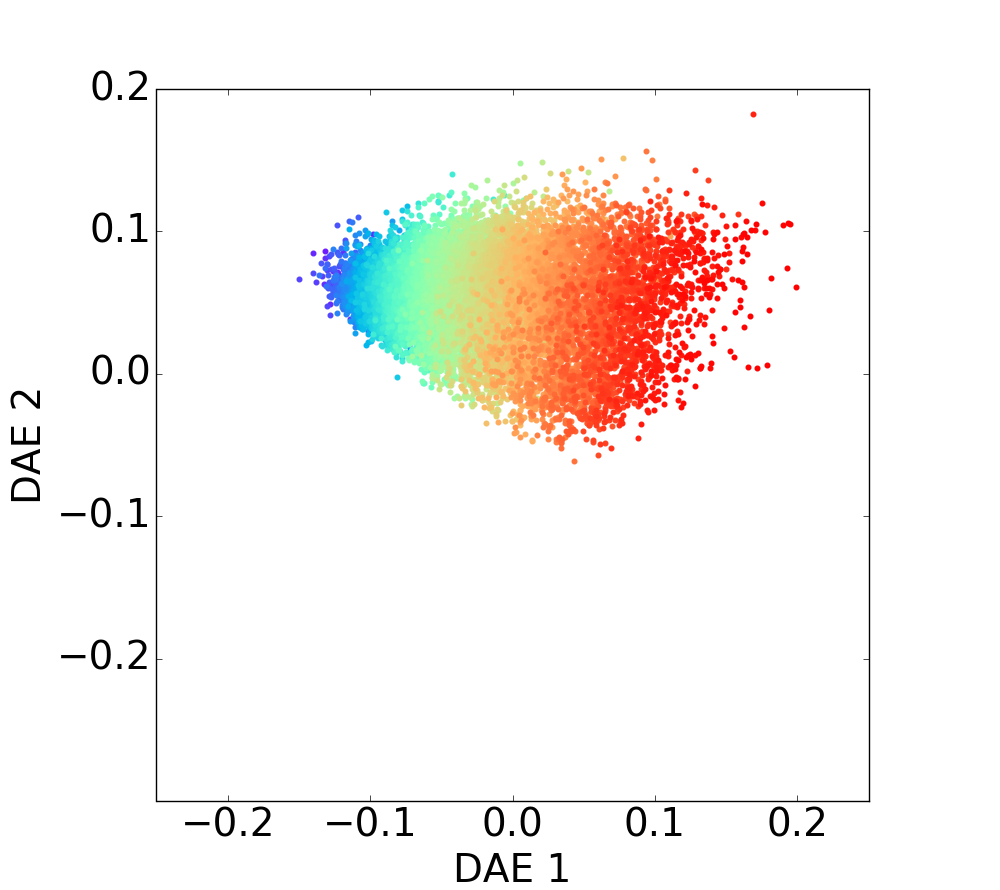}\\ 
\small{(a)  $z \in (0, 0.2]$} & \small{(b) $z \in (0.2, 0.4]$} &  \small{(c) $z \in (0.4, 0.8]$} & \small{(d) $z \in (0.8, 1.0]$}\\
\includegraphics[width=0.22\textwidth]{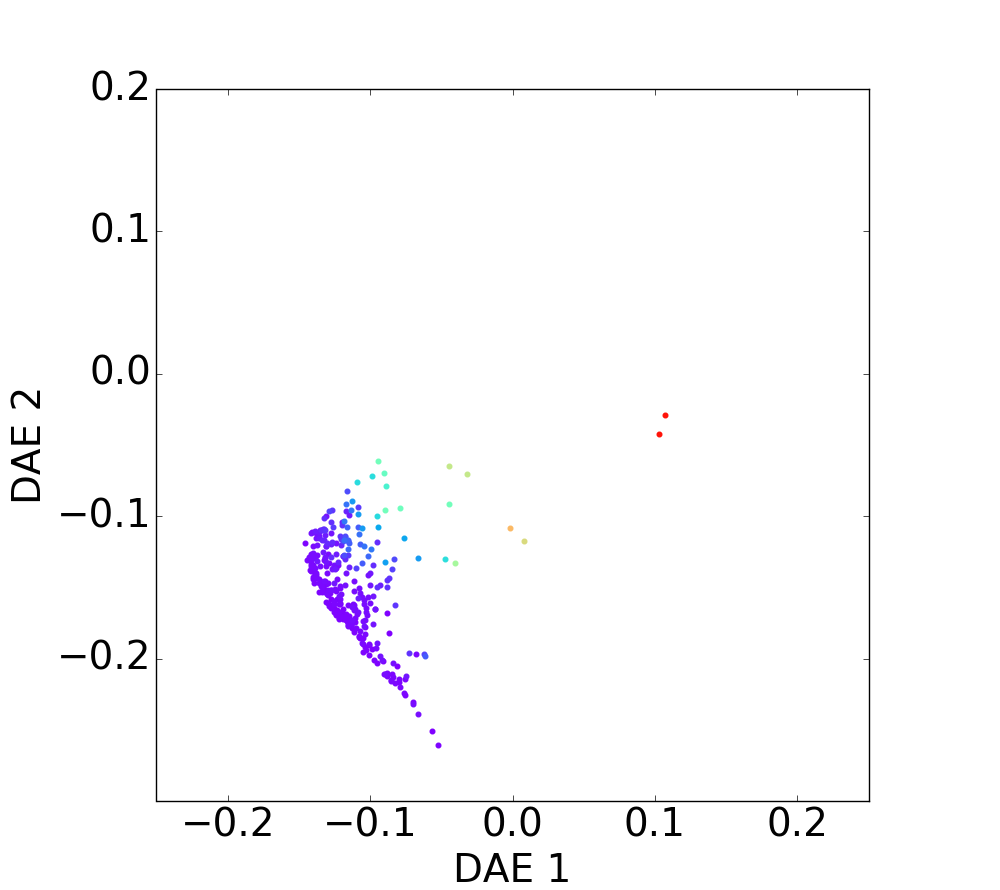}&
\includegraphics[width=0.22\textwidth]{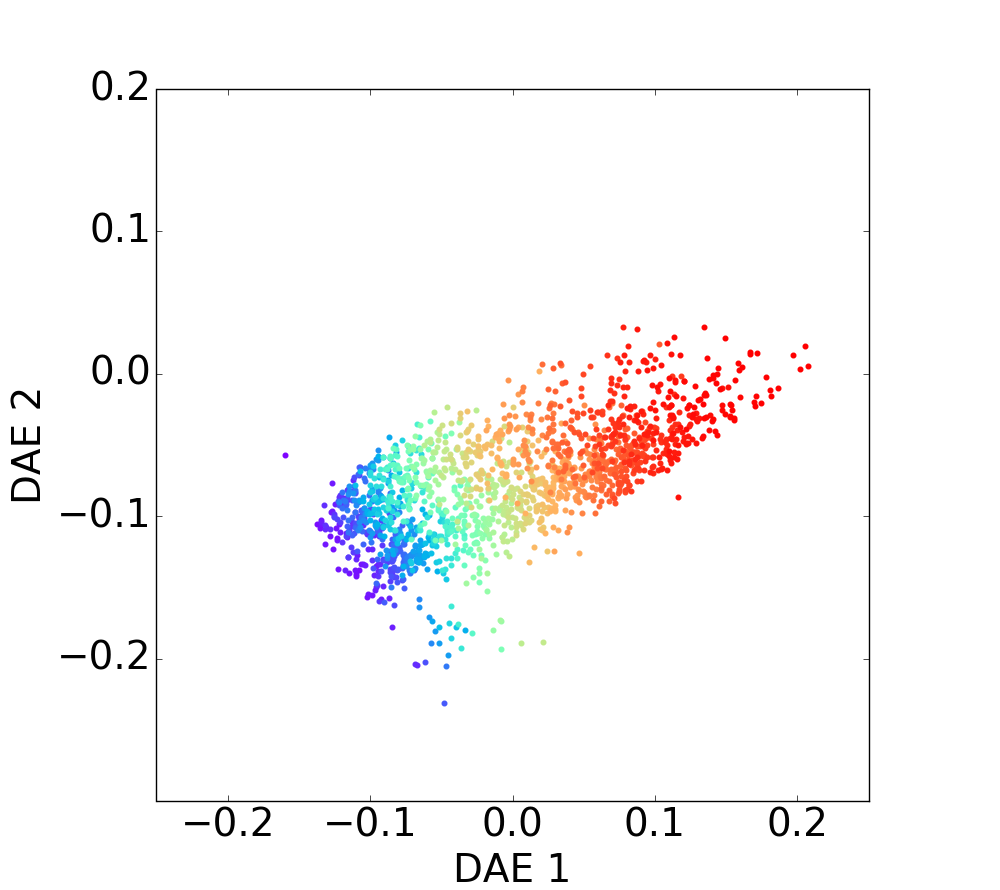}&
\includegraphics[width=0.22\textwidth]{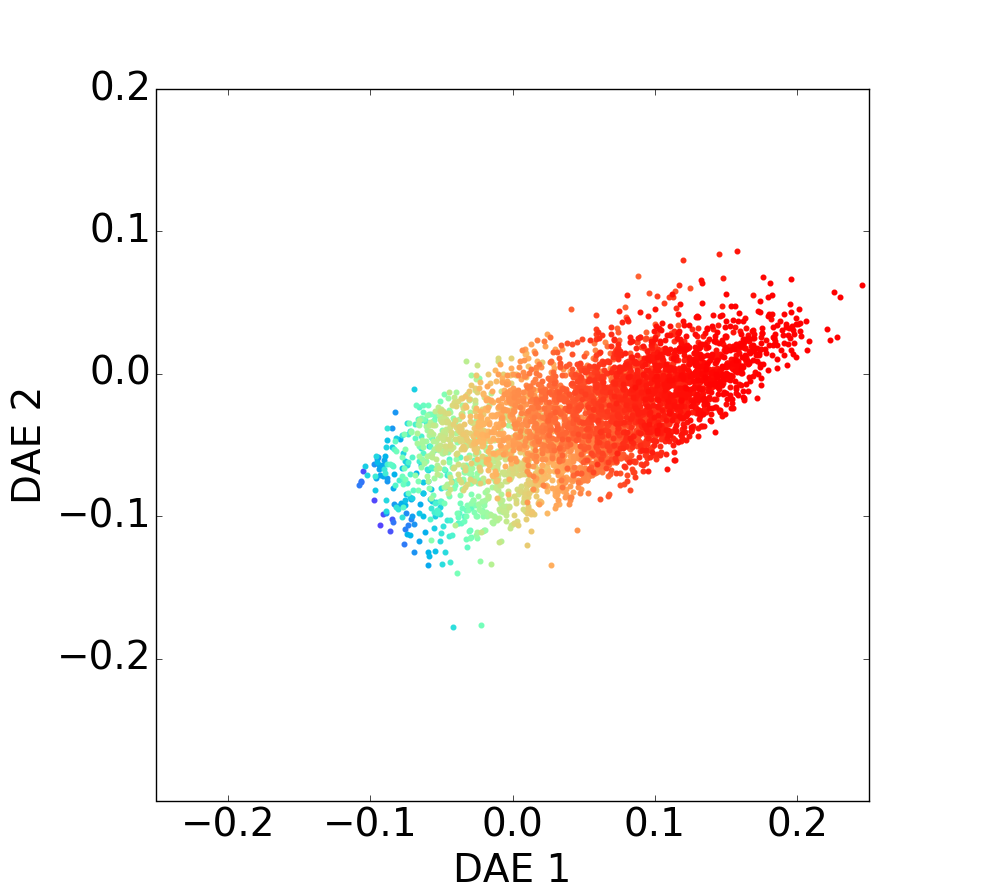}&
\includegraphics[width=0.22\textwidth]{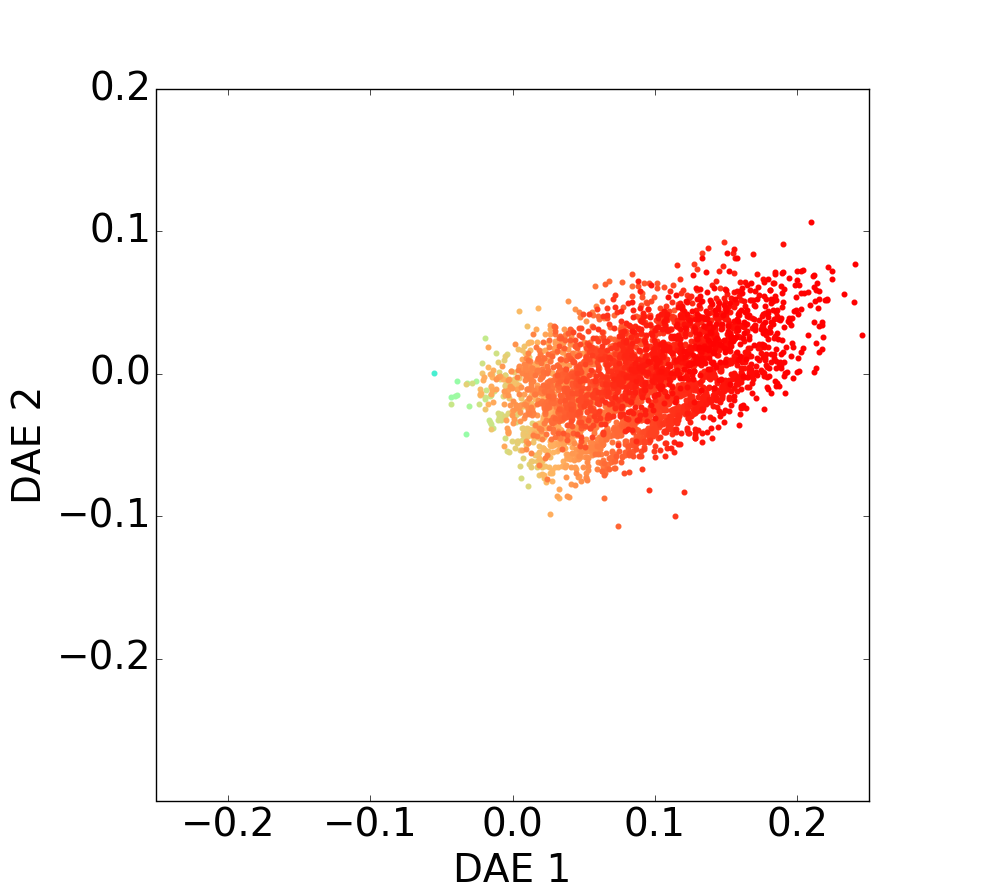}\\
\small{(e)  $z \in (0, 0.2]$} & \small{(f) $z \in (0.2, 0.4]$} &  \small{(g) $z \in (0.4, 0.8]$} & \small{(h) $z \in (0.8, 1.0]$} \\& 
\end{tabular}
\end{tabular}
\caption{Distribution of the mass of the galaxies on the DAE diagram through the different redshift bins. Figures on top row correspond to star-forming populations for different redshift bins and figures on the bottom to quiescent populations for the same redshift bins. The color code matches the galaxy stellar mass taken from the COSMOS2015 catalog. From these figures, it is worth pointing out the relationship between the proposed DAE diagram and the stellar mass. Notably, the value of the mass increases with the first component of the diagram.}
\label{fig:mass_AE}
\end{figure*}

\begin{figure*}[htb]
\centering
\begin{tabular}{cc}
\includegraphics[width=0.03\textwidth]{colorbar.png}&
\begin{tabular}{cccc}
\includegraphics[width=0.22\textwidth]{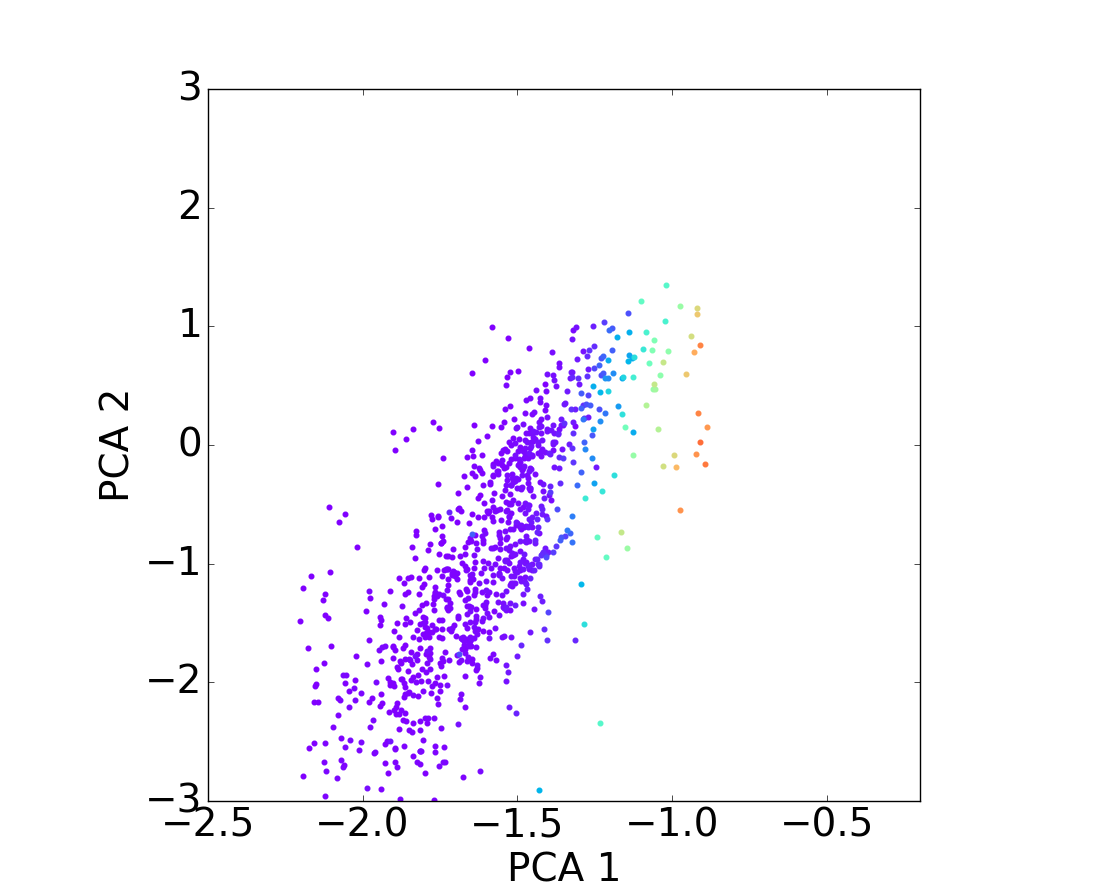}&
\includegraphics[width=0.22\textwidth]{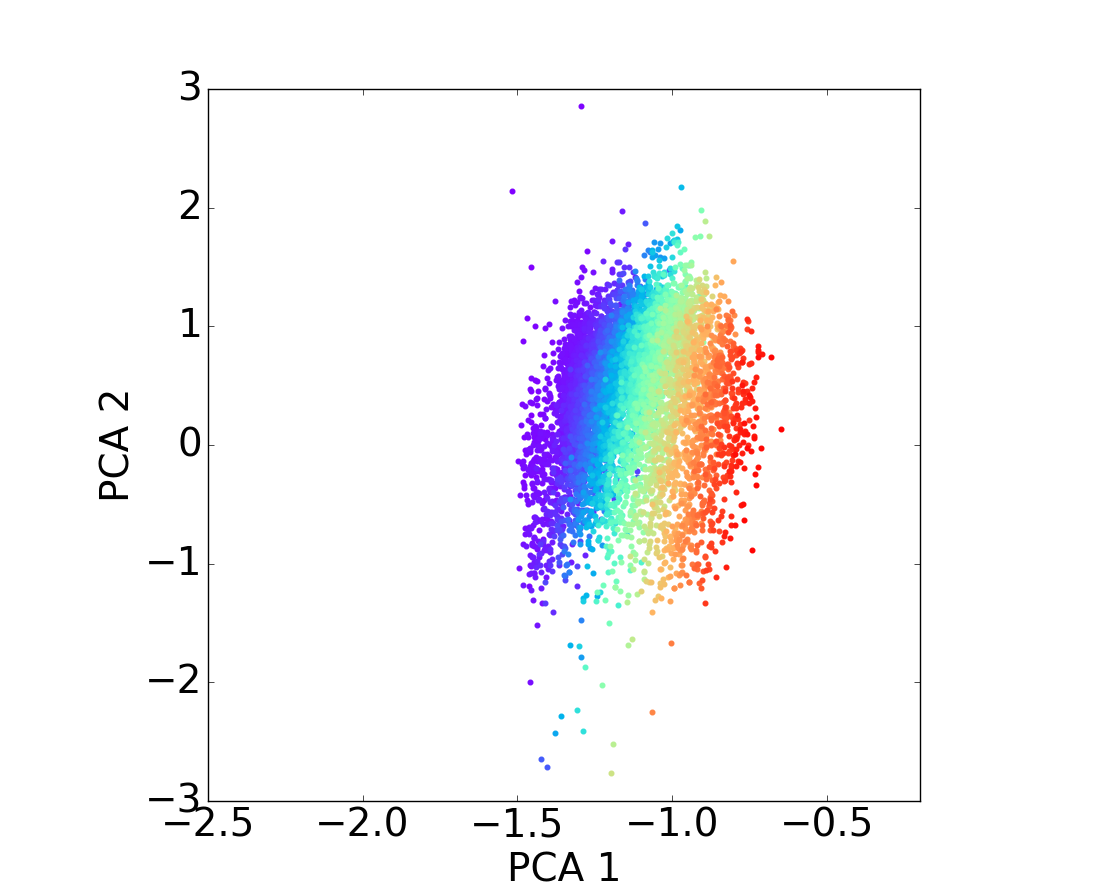}&
\includegraphics[width=0.22\textwidth]{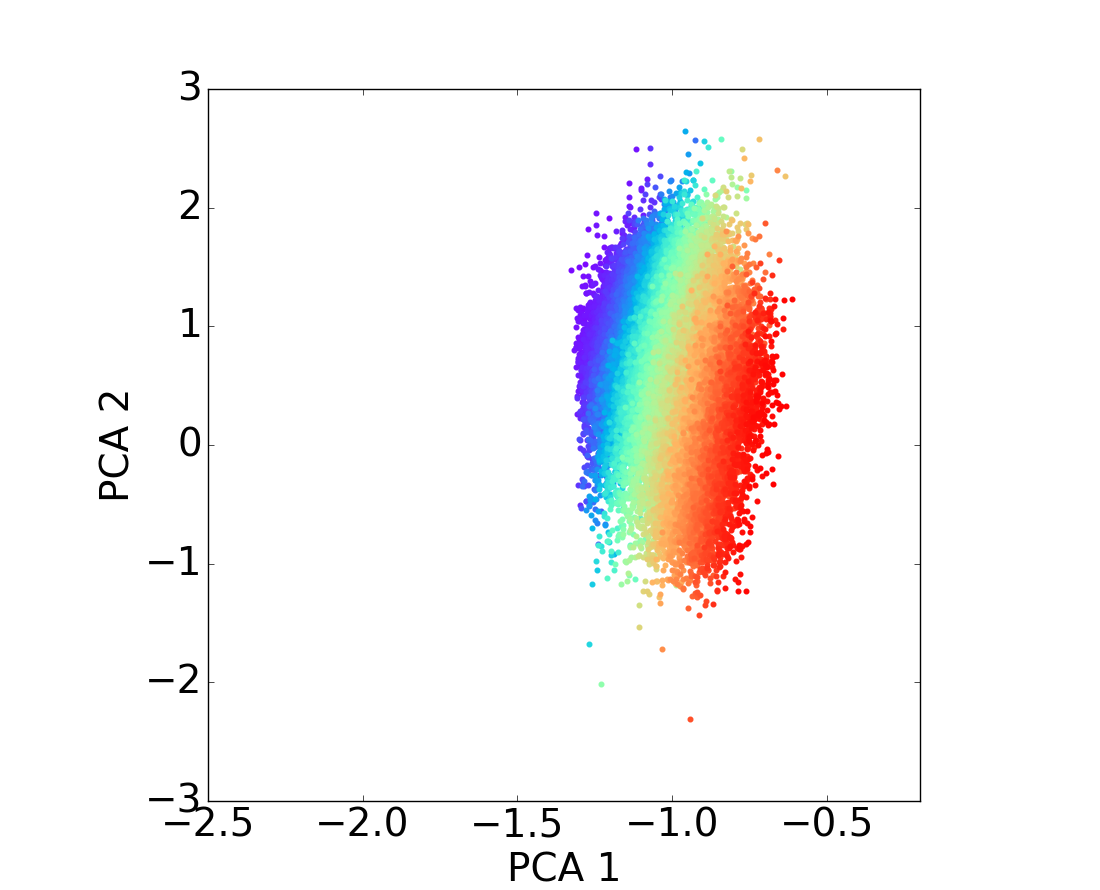}&
\includegraphics[width=0.22\textwidth]{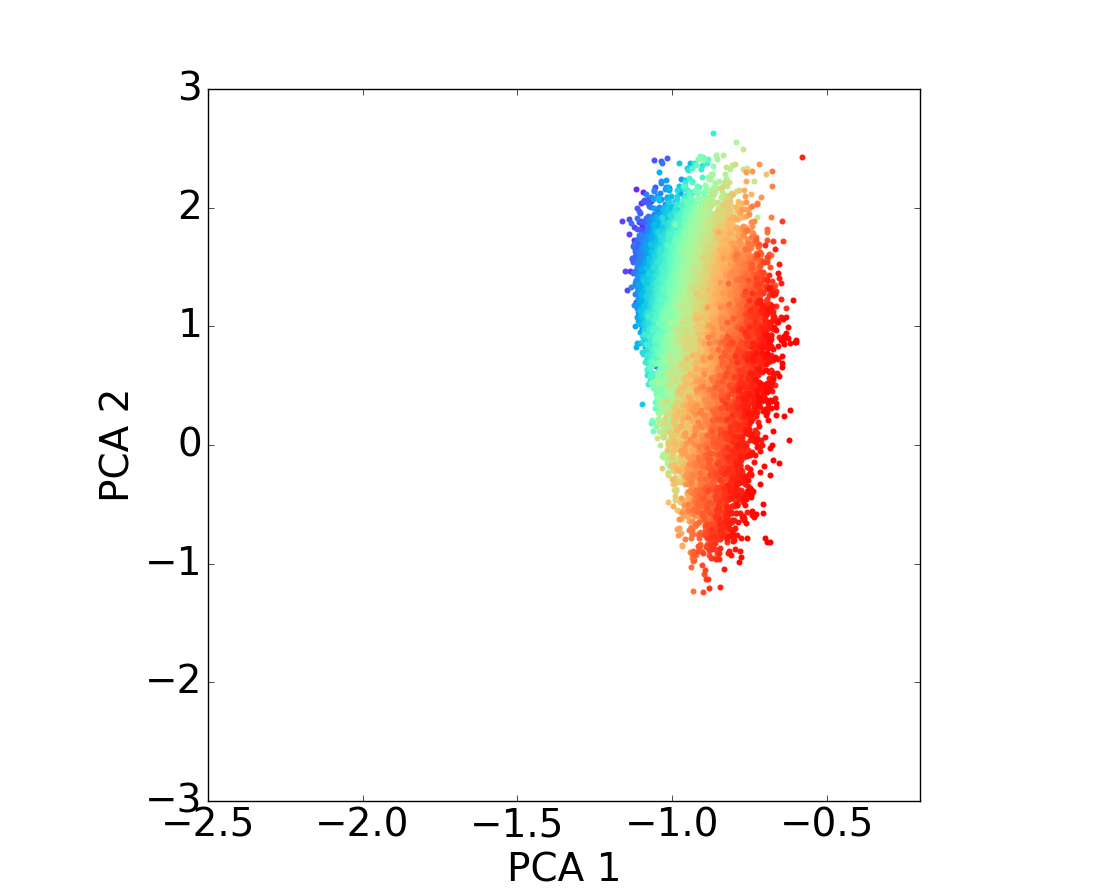}\\ 
\small{(a)  $z \in (0, 0.2]$} & \small{(b) $z \in (0.2, 0.4]$} &  \small{(c) $z \in (0.4, 0.8]$} & \small{(d) $z \in (0.8, 1.0]$}\\
\includegraphics[width=0.22\textwidth]{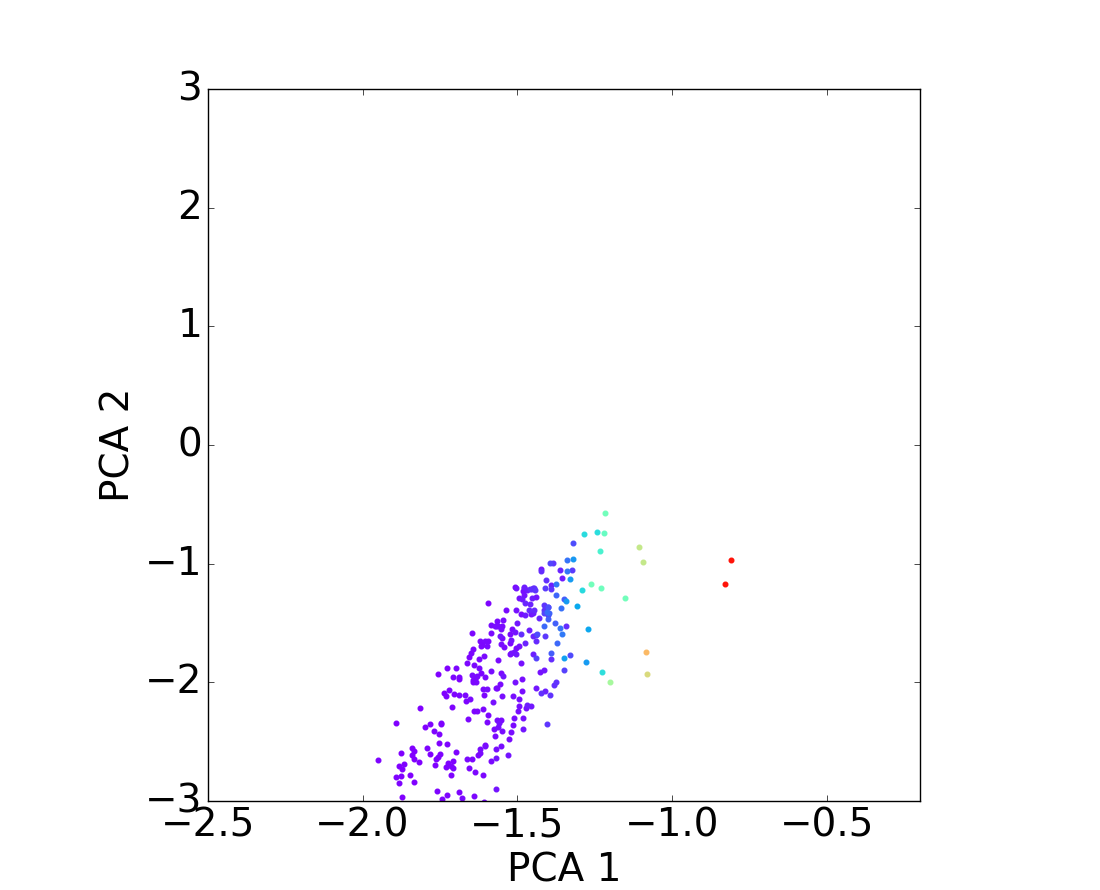}&
\includegraphics[width=0.22\textwidth]{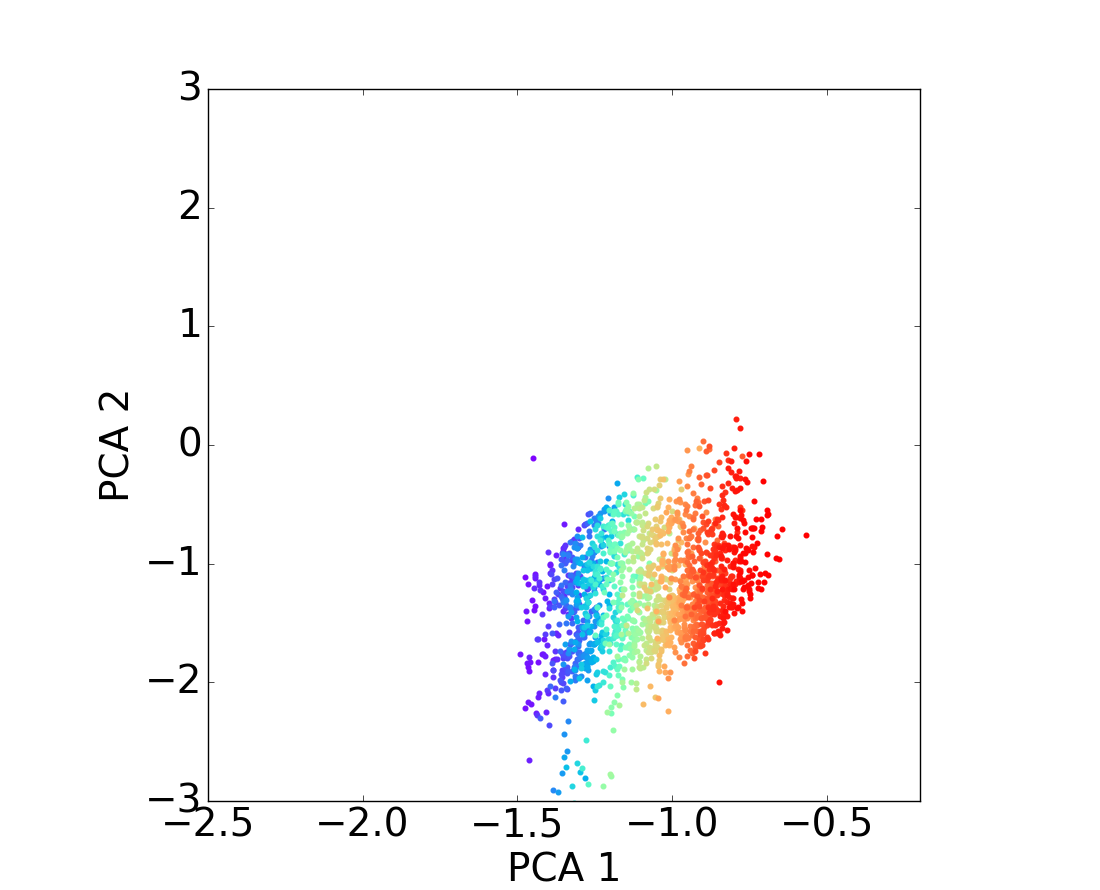}&
\includegraphics[width=0.22\textwidth]{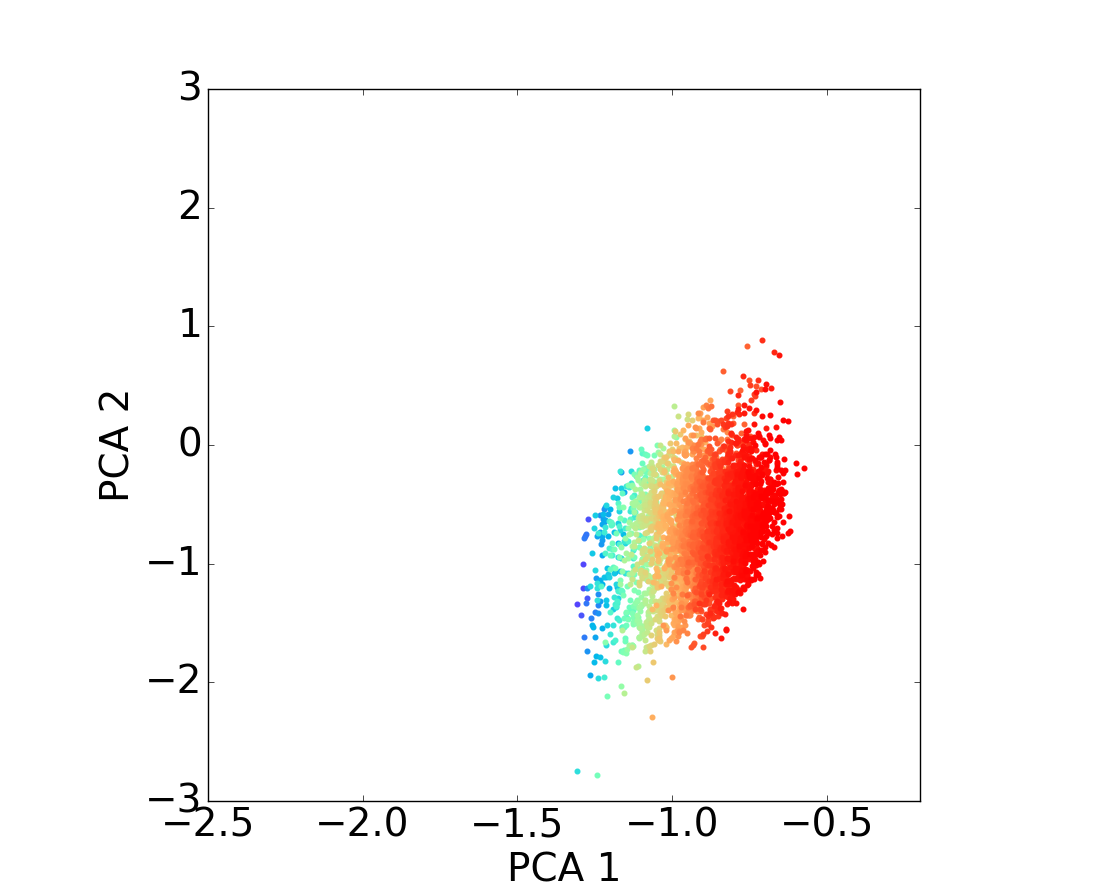}&
\includegraphics[width=0.22\textwidth]{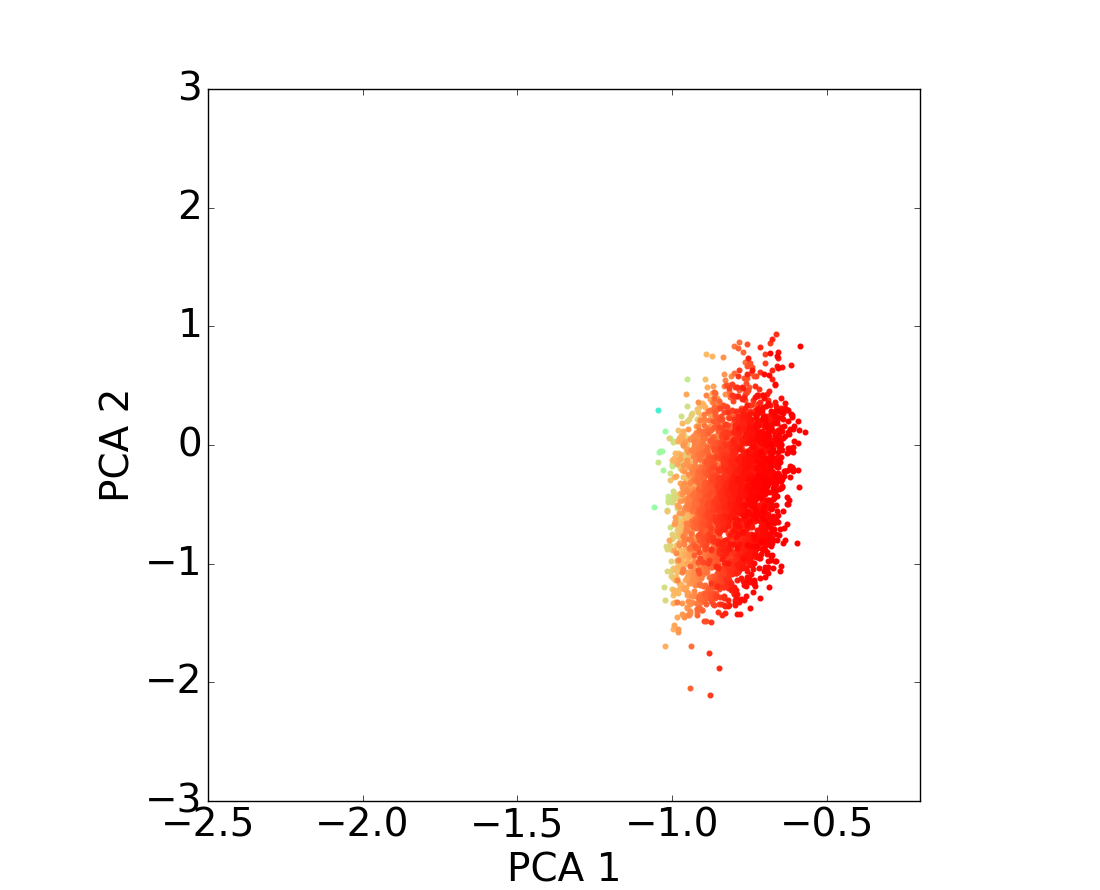}\\
\small{(e)  $z \in (0, 0.2]$} & \small{(f) $z \in (0.2, 0.4]$} &  \small{(g) $z \in (0.4, 0.8]$} & \small{(h) $z \in (0.8, 1.0]$} \\& 
\end{tabular}
\end{tabular}
\caption{Distribution of the mass of the galaxies on the PCA diagram through the different redshift bins. As above, figures on top row correspond to star-forming populations for different redshift bins and figures on the bottom to quiescent populations for the same redshift bins; and the color code matches the galaxy stellar mass taken from the COSMOS2015 catalog. From these figures, it is easy to notice a strong relationship between the first PCA component and the stellar mass.}
\label{fig:mass_PCA}
\end{figure*}

\begin{figure*}[h]
\centering
\begin{tabular}{cc}
\includegraphics[width=0.03\textwidth]{colorbar.png}&
\begin{tabular}{cccc}
\includegraphics[width=0.22\textwidth]{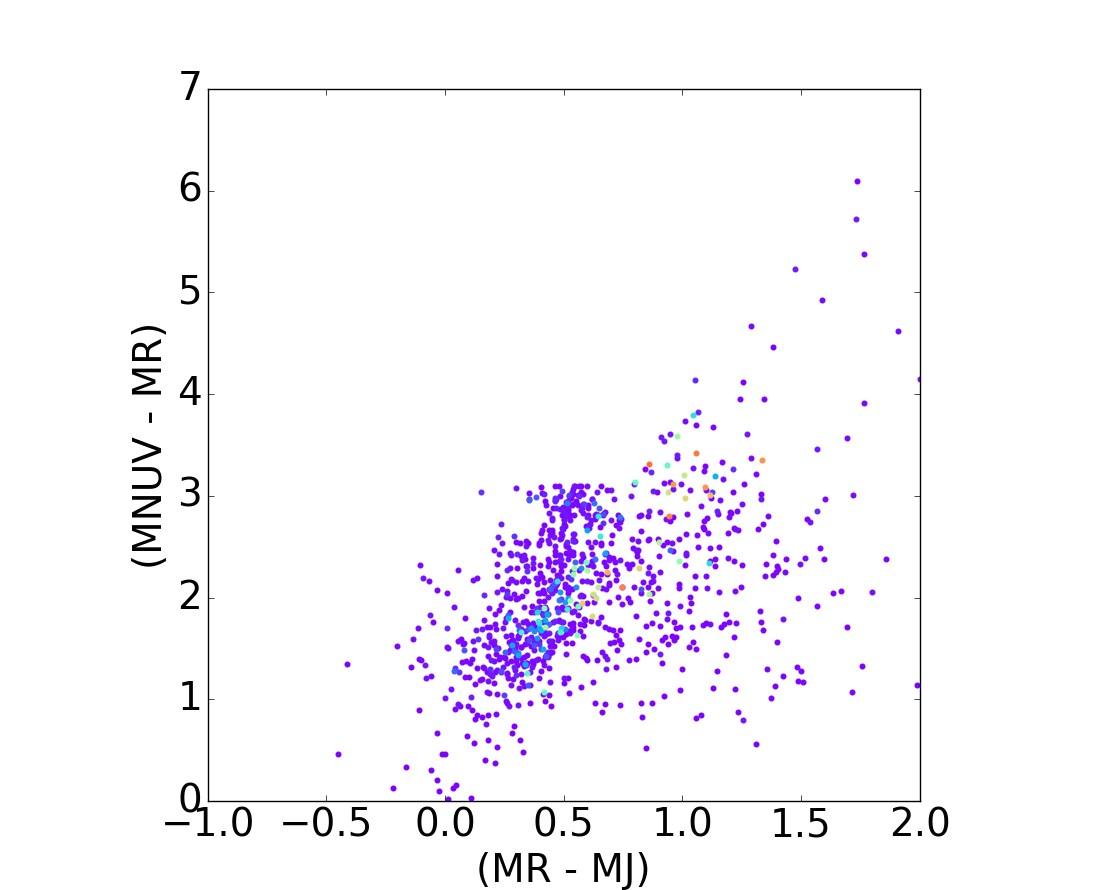}&
\includegraphics[width=0.22\textwidth]{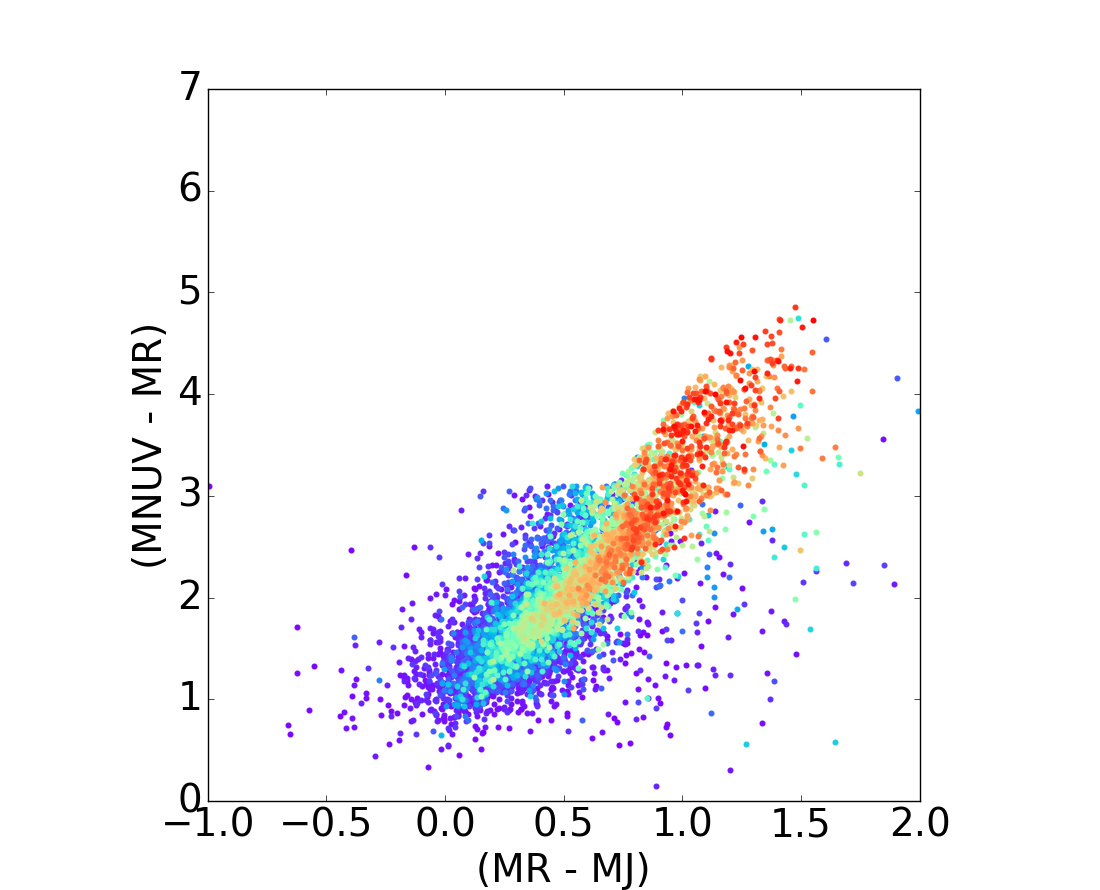}&
\includegraphics[width=0.22\textwidth]{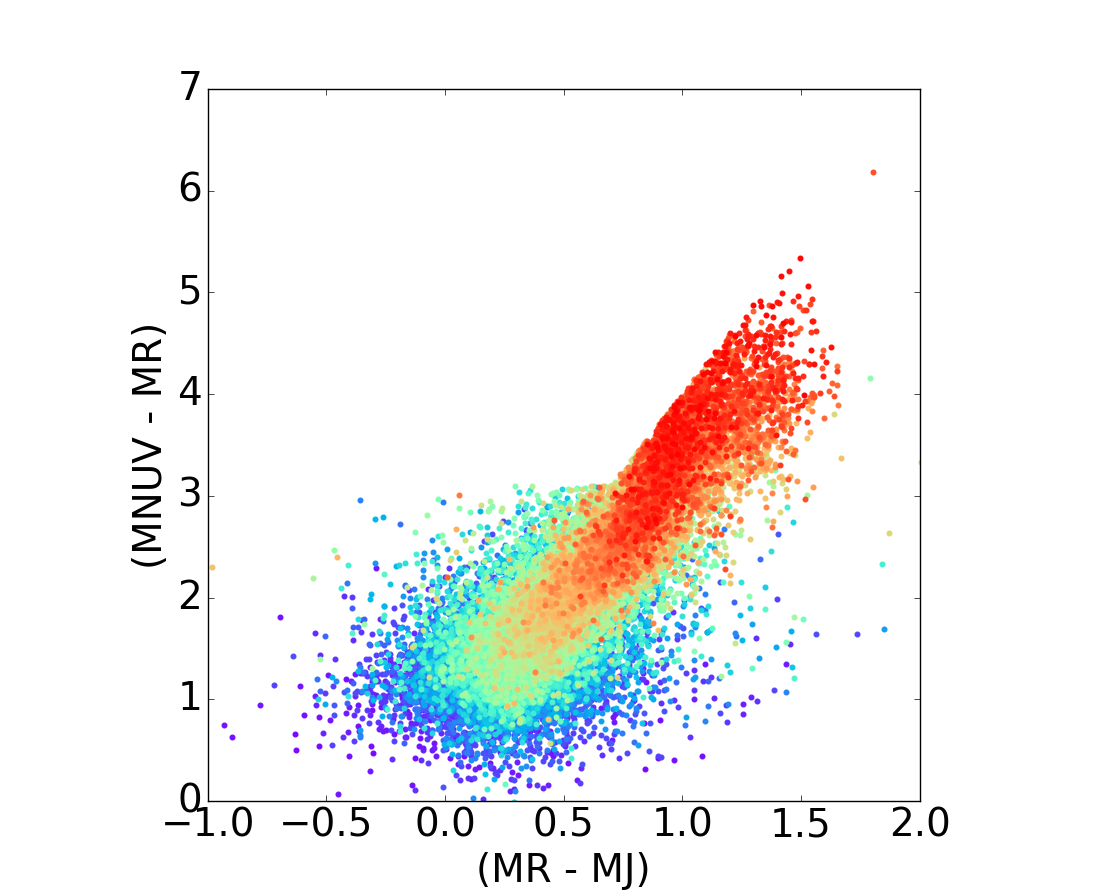}&
\includegraphics[width=0.22\textwidth]{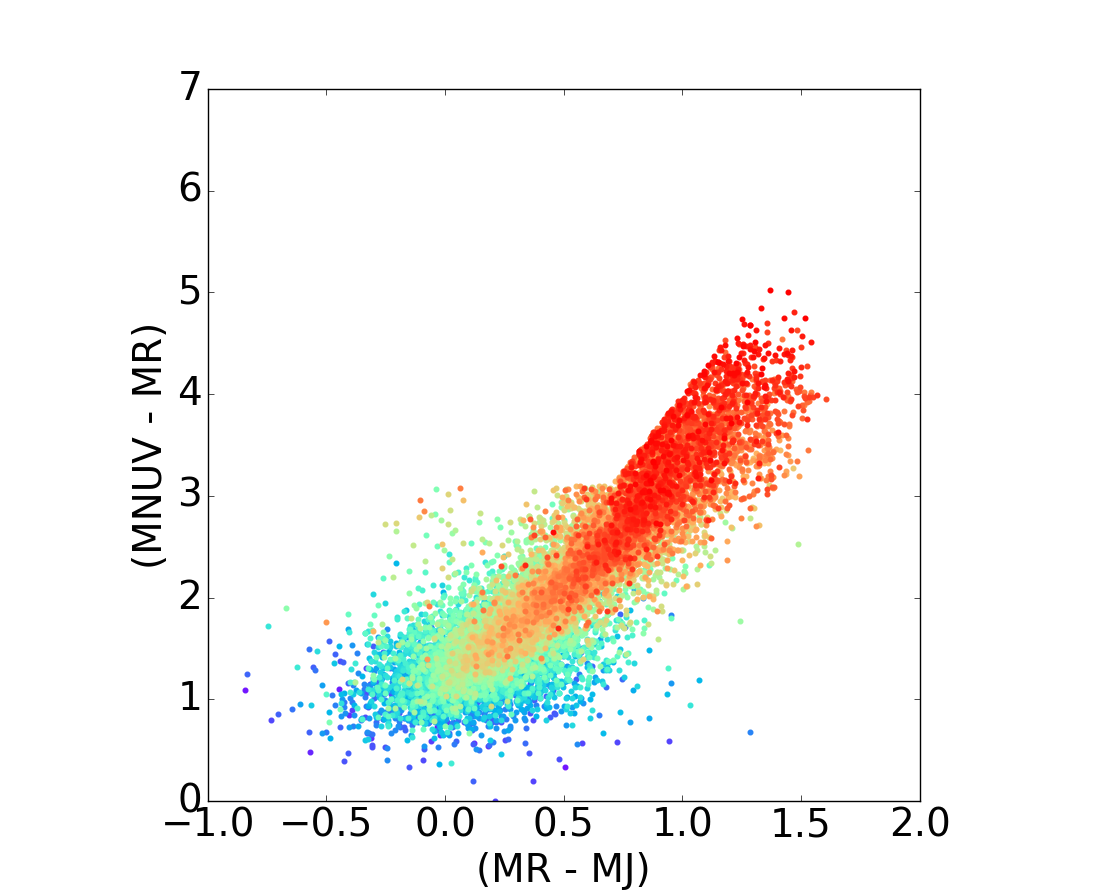}\\
\small{(a)  $z \in (0, 0.2]$} & \small{(b) $z \in (0.2, 0.4]$} &  \small{(c) $z \in (0.4, 0.8]$} & \small{(d) $z \in (0.8, 1.0]$}\\
\includegraphics[width=0.22\textwidth]{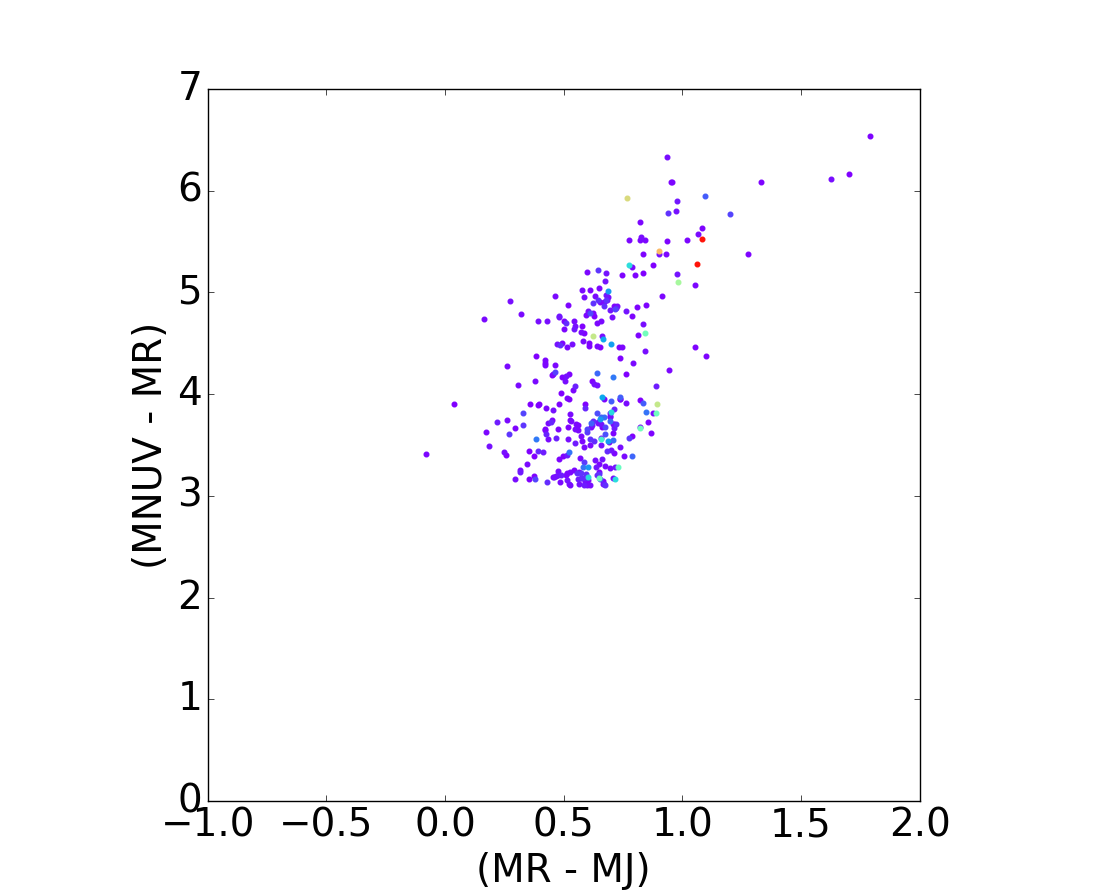}&
\includegraphics[width=0.22\textwidth]{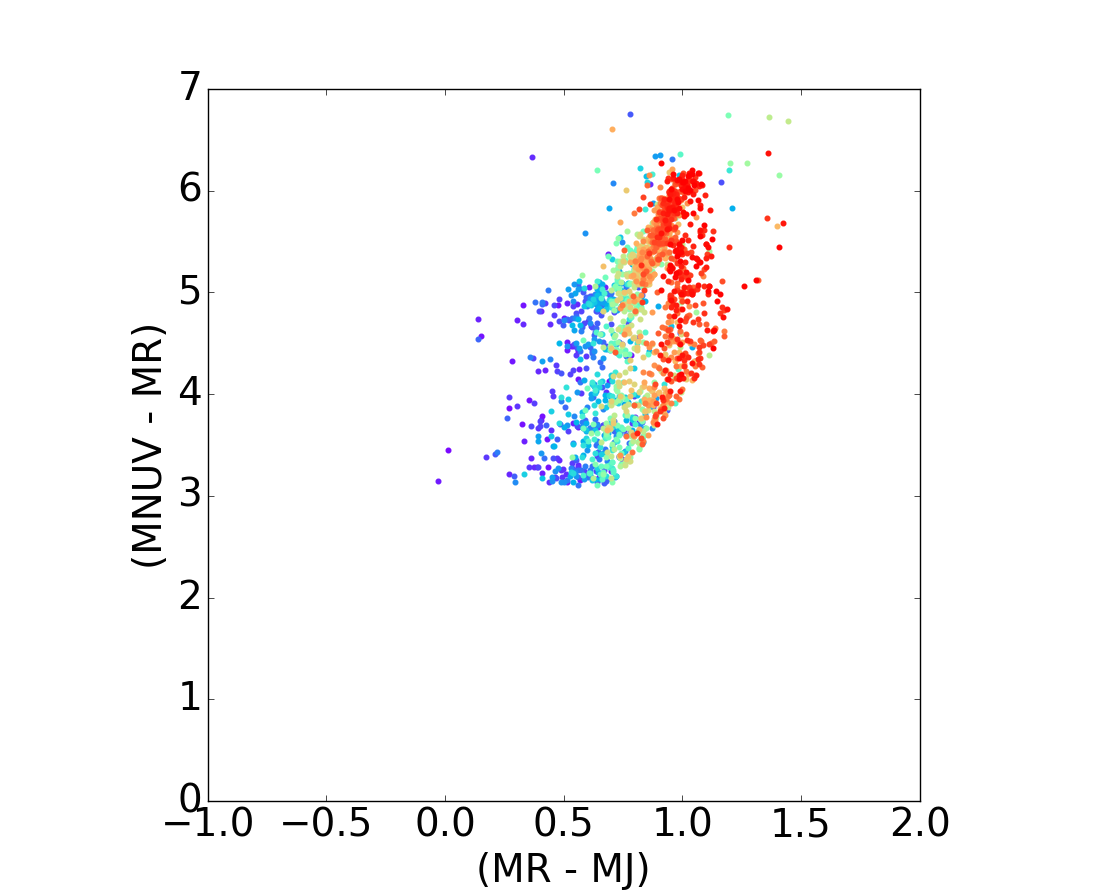}&
\includegraphics[width=0.22\textwidth]{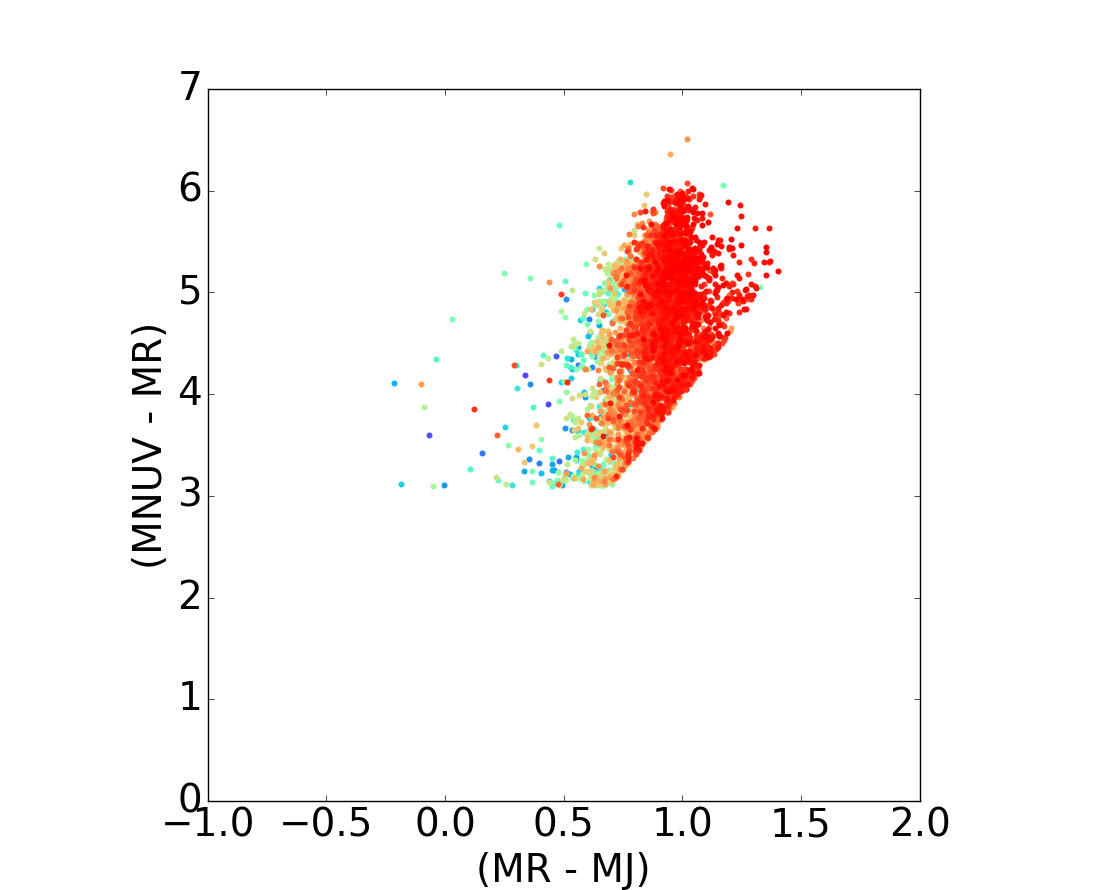}&
\includegraphics[width=0.22\textwidth]{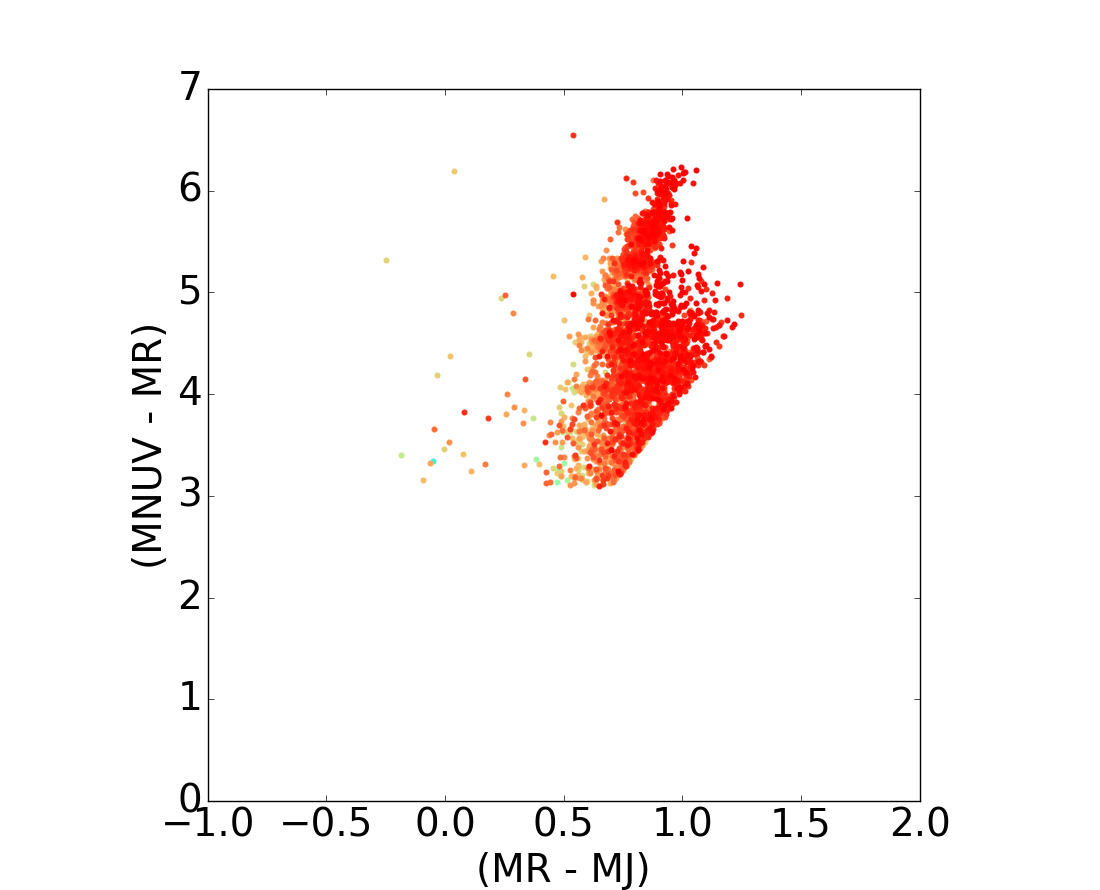}\\
\small{(e)  $z \in (0, 0.2]$} & \small{(f) $z \in (0.2, 0.4]$} &  \small{(g) $z \in (0.4, 0.8]$} & \small{(h) $z \in (0.8, 1.0]$}\\
\end{tabular}
\end{tabular}
\caption{Distribution of the mass of the galaxies on the RJ - NUVR diagram through the different redshift bins. As for the previous figure, top row corresponds to star-forming galaxies for different redshift bins and the bottom to quiescent populations for the same redshift bins. Once again, the color code matches the galaxy stellar mass taken from the COSMOS2015 catalog. The RJ - NUVR plane does not exhibit any special disposition related to the stellar mass.}
\label{fig:mass_cc}
\end{figure*}

\begin{figure}[h!]
\begin{tabular}{cc}
\centering
\includegraphics[width=0.24\textwidth]{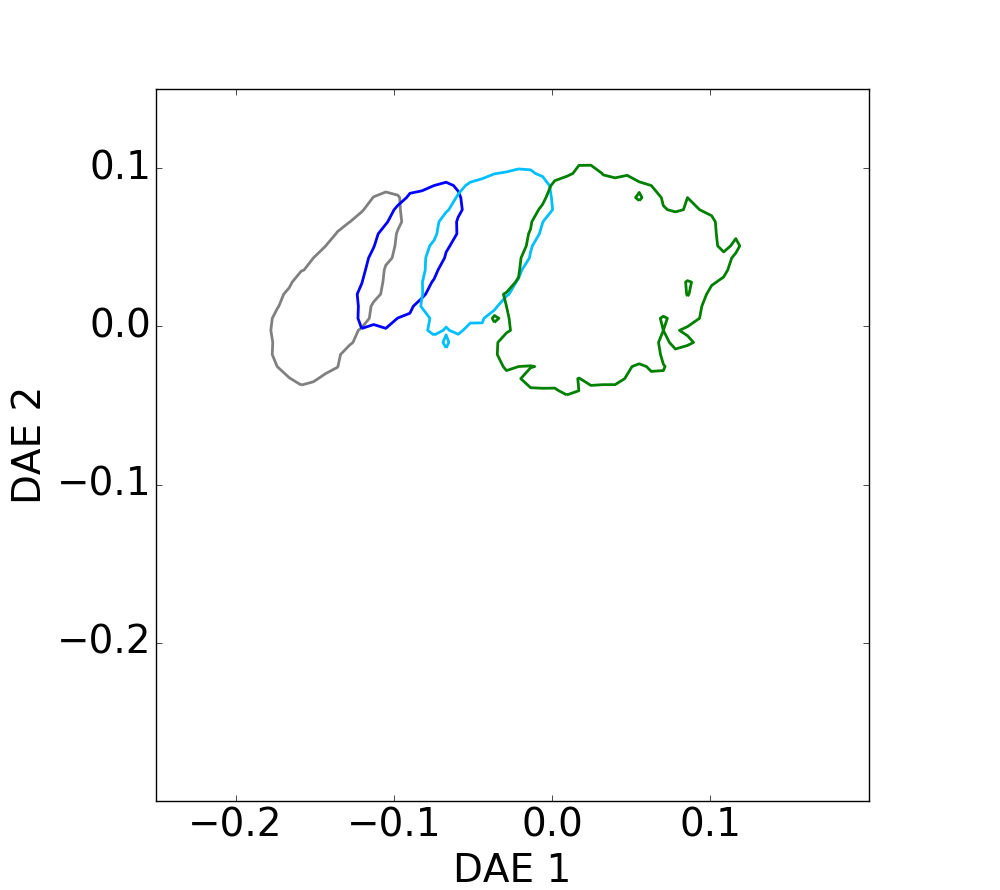}&
\includegraphics[width=0.24\textwidth]{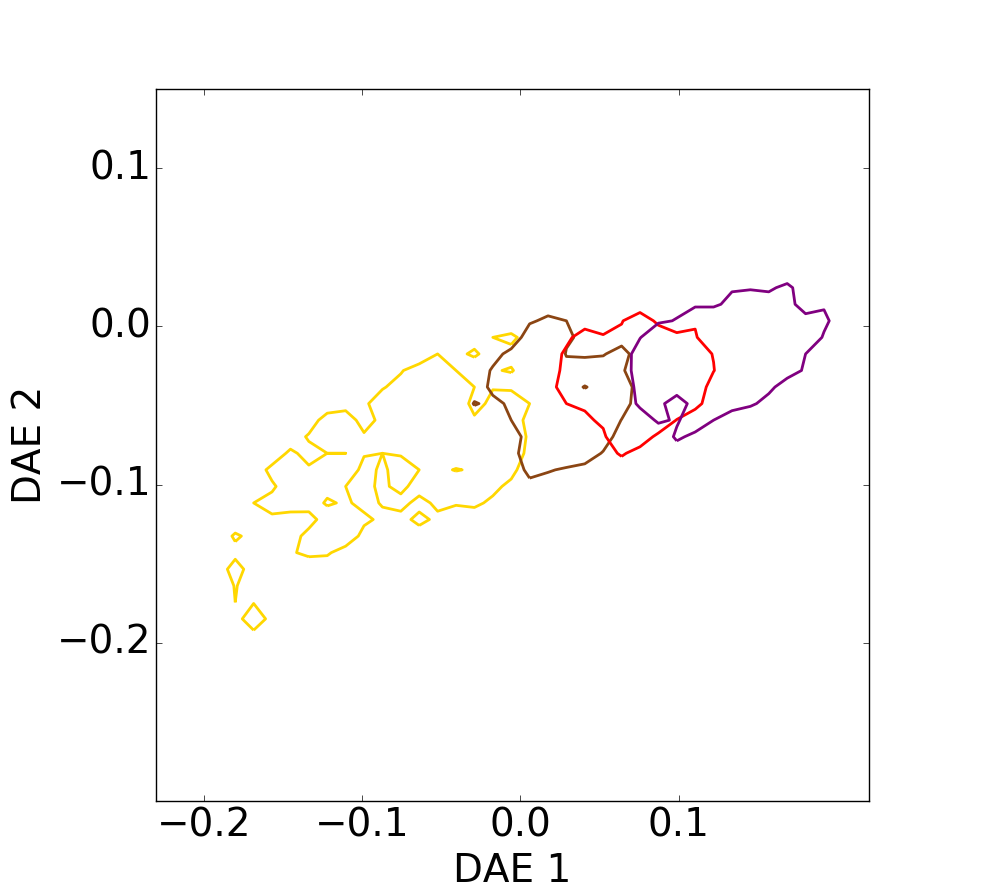}\\
\small{(a) Star forming - DAE}& \small{(b) Quiescent - DAE} \\
\includegraphics[width=0.24\textwidth]{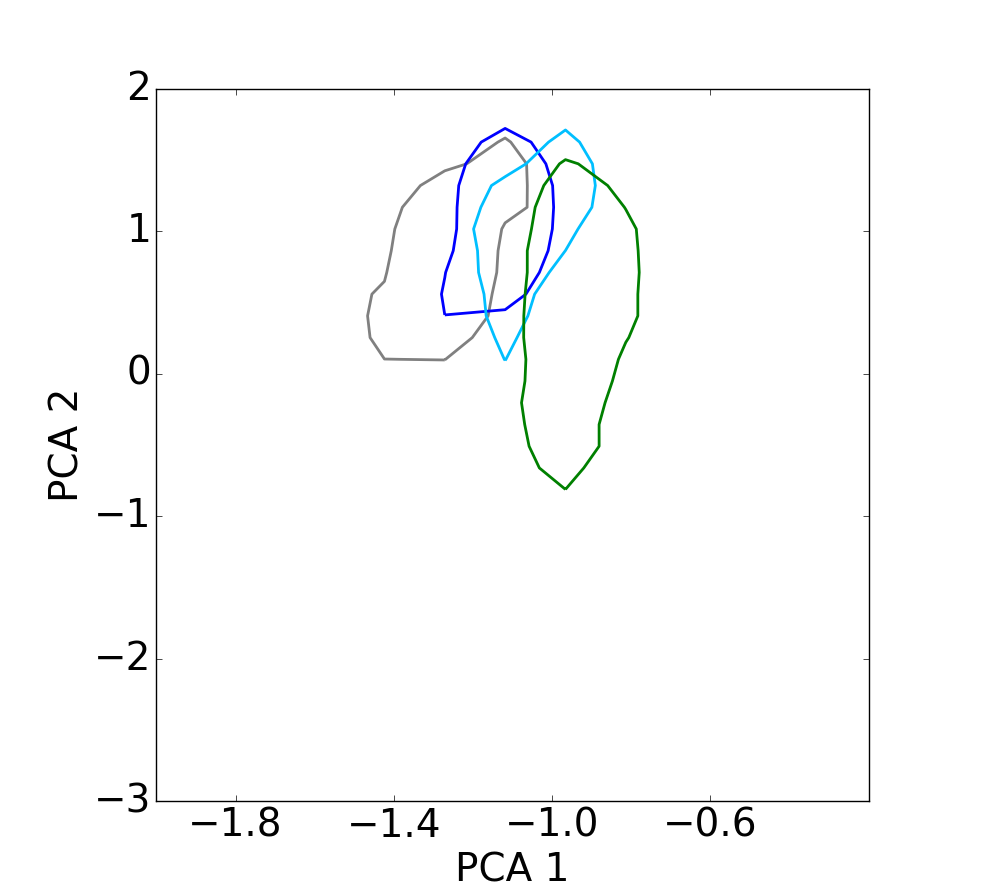}&
\includegraphics[width=0.24\textwidth]{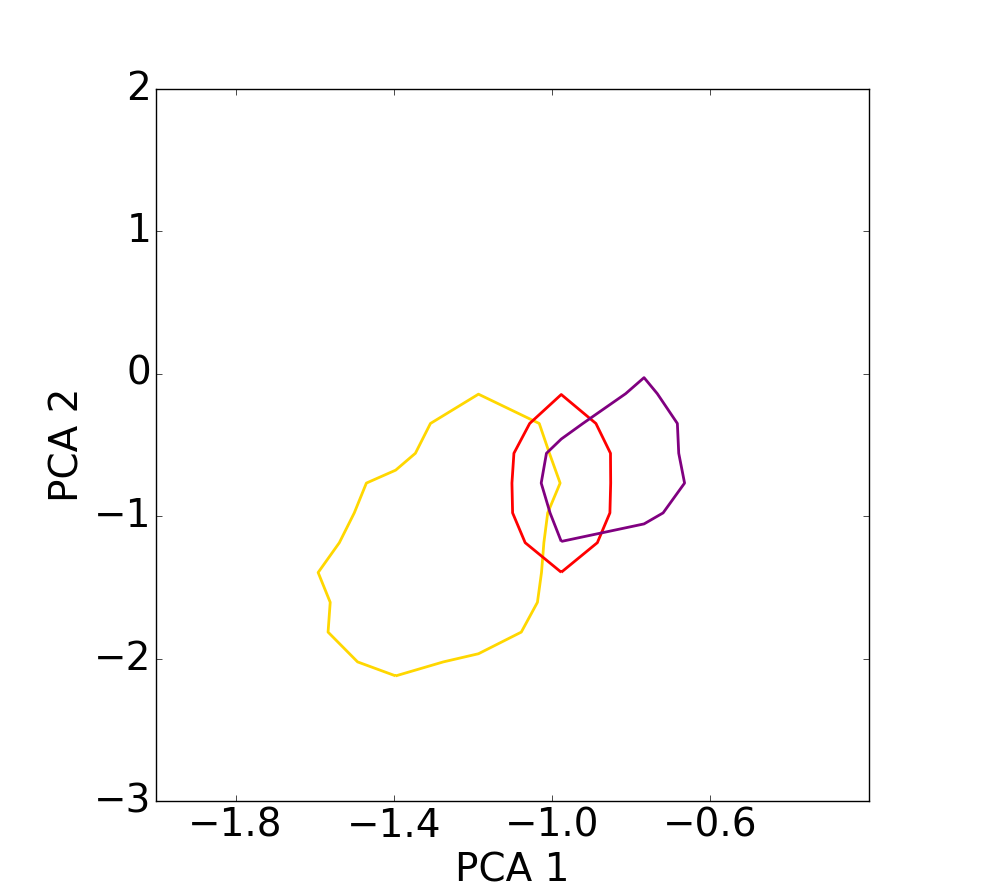}\\
 \small{(c) Star forming - PCA} & \small{(d) Quiescent - PCA}\\
\includegraphics[width=0.24\textwidth]{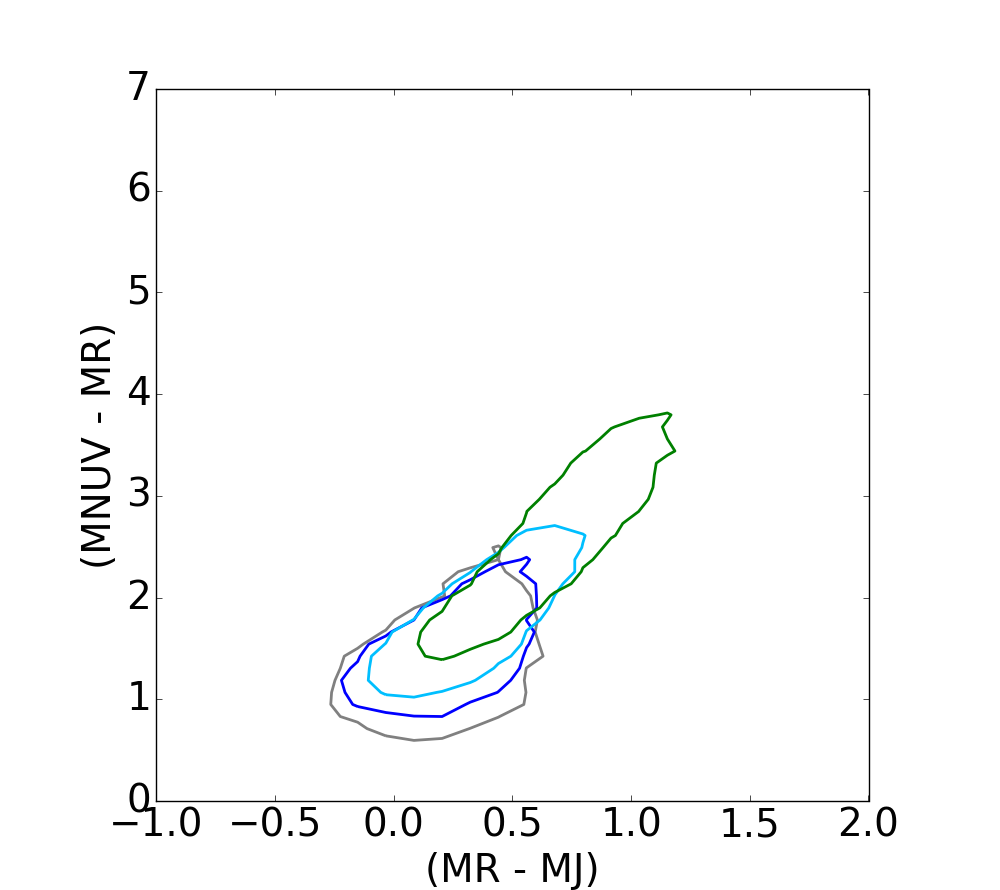}&
\includegraphics[width=0.24\textwidth]{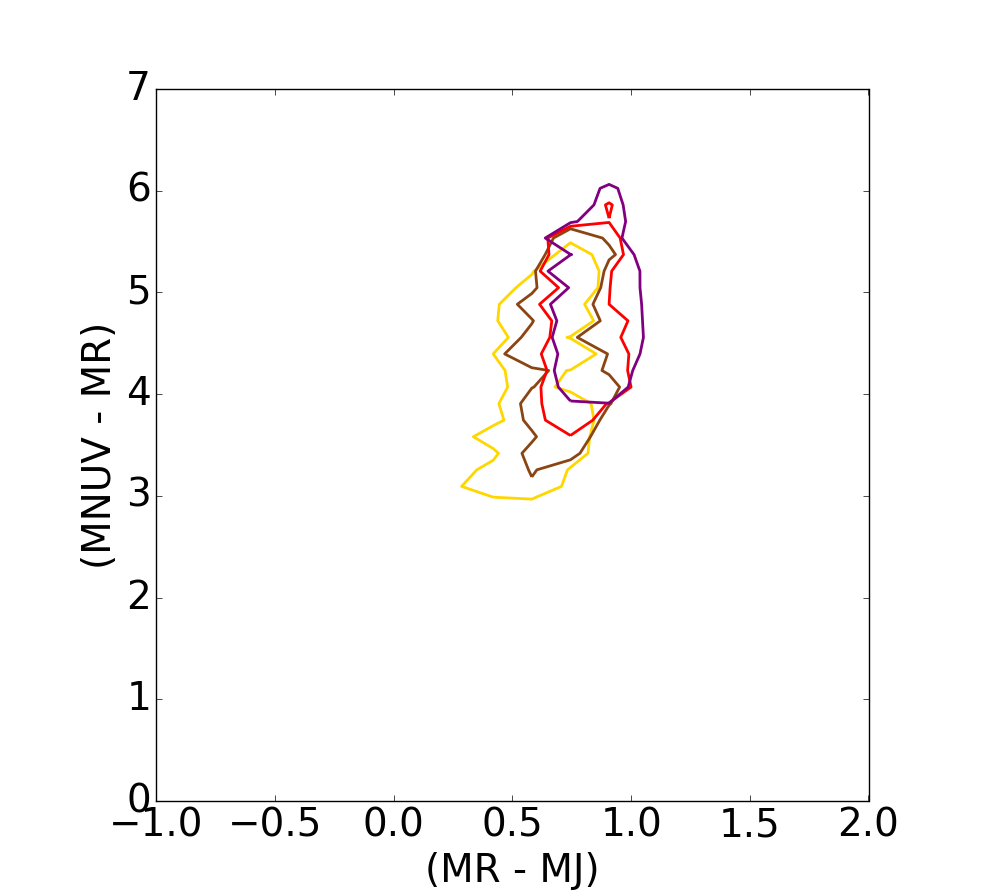}\\
 \small{(e) Star forming - RJNUV} & \small{(f) Quiescent - RJNUV}\\
\end{tabular}
\caption{Contour plots for different mass bins on the DAE diagram, on the PCA diagram and on the RJ - NUVR plane. The mass range is split over four bins and the contour plots are displayed at $75\%$ of the samples in the different mass bins. The DAE diagram is displayed on the top, PCA diagram at the center, and the RJ-NUVR diagram on the bottom. The contour's colors grow from grey to green for star-forming galaxies, and from yellow to purple for quiescent galaxies, respectively. These plots attest to the relationship with the stellar mass captured by the DAE and PCA diagrams for both star-forming and quiescent populations, while the overlap in the RJ - NUVR diagram is considerably higher.}
\label{fig:mass_bins}
\end{figure}

%From the COSMOS2015 catalog, first we use the best-fit model for each galaxy and obtain the corresponding mass.
Figure\,\ref{fig:mass_AE} shows the distribution of the stellar mass of the galaxies over the DAE diagram for the different redshift bins that were considered earlier. Here, the galaxy stellar mass is taken from the COSMOS2015 catalog and was obtained as  the  mass of the best SED fit model obtained for each galaxy. 
The top row corresponds to star-forming galaxies and the bottom row to quiescent populations.In addition, the same galaxies are displayed in  Fig.\,\ref{fig:mass_PCA} for the PCA diagram and in Fig. \ref{fig:mass_cc} for the classical RJ-NUVR planes for comparison purposes. From Figs.\,\ref{fig:mass_AE} and \ref{fig:mass_PCA}, the relationship between the representations and the mass can naturally be  established, and it is clearly observed for all the redshift bins explored here. It is  noticeable that the value of the mass increases with the first DAE and the first PCA component. On the other hand, the classical RJ-NUVR diagram does not present such a clear disposition for the mass as shown in Fig.\,\ref{fig:mass_cc}.\\
This fact is also illustrated in Fig.\,\ref{fig:mass_bins}, where the contour plots at 75\%  of the population are depicted. Fig.\,\ref{fig:mass_bins} shows the mass range split over four bins for DAE diagram on the top row, for the PCA at the center, and for the RJ-NUVR diagram on the bottom row. We note that the overlap is significantly lower for the DAE diagram, compared to the PCA diagram and to the RJ-NUVR plane. The data-driven representations present a clear evolution over the axis in terms of mass for both star-forming and quiescent populations, while this evolution is barely observed in 
%however no specific disposition related to the mass can be identified on 
the classical  RJ-NUVR diagram.
These distinct behaviors of the stellar mass distributions in the diagrams thus reveal variations of the galaxy SED properties that are again more efficiently extracted with the unsupervised learning techniques than when just taking  the  $NUV$$-$$r^+$ and $r^+-J$ rest-frame colors into consideration. 

Given that the stellar mass of galaxies tightly correlates with their near-infrared luminosities, the mass distributions observed in Fig.\,\ref{fig:mass_AE} may suggest that the PCA and the DAE results at a given redshift are partially driven by the variations of the near-IR magnitudes that can be probed at this kind of redshift. On the other hand, the stellar mass also correlates with other physical properties such as the dust extinction and the activity of star formation, which both modify the overall shape of galaxy SEDs and necessarily also impact  the DAE outputs. For instance, galaxies with enhanced recent star formation display SEDs characterized by a more important contribution of young and massive stars, which  produces intense radiations of UV photons  boosting the overall galaxy emission at short wavelengths. The star formation rate (SFR) of galaxies also correlates  with their stellar mass, but in a non-linear relation such that the SFR per unit of mass mildly decreases in more massive sources \citep{Speagle14}. The relative level of star-forming activity in galaxies  could thus also  contribute to the apparent variation of the DAE and PCA distributions with the stellar mass.

%\clearpage

\subsection{Specific star formation rate}

\begin{figure}[h!]
\begin{tabular}{cc}
\centering
\includegraphics[width=0.24\textwidth]{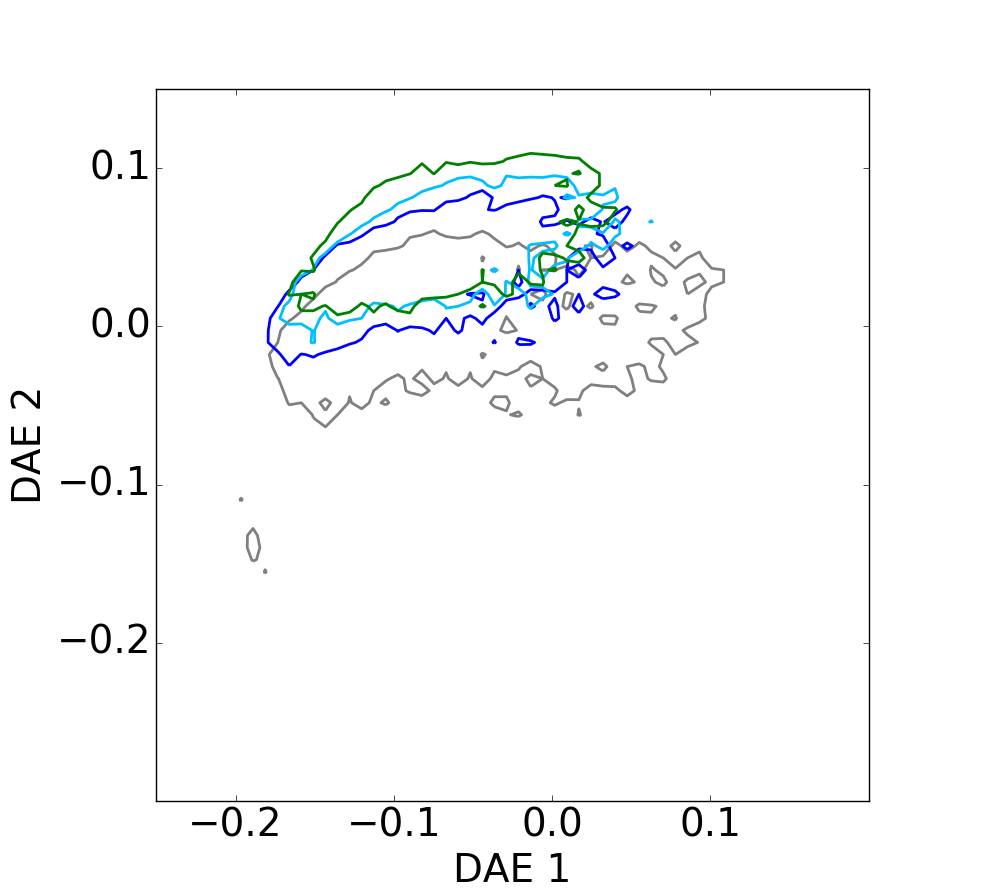}&
\includegraphics[width=0.24\textwidth]{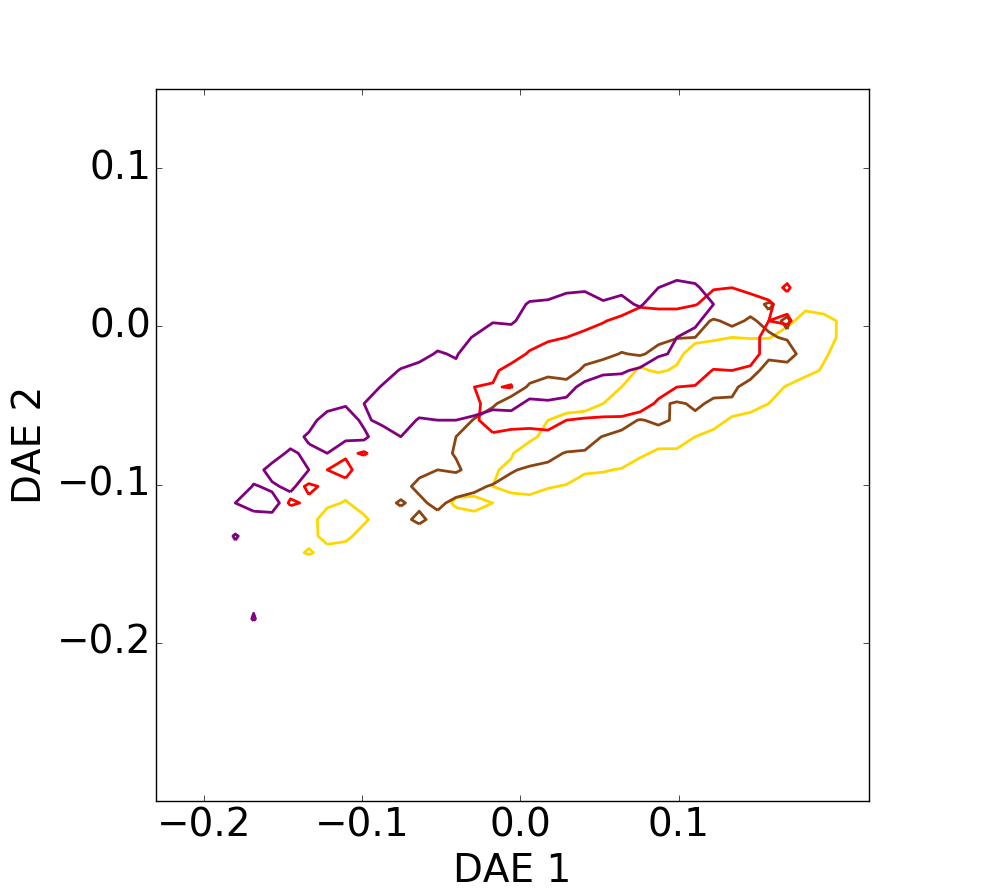}\\
\small{(a) Star forming - DAE}& \small{(b) Quiescent - DAE} \\
\includegraphics[width=0.24\textwidth]{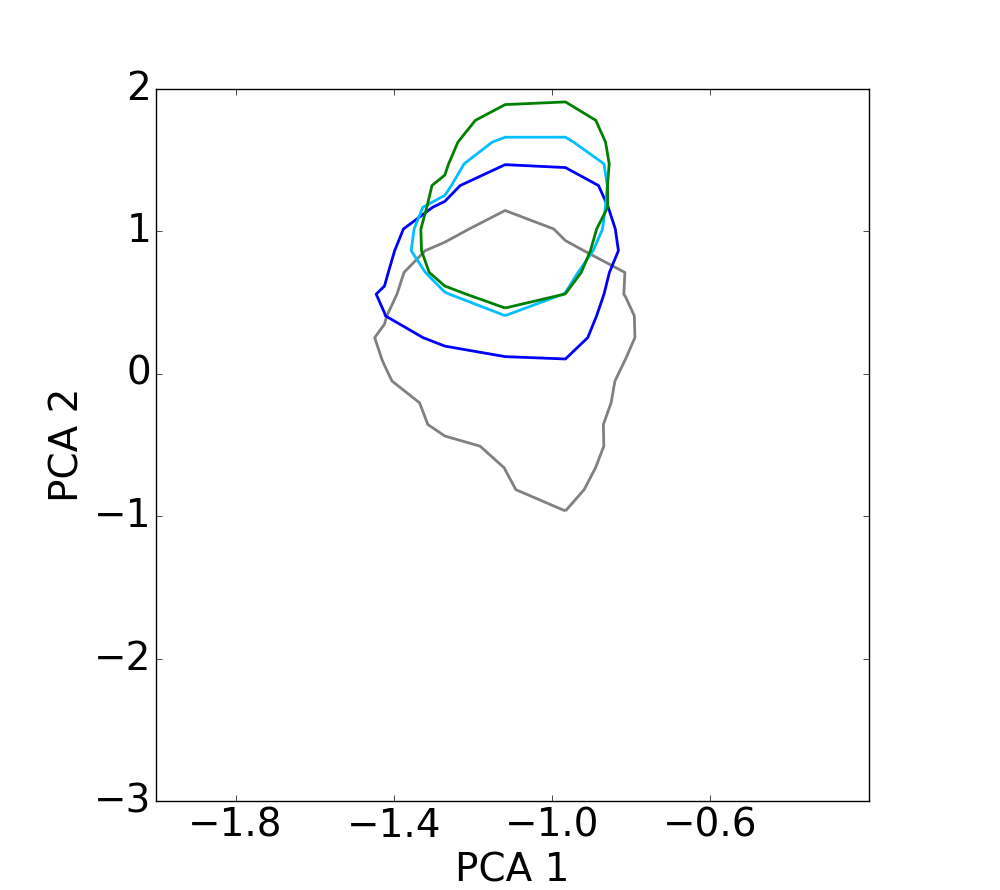}&
\includegraphics[width=0.24\textwidth]{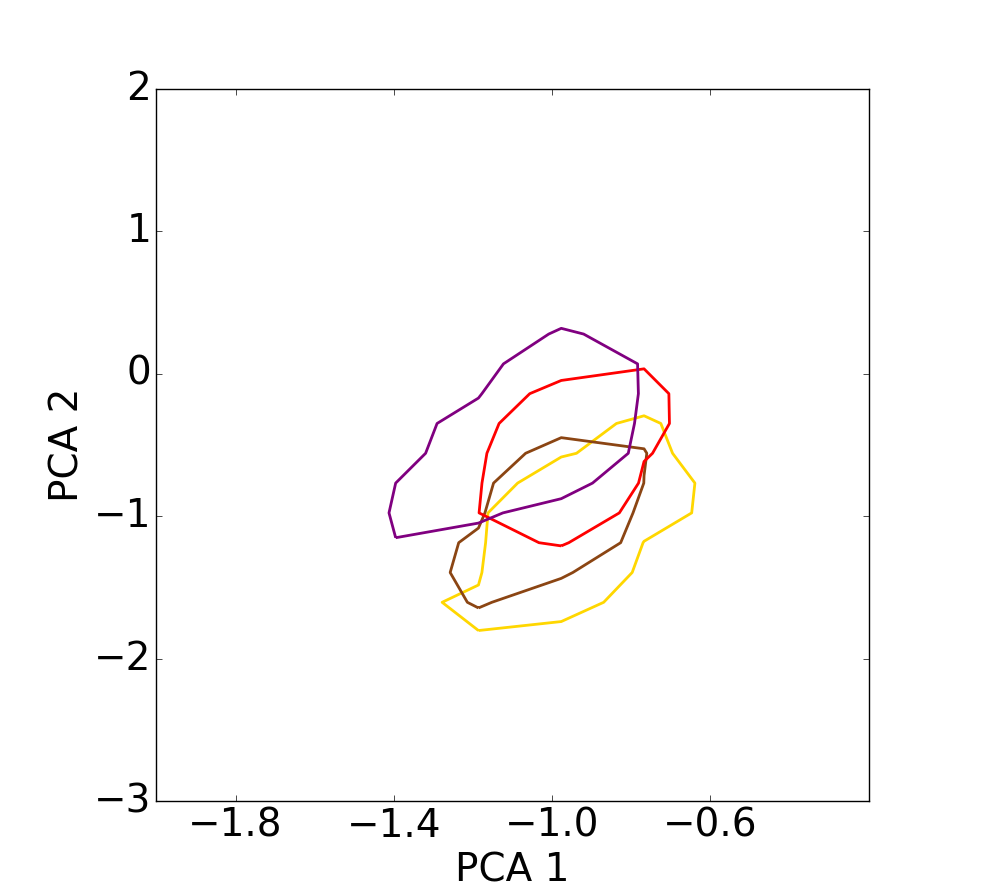}\\
 \small{(c) Star forming - PCA} & \small{(d) Quiescent - PCA}\\
\includegraphics[width=0.24\textwidth]{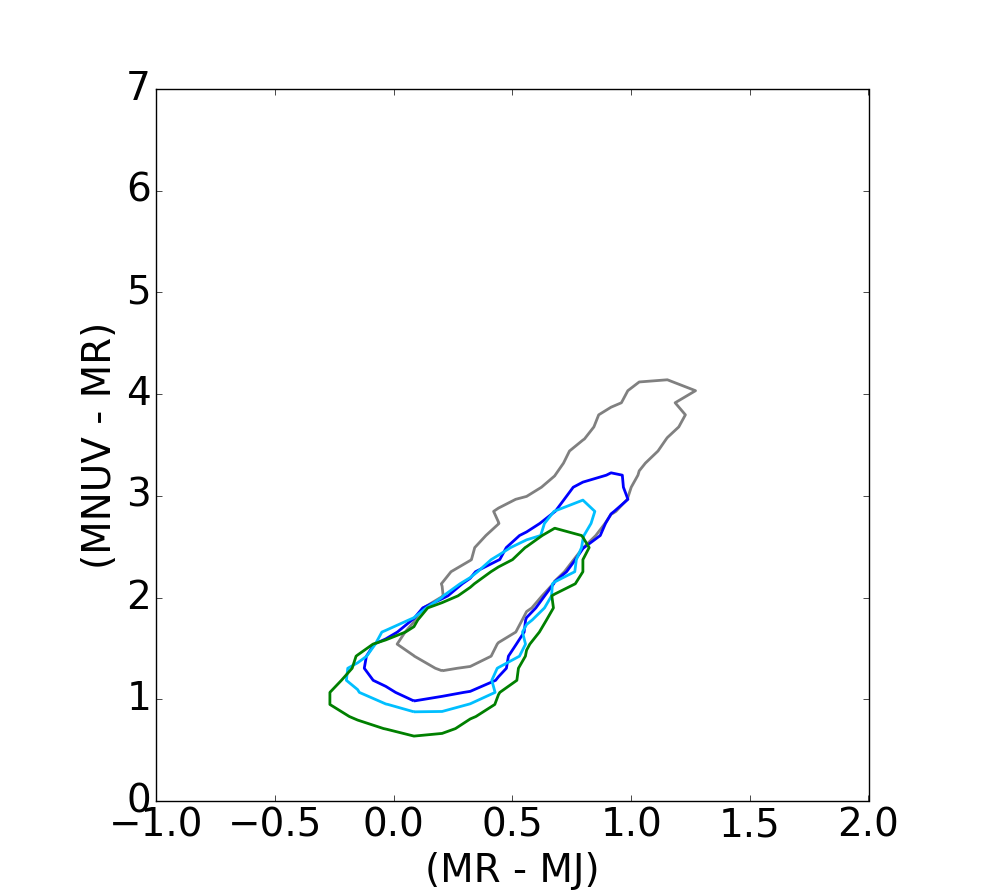}&
\includegraphics[width=0.24\textwidth]{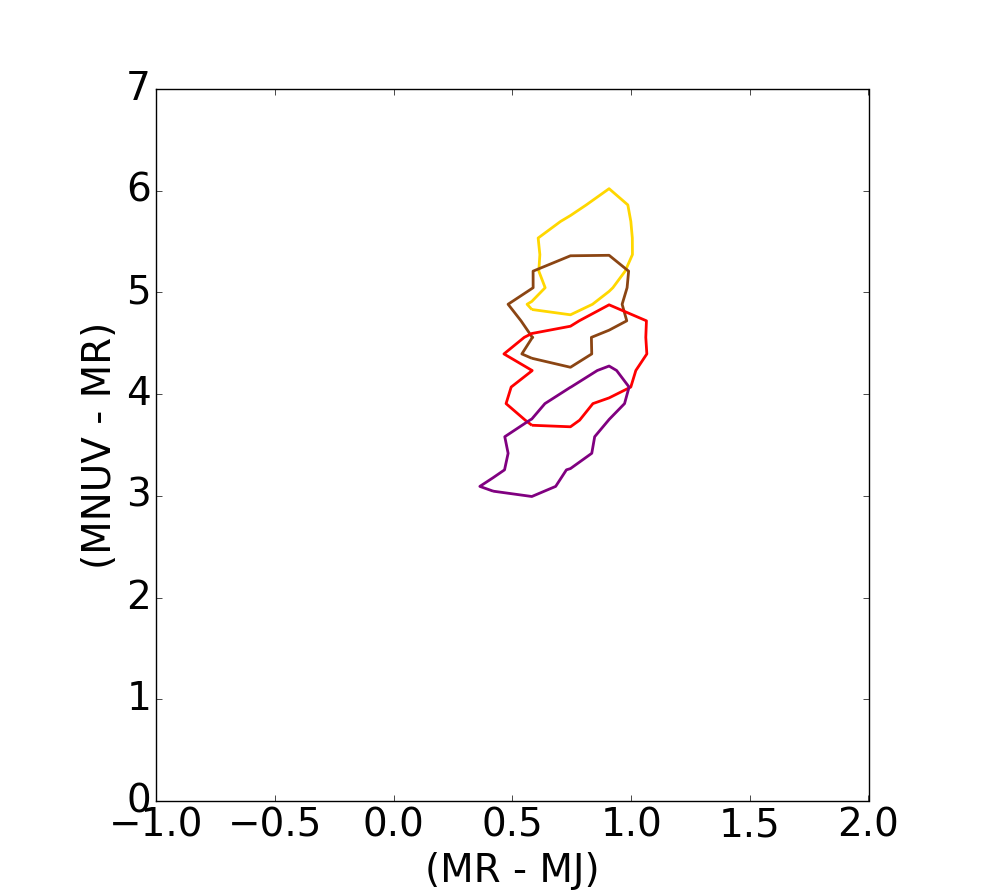}\\
 \small{(e) Star forming - RJNUV} & \small{(f) Quiescent - RJNUV}\\
\end{tabular}
\caption{Contour plots for different sSFR bins on the DAE diagram, on the PCA projection and on the RJ - NUVR plane.  The sSFR values have been split over four bins and the contour plots are depicted at $75\%$ of the samples in the different sSFR bins. The colors grow from grey to green for star-forming galaxies, and from yellow to purple for quiescent galaxies respectively. These figures highlight the trend observed in the DAE diagram, the sSFR values increase in the direction orthogonal to the mass. Moreover, the second component of the PCA representation strongly correlates with the sSFR.}
\label{fig:sSFR_bins}
\end{figure}

To further explore the impact of the galaxy star-forming activity  in the different diagrams, we conducted a similar analysis, taking into 
consideration this time the influence of the
%has been conducted for the
 specific star formation rate (sSFR), which measures  the SFR of galaxies normalized to  their  stellar mass. {As discussed above, the sSFR of star-forming sources at a fixed redshift slightly decreases with their mass, which could contribute to the variations observed in Fig.\,\ref{fig:mass_AE}.
 %the DAE as a function of stellar mass. 
 Furthermore, at fixed stellar mass we know that the average sSFR of star-forming sources has decreased by a factor $\sim$\,10 since redshift $z$\,$\sim$\,1  \citep[{e.g.},][]{Speagle14}, which may also influence how the galaxy properties evolve with cosmic time and how they transpose on the different planes}.
 Figure \ref{fig:sSFR_bins} displays the corresponding contour plots for the sSFR values taken from the COSMOS2015 catalog, split over four different bins on the DAE diagram, on the PCA diagram and on the RJ-NUVR diagram.  
Although the corresponding progression of the sSFR in the DAE and in the PCA diagrams is not as smooth as the observed relationship with the mass, one can notice a trend on the direction orthogonal to the evolution of the mass for the DAE and on the second component for the PCA. 

Unlike the galaxy stellar mass, the sSFR parameter is  relatively hard to constrain accurately when using only optical SED fitting as done in the COSMOS2015 catalog. This result in the representations thus has to be taken with caution. Yet, the observed trend could imply that the learning methods can also extract features linked to the on-going activity of star formation in galaxies as well as its relative importance over the stellar mass already assembled.

\subsection{Dust extinction}
\begin{figure}[h]
\begin{tabular}{cc}
\centering
\includegraphics[width=0.24\textwidth]{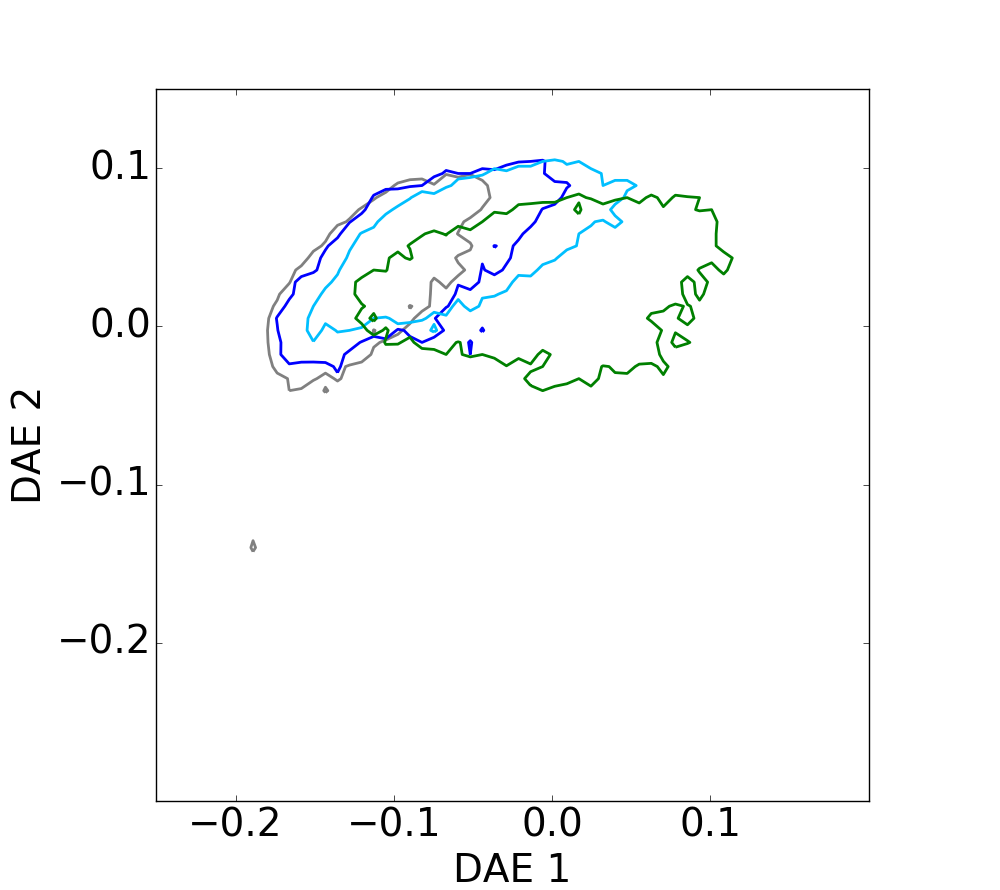}&
\includegraphics[width=0.24\textwidth]{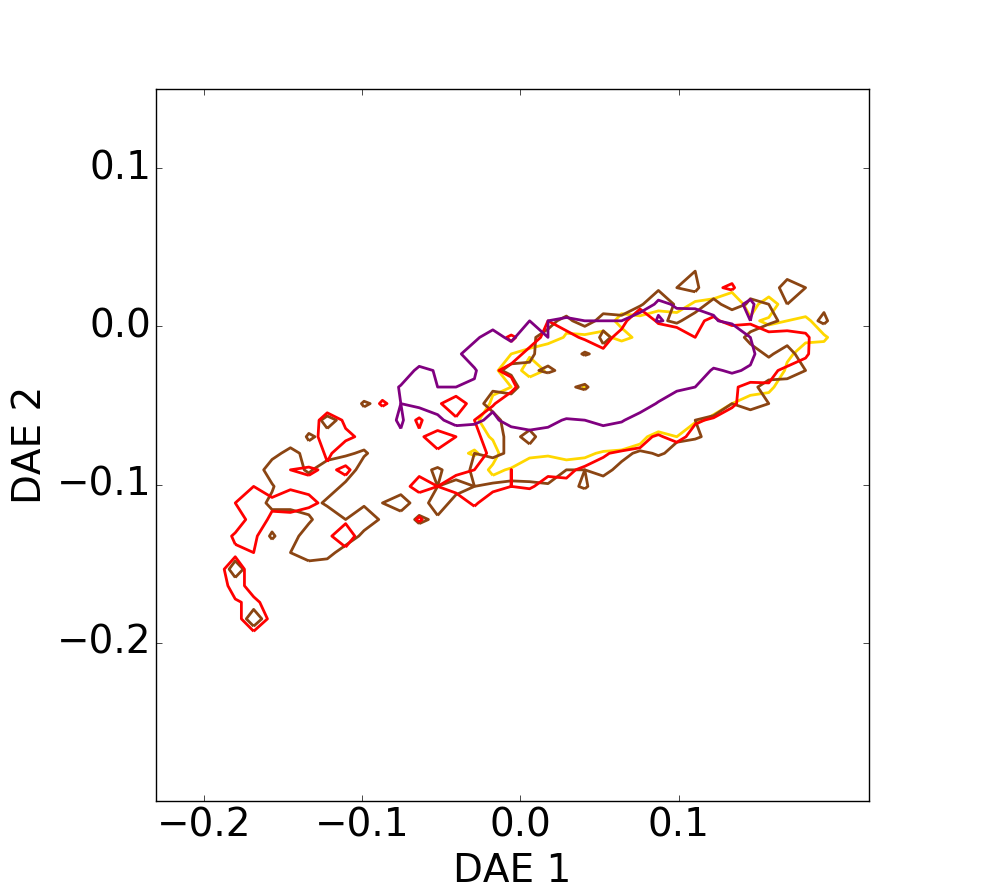}\\
\small{(a) Star forming - DAE}& \small{(b) Quiescent - DAE} \\
\includegraphics[width=0.24\textwidth]{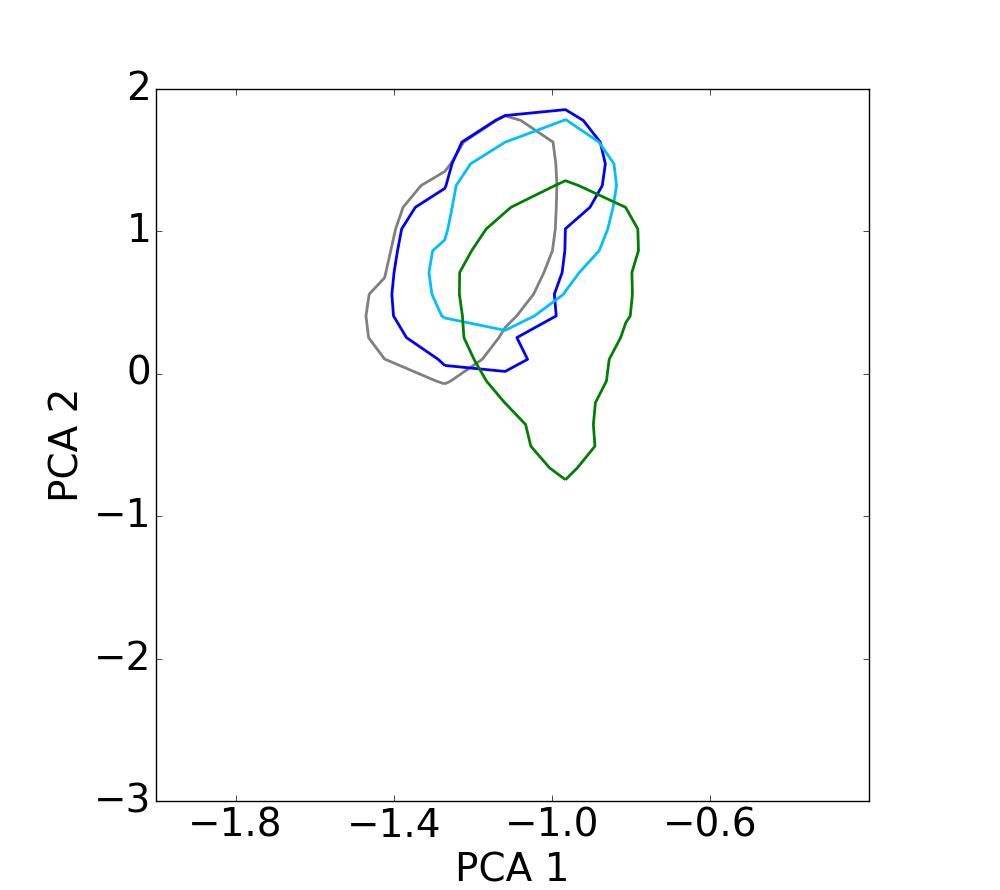}&
\includegraphics[width=0.24\textwidth]{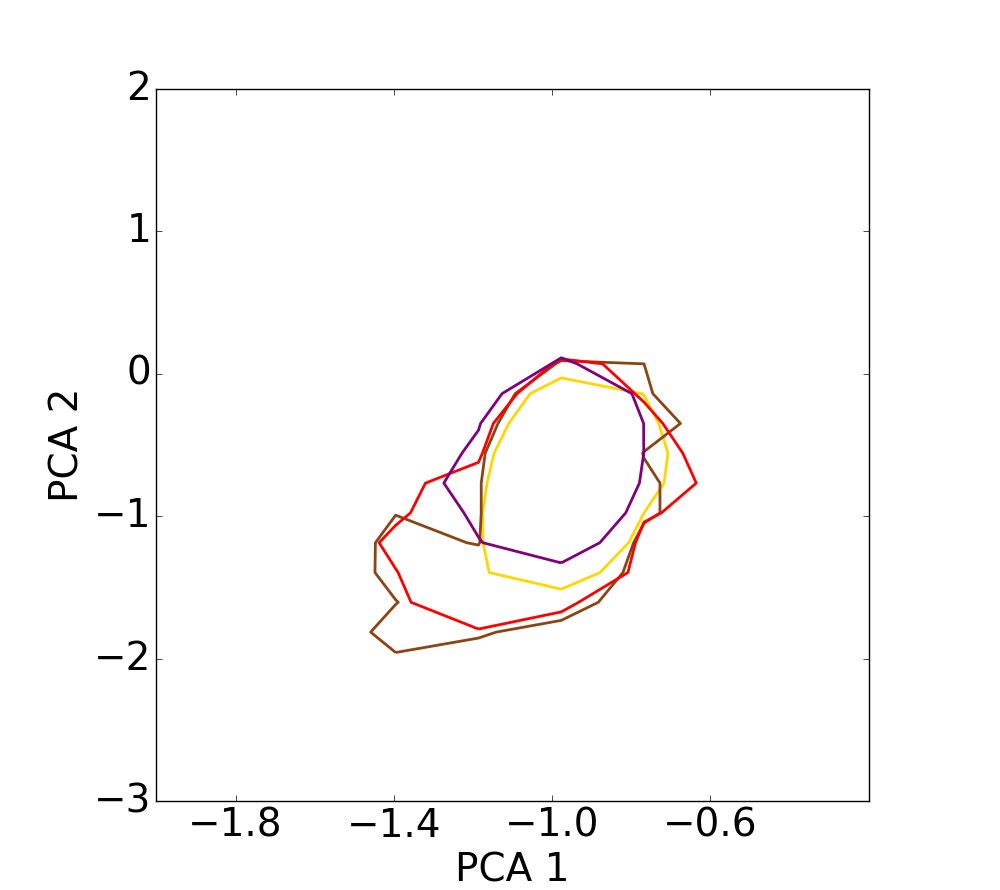}\\
 \small{(c) Star forming - PCA} & \small{(d) Quiescent - PCA}\\
\includegraphics[width=0.24\textwidth]{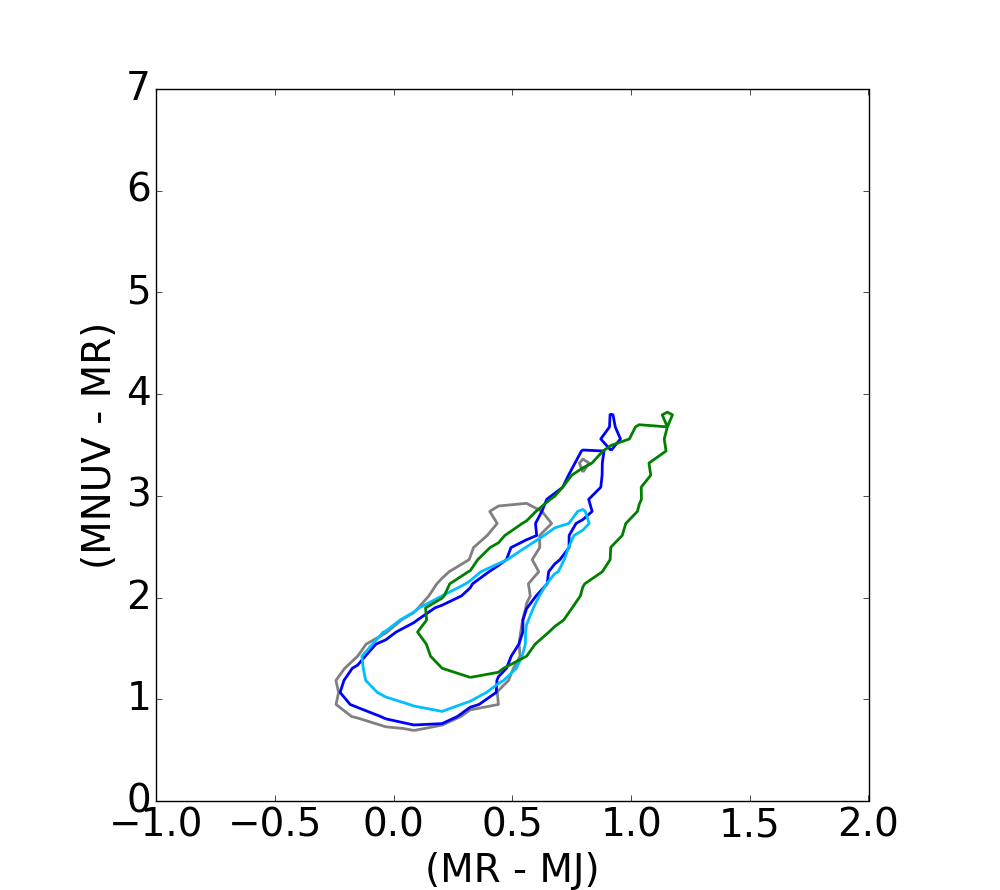}&
\includegraphics[width=0.24\textwidth]{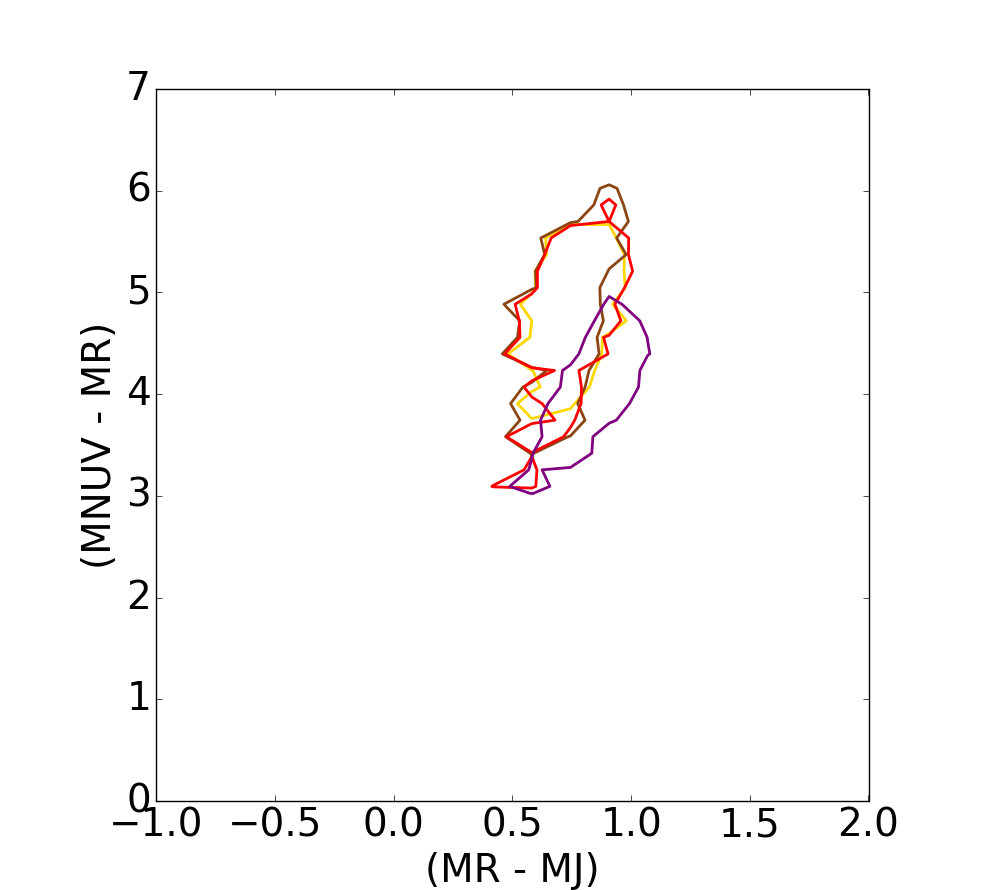}\\
 \small{(c) Star forming - RJNUV} & \small{(d) Quiescent - RJNUV}\\
\end{tabular}
\caption{Contour plots for different extinction bins on the DAE diagram, on the PCA projection and on the RJ - NUVR plane. This figure illustrates the distribution of the extinction range evolution. The values have been split over four bins and displayed from grey to green for star-forming populations, and from yellow to purple for quiescent galaxies. The contour plots are depicted at $75\%$ of the samples at each bin for the DAE and the RJ - NUVR diagrams.}
\label{fig:extinction_bins}
\end{figure}

Extinction characterizes the changes in the observed spectrum from the one that was originally emitted. Blue light is more attenuated owing to the light scattering of dust modifying the spectrum of galaxies to make them look redder. Figure\,\ref{fig:extinction_bins} illustrates {the DAE galaxy distribution split over four bins of dust extinction, as derived from the best SED fit in the COSMOS2015 catalog.}  The contour plots are depicted at 75\% for the DAE diagram, the PCA, and the RJ-NUVR diagram. Dust  extinction is very often negligible in passive sources, which explains why no clear trend neither appears on  the DAE (Fig.\,\ref{fig:extinction_bins}b) nor on the RJ-NUVR diagrams (Fig.\,\ref{fig:extinction_bins}d). However, in the case of star-forming sources, this parameter seems to correlate much more clearly  with the first DAE component. This is similar to the trend observed with the mass in Sect.\,4.2, and  is probably due to the known correlation linking dust attenuation and stellar mass in star-forming galaxies \citep{Zahid13}. The evolution suggested by the PCA diagram is very similar to the one conjectured for the DAE diagram.

 In the RJ-NUVR diagram, dust attenuation stretches the star-forming galaxy distribution to redder colors, hence to higher $NUV$$-$$r^+$ and $r^+-J$ values \citep{Williams09}. It  explains the elongated structure observed in Fig.\,\ref{fig:extinction_bins}c, especially for galaxies in the highest extinction bin.  However, 
the light blue curve corresponding to the third bin of the distribution is somehow more outspread on the DAE than in the RJ-NUVR diagram, which may again suggest  a higher efficiency of the DAE to discriminate galaxies, depending on their dust extinction.

\section{Discusion and conclusion}

In this article, we studied a radically new approach for the classification of galaxy SEDs, which is based on modern-day unsupervised machine-learning methods. Our first objective was to investigate the ability of these methods, namely denoising autoencoders, to retrieve information for galaxy rest-frame SEDs. Keeping in mind that, unlike currently used color/color diagrams, the DAE performs in an unsupervised manner and allows defining a data-driven diagram for galaxy classification. Consequently, it is important to note that the proposed DAE diagram already provides the following results:
\begin{itemize}
\item{Recovering the galaxy bimodality:} we showed that similarly to the standard color/color diagram, the DAE diagram features a clear separation between the star-forming and quiescent galaxy populations.\\
\item{Unveiling the redshift evolution:} unlike the standard diagrams, we point out that the DAE diagram provides a very neat continuous evolution of these galaxy populations with respect to their redshift. \\
\item{Extracting the mass dependency:} very interestingly, the DAE diagram exhibits a clear sequence of the galaxy distribution according to their mass.\\
\end{itemize}
These results highlight that the DAE is a promising tool for the classification of galaxy SEDs, and therefore could be a data-driven alternative to the standard color/color diagrams. Compared to DAE, PCA also results in an unsupervised representation with physical properties, such as mass and sSFR encoded in the first components, albeit this representation separates less other features (bimodality, redshift evolution) than DAE. The performances of the method are in good agreement with our current understanding of unsupervised machine learning methods: their ability to unfold complex non-linear features from the data. In the current application, this ability translates into the seeming capacity to unveil features that are associated with astrophysical quantities such as the galaxies' redshift or their mass.\\

\begin{figure}[h]
\centering
\includegraphics[width=0.55\textwidth]{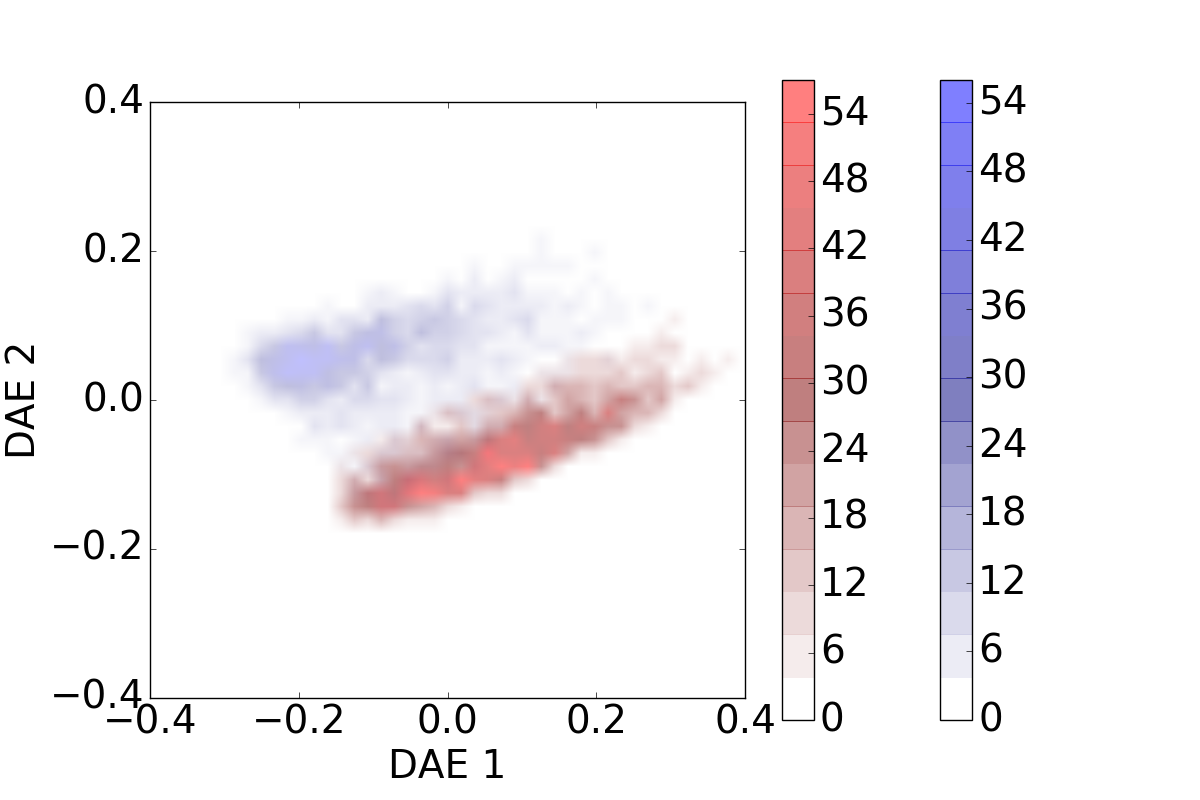}
\caption{Histogram of the test set on the DAE diagram built from apparent magnitudes. The proposed DAE diagram is capable of discriminating between star-forming from quiescent galaxies directly from the observed galaxy SEDs.}
\label{fig:observed}
\end{figure}

The proposed method is strongly sensitive to an accurate estimation of the redshift and to the SED-fitting stage. Since the DAE has  customarily been applied to the rest-frame SEDs  defined by best-fit parameterized SED models, some structures may appear on the DAE diagram owing to the modelling of the SEDs. This hinders drawing strong astrophysical conclusions based on our findings in the DAE diagram. For instance, the DAE diagram extracts a very specific population of low-redshift galaxies both in quiescent and star-forming galaxy populations, which does not specifically emerge in the color/color diagram. To the best of our understanding, we found no clear astrophysical explanations for the extraction of this particular low-redshift galaxy population. Further investigations will focus on applying the DAE directly on the observed galaxy SEDs. However, owing to the increased complexity of the data, this requires adapting the learning method to perform on observed SEDs and other models should be explored to take into account the SED variation with the redshift. Preliminary results on apparent magnitudes are presented in Fig. \ref{fig:observed}. To reduce the variability of the SEDs introduced by the redshift dependence, the denoising autoencoders were  first applied to a small subset of galaxies from the COSMOS catalog within a redshift between 0.8 and 1.1. The choice of this particular bin is due to the presence of a larger number of galaxy samples. We note that the DAE diagram is still capable of discriminating between the two different classes of populations, as illustrated in Fig. \ref{fig:observed}.\\

In addition, autoencoders are one of the building blocks of the so-called deep-learning methods. In this  article, we restrict our investigations to a single-layer autoencoder, which already provides a proof-of-concept of the applicability of DAE to galaxy SED classification. It is already known that increasing the number of autoencoder layers enables more flexibility for better capturing the underlying structures of the data. This will be investigated in our future work.

\begin{acknowledgements}
This work is funded by the DEDALE project (contract no. 665044) and LENA (ERC StG no. 678282) within the H2020 Framework Program of the European Commission.
\end{acknowledgements}

{\small 
\bibliographystyle{aa}
\bibliography{refs}}

\end{document}